\RequirePackage[l2tabu]{nag}	% Warns for incorrect (obsolete) LaTeX usage
\documentclass[a4paper,11pt,leqno,openbib]{memoir} %add 'draft' to turn draft option on (see below)

\usepackage{url}            % simple URL typesetting
\usepackage{booktabs}       % professional-quality tables
\usepackage{amsfonts}       % blackboard math symbols
\usepackage{nicefrac}       % compact symbols for 1/2, etc.
\usepackage{microtype}      % microtypography
\usepackage{multirow}
\usepackage{setspace}

\nonzeroparskip

\usepackage{datetime}
\usepackage{ifpdf}
\ifpdf
\fi
\newsubfloat{figure}
\newsubfloat{table}
\settrimmedsize{297mm}{210mm}{*}
\settypeblocksize{634pt}{448.13pt}{*} 
\setulmargins{4cm}{*}{*} 
\setlrmargins{*}{*}{1.5} 
\setmarginnotes{17pt}{51pt}{\onelineskip} 
\setheadfoot{\onelineskip}{2\onelineskip} 
\setheaderspaces{*}{2\onelineskip}{*} 
\checkandfixthelayout
\usepackage{fouriernc}
\usepackage[T1]{fontenc}

\OnehalfSpacing 
\setsecnumdepth{subsection} 
\maxsecnumdepth{subsubsection}
\usepackage{calc,soul,fourier}
\makeatletter 
\newlength\dlf@normtxtw 
\newsavebox{\feline@chapter} 
\newcommand\feline@chapter@marker[1][6cm]{%
	\sbox\feline@chapter{% 
		\resizebox{!}{#1}{\fboxsep=1pt%
			\colorbox{gray}{\color{white}\thechapter}% 
		}}%
		\rotatebox{90}{% 
			\resizebox{%
	        	\heightof{\usebox{\feline@chapter}}+\depthof{\usebox{\feline@chapter}}}% 
			{!}{\scshape\so\@chapapp}}\quad%
		\raisebox{\depthof{\usebox{\feline@chapter}}}{\usebox{\feline@chapter}}%
} 
\newcommand\feline@chm[1][6cm]{%
	\sbox\feline@chapter{\feline@chapter@marker[#1]}% 
	\makebox[0pt][c]{% aka \rlap
		\makebox[0cm][r]{\usebox\feline@chapter}%
	}}
\makechapterstyle{daleifmodif}{

	\renewcommand\printchapternum{\null\hfill\feline@chm[2.5cm]\par}

} 
\makeatother 
\chapterstyle{daleifmodif}
\makepagestyle{myvf} 
\makeoddfoot{myvf}{}{\thepage}{} 
\makeevenfoot{myvf}{}{\thepage}{} 
\makeheadrule{myvf}{\textwidth}{\normalrulethickness} 
\makeevenhead{myvf}{\small\textsc{\leftmark}}{}{} 
\makeoddhead{myvf}{}{}{\small\textsc{\rightmark}}
\newcommand{\clearemptydoublepage}{\newpage{\thispagestyle{empty}\cleardoublepage}}
\makeindex
\usepackage{import}

\usepackage{lipsum}					%Needed to create dummy text
\usepackage{amsfonts} 					%Calls Amer. Math. Soc. (AMS) fonts
\usepackage[centertags]{amsmath}			%Writes maths centred down
\usepackage{stmaryrd}					%New AMS symbols
\usepackage{amssymb}					%Calls AMS symbols
\usepackage{amsthm}					%Calls AMS theorem environment
\usepackage{newlfont}					%Helpful package for fonts and symbols
\usepackage{layouts}					%Layout diagrams
\usepackage{graphicx}					%Calls figure environment
\usepackage{longtable,rotating}			%Long tab environments including rotation. 
\usepackage[applemac]{inputenc}			%Needed to encode non-english characters 
\usepackage{colortbl}					%Makes coloured tables
\usepackage{wasysym}					%More math symbols
\usepackage{mathrsfs}					%Even more math symbols
\usepackage{float}						%Helps to place figures, tables, etc. 
\usepackage{verbatim}					%Permits pre-formated text insertion
\usepackage{upgreek }					%Calls other kind of greek alphabet
\usepackage{latexsym}					%Extra symbols

\usepackage{url}						%Supports url commands
\usepackage[spanish,english]{babel}		%For languages characters and hyphenation
\usepackage{color}                    				%Creates coloured text and background
\usepackage[colorlinks=true,
            urlcolor=blue,
		    linkcolor=black,
		    citecolor=black]{hyperref}              %Creates hyperlinks in cross references
\usepackage{apacite}                    % bib format

\usepackage{memhfixc}					%Must be used on memoir document 
\usepackage{enumerate}					%For enumeration counter

\usepackage{footnote}					%For footnotes
\usepackage{microtype}					%Makes pdf look better.
\usepackage{rotfloat}					%For rotating and float environments as tables, 
\usepackage{alltt}

\usepackage[version=0.96]{pgf}			%PGF/TikZ is a tandem of languages for producing vector graphics from a 
\usepackage{tikz}						%geometric/algebraic description.
\usetikzlibrary{arrows,shapes,snakes,
		       automata,backgrounds,
		       petri,topaths}				%To use diverse features from tikz		
\usepackage{etoolbox}
\newrobustcmd{\nunimg}{\includegraphics[height=\fontcharht\font`\B]{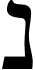}}

\newcommand{\pgftextcircled}[1]{                                                                    %Defines encircled text
    \setbox0=\hbox{#1}%
    \dimen0\wd0%
    \divide\dimen0 by 2%
    \begin{tikzpicture}[baseline=(a.base)]%
        \useasboundingbox (-\the\dimen0,0pt) rectangle (\the\dimen0,1pt);
        \node[circle,draw,outer sep=0pt,inner sep=0.1ex] (a) {#1};
    \end{tikzpicture}
}
\newcommand{\blackged}{\hfill$\blacksquare$}
\newcommand{\whiteged}{\hfill$\square$}
\newcounter{proofcount}

\let\oldsqrt\sqrt
\def\sqrt{\mathpalette\DHLhksqrt}
\def\DHLhksqrt#1#2{%
\setbox0=\hbox{$#1\oldsqrt{#2\,}$}\dimen0=\ht0
\advance\dimen0-0.2\ht0
\setbox2=\hbox{\vrule height\ht0 depth -\dimen0}%
{\box0\lower0.4pt\box2}}
\newcommand{\mycaption}[2][\@empty]{
	\captionnamefont{\scshape} 
	\changecaptionwidth
	\captionwidth{0.9\linewidth}
	\captiondelim{.\:} 
	\indentcaption{0.75cm}
	\captionstyle[\centering]{}
	\setlength{\belowcaptionskip}{10pt}
	\ifx \@empty#1 \caption{#2}\else \caption[#1]{#2}
}
\newcommand{\mysubcaption}[2][\@empty]{
	\subcaptionsize{\small}
	\hangsubcaption
	\subcaptionlabelfont{\rmfamily}
	\sidecapstyle{\raggedright}
	\setlength{\belowcaptionskip}{10pt}
	\ifx \@empty#1 \subcaption{#2}\else \subcaption[#1]{#2}
}
\usepackage{lettrine}
\newcommand{\initial}[1]{%
	\lettrine[lines=3,lhang=0.33,nindent=0.0em]{
	\color{gray}
    {\textsc{#1}}}{}}
            
\newcommand{\mycomment}[1]  {}
\theoremstyle{plain}

\theoremstyle{plain}

\theoremstyle{plain}
\theoremstyle{definition}

\theoremstyle{plain}

\theoremstyle{plain}

\theoremstyle{plain}

\begin{document}
%\tracingall
	
% UoB guidlines:
%
% Preliminary pages
% 
% The five preliminary pages must be the Title Page, Abstract, Dedication
% and Acknowledgements, Author's Declaration and Table of Contents.
% These should be single-sided.
% 
% Table of contents, list of tables and illustrative material
% 
% The table of contents must list, with page numbers, all chapters,
 % sections and subsections, the list of references, bibliography, list of
% abbreviations and appendices. The list of tables and illustrations
% should follow the table of contents, listing with page numbers the
% tables, photographs, diagrams, etc., in the order in which they appear
% in the text.
% 
\frontmatter
\pagenumbering{roman}
%
%
% File: Title.tex
% Author: Tomer Nussbaum
% Description: Contains the title page
%
% UoB guidelines:
% 
% At the top of the title page, within the margins, the dissertation should give the title and, if 
% necessary, sub-title and volume number. If the dissertation is in a language other than English, the 
% title must be given in that language and in English. The full name of the author should be in the 
% centre of the page. At the bottom centre should be the words ?A dissertation submitted to the 
% University of Bristol in accordance with the requirements for award of the degree of ? in the 
% Faculty of ...?, with the name of the school and month and year of submission. The word count of 
% the dissertation (text only) should be entered at the bottom right-hand side of the page.
%
%
\begin{titlingpage}
\begin{SingleSpace}
\calccentering{\unitlength} 
\begin{adjustwidth*}{\unitlength}{-\unitlength}
\vspace*{0mm}
\begin{center}
% \rule[0.5ex]{\linewidth}{2pt}\vspace*{-\baselineskip}\vspace*{3.2pt}
% \rule[0.5ex]{\linewidth}{1pt}\\[\baselineskip]
{\HUGE  The Nun Path \texorpdfstring{(\nunimg)}- } \\[\baselineskip]
{\huge  The Evolution and Quenching of Satellite Galaxies}\\[\baselineskip]
% \rule[0.5ex]{\linewidth}{1pt}\vspace*{-\baselineskip}\vspace{3.2pt}
% \rule[0.5ex]{\linewidth}{2pt}\\

\vspace{10mm}
{\normalsize By} \\
\vspace{1.5mm}
{\normalsize\textsc{Tomer Nussbaum}}\\
\vspace{4mm}
\normalsize Supervisor\\
\vspace{1.5mm}
\normalsize\textsc{Prof. Avishai Dekel}\\
\vspace{9mm}
\includegraphics[scale=0.15]{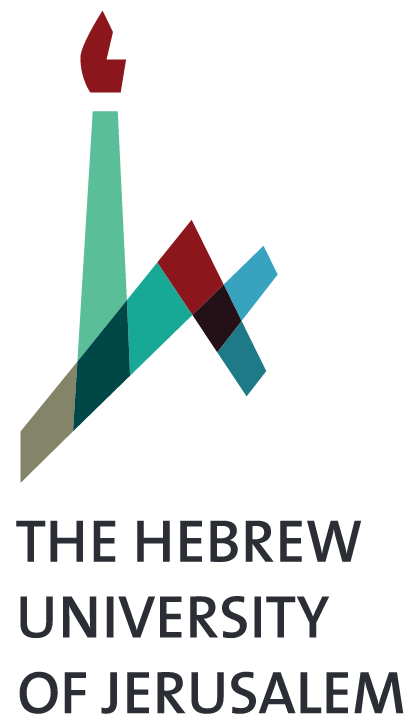}\\

\vspace{17mm} \scriptsize
In Memory of the great Nussbaum, \\
a Holocaust survivor, \\
badly wounded in his freedom fight for the Jewish nation, \\
and the establishment of the miracle of the middle east, the state of Israel. \\
Nurtured and cared for orphans and persecuted Jewish survivor youth at the Youth Aliyah. Manager of Mount Herzl. \\
A true Zionist, a family man, and a kind man to all (with plenty of jokes for any event). \\
\vspace{2mm}
A good husband, a great father, and a wonderful grandpa - my grandpa. \\
In our hearts you remain, \\
Kalman Nussbaum (1928-2020) \\

\vspace{12mm}
{\normalsize The Racah Institute of Physics\\
\textsc{The Hebrew University of Jerusalem}}\\
\vspace{11mm}
\begin{minipage}{11cm}
\scriptsize A dissertation submitted to the Hebrew University of Jerusalem as a partial fulfillment of the requirements of the degree of \textsc{Master of Science} at The Racah Institute of Physics.
\end{minipage}\\
\vspace{5mm}
{\large\textsc{December 2018}}
\vspace{12mm}
\end{center}
% \begin{flushright}
% % {\small Word count: nine thousand eight hundred and fifty six}
% \end{flushright}
\end{adjustwidth*}
\end{SingleSpace}
\end{titlingpage}
\clearemptydoublepage
%
%
% File: abstract.tex
% Author: Tomer Nussbaum
% Description: Contains the text for thesis abstract
%
% UoB guidelines:
%
% Each copy must include an abstract or summary 
% of the dissertation in not more than 300 words, 
% on one side of A4, which should be single-spaced in a
% font size in the range 10 to 12. 

% If the dissertation is in a language other than English,
% an abstract in that language and an abstract in English must
% be included.

\chapter*{Abstract}

\initial{T}he galaxy quenching process, a process in which a galaxy stops forming stars is a crucial stage in galaxy life. Two primary mechanisms for quenching are possible: halo mass quenching of central galaxies in their dark-matter halos and environmental quenching of satellite galaxies.
This thesis will describe the satellite galaxies (SG) quenching process and its primary causes. The analysis contains a study of a large sample of 118 SGs within the Vela zoom-in cosmological simulations, identified with a specific SG merger tree algorithm. \\
\\
We consider a SG stellar mass sample which ranges between $10^6-10^{9.5}\ M_{\odot}$, orbiting and accreting onto host galaxies with stellar masses between $10^{9.5}-10^{11}\ M_{\odot}$ at low redshift (z<1). 
We find that the quenching evolves through a typical path in the diagram of specific star formation (sSFR) vs. inner stellar surface density ($\Sigma_{\star, 0.5\textrm kpc}$), 
with the inner surface density defined as the density within 0.5kpc from the center of mass of the SG. 
Three discrete phases characterize this path: 
1. Halt in gas accretion with SG compaction at high sSFR as the SG keeps forming stars 
2. Gas removal and rapid drop in the sSFR at the peri-center of its orbit within the host halo
3.  Stellar heating and stripping that may lead to coalesce with the halo center, an ultra-diffuse galaxy (UDG), a compact elliptical galaxy (eCg) or a globular cluster (GC).\\
\\
We find that the main drivers of SG evolution are ram pressure stripping at the initial stages and tidal forces at the final stages, aided by starvation, suppression of gas accretion.
We also show that other processes, such as gas depletion by star formation or stellar feedback are secondary. 
We construct an analytic model that successfully reproduces this SG evolution. \\
\\
Moreover, we present an innovative method for constructing galaxy stellar merger trees in simulations which solves systemic errors in both identification and follow up of galaxies.
Two new significant results of this method presented here: 
a. Decoupling of dark-matter from the stellar component in SGs.
b. Compact elliptical SG formation. 
We present an assembly of these results and others in a fast and robust catalog for further research.

\clearpage
\clearemptydoublepage
%
%
% file: dedication.tex
% author: Tomer Nussbaum
% description: Contains the text for thesis dedication
%

\chapter*{Dedication and acknowledgements}
\begin{SingleSpace}
\setlength{\parindent}{0pt}
\initial{I} would like to thank the universe for existing and giving me the chance to study it.

\begin{center}
Thank you oh mighty universe!
\end{center}

\vspace{.707106781cm}
\vspace{.707106781cm}

Equivalent in importance! I wish to thank my supervisor, Prof. Avishai Dekel, who taught me core physics work (``You should make the right, simplest assumptions''). It was a privilege. 

For my awesome group co-researchers: Dr. Yuval Birnboim, Sharon Lapiner, Dr. Nir Mandelker, Omry Ginzburg, 
Dr. Jonathan Freundlich, Dr. Nicolas Cornuault, Dr. Fangzhou Arthur Jiang, Dr. Sarkar Kartick and Dr. Yakov Faerman,
As well as the big Astro-center, many people from many backgrounds with many adventures together, Thank you guys.

To my Alpha program pupils and guests, dearest galactic hunters! Naftali Deutch, Zohar Milman, Oz Weizer, Raphael Buzaglo, Noam Chouchena and Adam Beili. Remember that, ``Superhuman effort isn't worth a damn unless it achieves results''. So where are they? ;)

\hspace{0.36787944117cm}

I want to thank my friends, especially Or Sharir, Odelia Teboul, Ragit Moshe and Yohai Devir.

\hspace{2.718281828459045cm}

To my dear family, my stunning siblings: Bar, Raz and Netta, \\
and to my mom and dad, Rachel and Ran.
\\
Thanx ;x) 

\vspace*{3.141592cm}

For all who are far beyond the known universe... I dedicate this work to you. :)

\end{SingleSpace}
\clearpage
\clearemptydoublepage
\renewcommand{\contentsname}{Table of Contents}
\maxtocdepth{subsection}
\tableofcontents*
\addtocontents{toc}{\par\nobreak \mbox{}\hfill{\bf Page}\par\nobreak}
\clearemptydoublepage
%
% \listoftables
% \addtocontents{lot}{\par\nobreak\textbf{{\scshape Table} \hfill Page}\par\nobreak}
% \clearemptydoublepage
% %
% \listoffigures
% \addtocontents{lof}{\par\nobreak\textbf{{\scshape Figure} \hfill Page}\par\nobreak}
% \clearemptydoublepage
%
%
% The bulk of the document is delegated to these chapter files in
% subdirectories.
\mainmatter
%
%
% File: chap01.tex
% Author: Tomer Nussbaum
% Description: Introduction and prior work
%
\let\textcircled=\pgftextcircled
\chapter{Physical Introduction}
\label{chap:p_intro}
\initial{C}osmology, the field of formation and evolution of the universe has been an existential need since the days of the first man across a variety of cultures, some independently growing oceans away from each other. The questions of ``who and what we are?'' coupled with ``where and in what we are?'' always appeared.

In the last century and especially these following 25 years, we witness a fantastic rise in our ability to answer the question ``In what we are?''. 
With the arrival of new mesmerizing observation techniques we can observe the early formation stages of the universe and watch galaxies at z $\sim$ 11.1, 400 Myr after the big bang and have high quality observations of enormous number of galaxies at z $\sim$ 2-3 (2-3 Gyr after the big bang) which is the peak of formation of galaxies in the universe.
These observations aid us to solve the core question regarding the galaxy and structure evolution of the universe and also learn about the initial conditions of the universe itself. From initial fluctuations in the cosmic background radiation to the breathtaking spiral galaxies vs. elliptical giants we see around our Milky Way galaxy. 

On the broad aspect, galaxies and clusters of galaxies are our only sign today to the presence of dark-matter or a change in the gravity force on large scales, and one of our signs to the presence of dark energy. Dark-matter and dark energy together hold 95\% of the energy of the universe! Therefore, today, those issues are part of the core questions in physics and science.

The dramatic change in observations does not stand alone, as hand in hand with the advancement in observation tech, came an exponential growth in computing power and numerical algorithm methods, allowing us simulating Megaparsecs wide cubes of the universe in details. These simulations became a useful tool, and with this tool, we explore an assortment of new theories and find new phenomena. A short description on our numeric methods is given at \S\ref{chap:n_intro}.

Finally, we wish to point out that the following work assumes $\Lambda CDM$ Model, the current cosmological standard model which is considered to be the simplest model that can explain the observed phenomena, from the cosmic microwave background (CMB) radiation properties to the structural evolution of the universe, including hydrogen abundance and universe expansion acceleration.

\section{Galaxy Bi-modality}
\label{sec:intro-Bi-modality}

\begin{figure}[h] \centering
\includegraphics[width=0.9\linewidth]{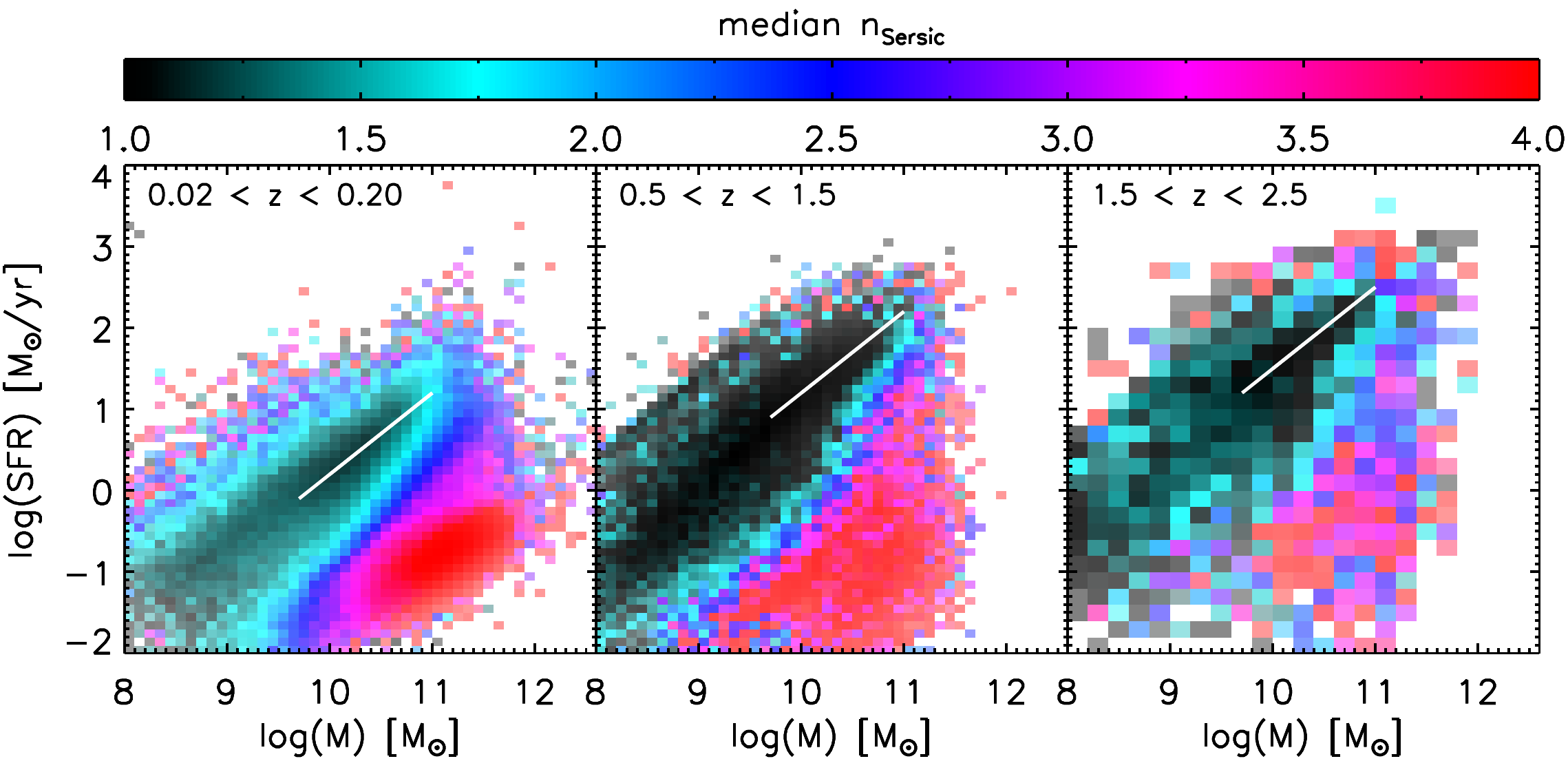}
\caption{\small \textbf{SFR - stellar Mass diagram colored by the surface brightness profile shape.\hspace{1cm} By \protect\citeA{Wuyts2011}.} A ``structural main sequence'' is present at all observed epochs. Star-forming galaxies on the main sequence (MS) are well characterized by exponential disks, and the quenched (QG) galaxies at all epochs are elliptical. \hspace{0.5cm}
The quenched population include galaxies below $10^{10.5} M_{\odot}$, the mass in which galaxies become quenched by the ``mass quenching'' process, showing another quenching mechanism is needed to explain the less massive galaxies quenching process.}
\label{fig:Wuyts2011_gal_main_seq}
\end{figure}

The primary ``hot'' subject today in the galaxy formation field can be shown in the riddle hiding of one plot. The ``Star formation rate (SFR) vs Stellar mass galaxy distribution colored by the galaxy shape'', shown in fig \ref{fig:Wuyts2011_gal_main_seq} (Sometimes there are variations of this plot by sSFR instead of SFR, when sSFR is the SFR divided by the stellar mass or by surface density instead of mass, in this work we will use sSFR vs. surface density of the galaxies). Going back to $z \sim 2$ we can already see two distinctive classes of galaxies: star-forming (SF) galaxies (``The main sequence'', MS) and not forming stars, quenched (QG) galaxies (Red \& dead region).

Today, the research focuses on ``field galaxies'' which galaxies that are alone in the sky (not in groups or clusters) or they are massive by ten times or more from their nearest neighbors. The field galaxy evolution from a star-forming stage to a quenched stage is often addressed as halo mass quenching.

\begin{figure}[H] \centering
\includegraphics[width=0.9\linewidth]{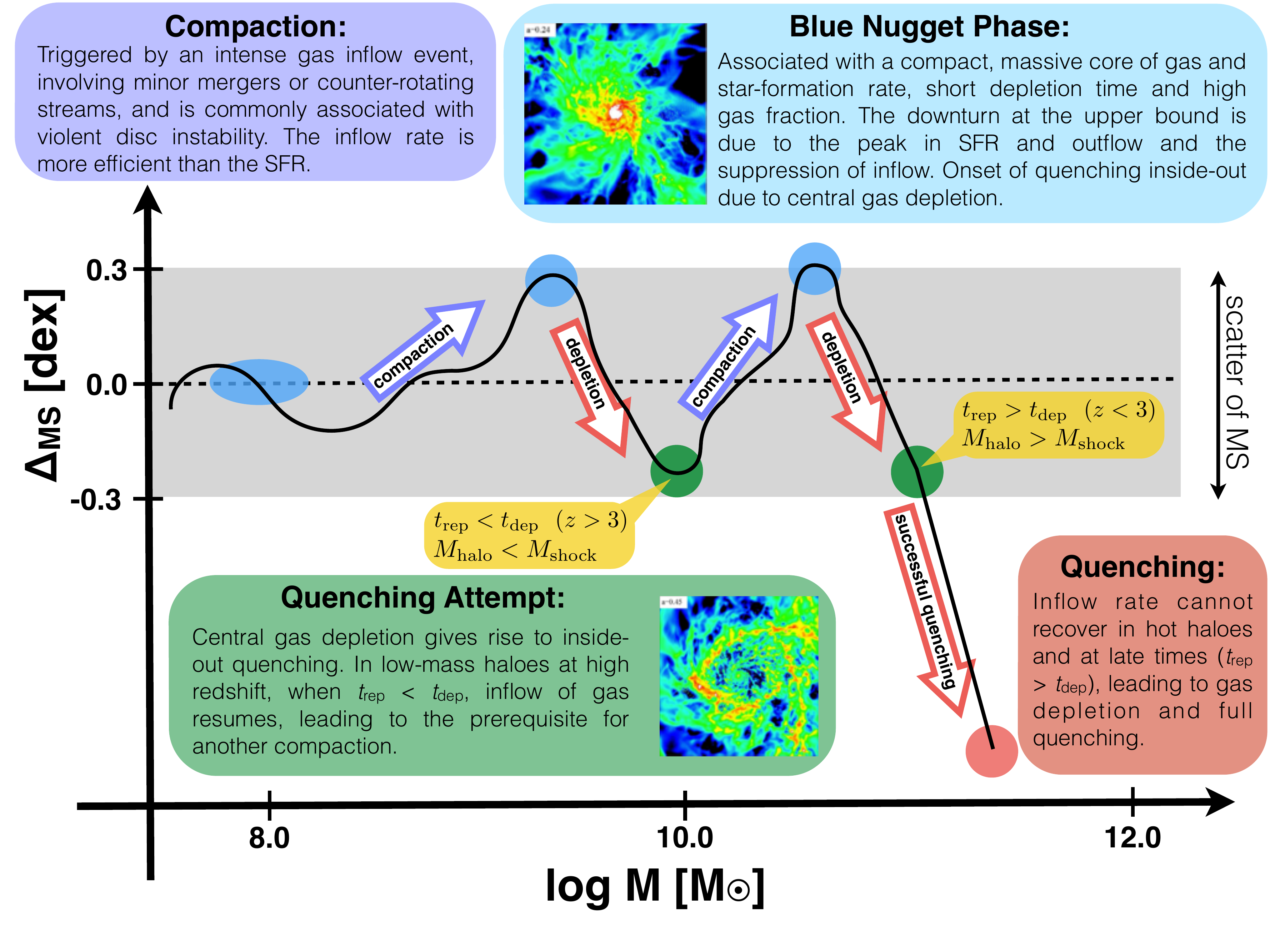}
	\caption{\small \textbf{Field Galaxies main sequence evolution and mass quenching, by \protect\citeA{Tacchella2016a}:} Sketch of the self-regulated evolution along the star-forming galaxies (SFG) main sequence (MS). SFG is confined to a narrow MS before they quench. During this evolution, the galaxy lives through one or more blue nugget phases during which a minimum in gas depletion time and a maximum in the gas fraction reached. The blue nugget phases are followed by gas depletion inside-out. These quenching attempts fail for low halo masses and at high redshifts since the recovered inflow rate triggers a new episode of compaction and high star formation. At high halo masses (hot halo), the inflow rate cannot recover, and the galaxies cease its star formation activity.}
 	\label{fig:Tacchella2016a_gal_MS_cartoon}
\end{figure}

One of the most advanced studies today \cite{Dekel_Lapiner, Tacchella2016a} suggests that a galaxy is having compacting fluctuations on the MS until it reaches a stellar mass of about $10^{10.5}M_\odot$, there the galaxy suffers a sharp drop in its gas fraction and therefore lacks the material to form new stars and becomes quenched. The process itself happens due to a heating of the gas by the stars, a powerful process when the stars are compact. A nice explanation can be seen at S\ref{fig:Tacchella2016a_gal_MS_cartoon}.

This process also changes the shape and color of the galaxy from a blue disc shape (blue as new stars a bluish) to a red elliptical shape galaxy (red as old stars are red).

Two other significant phenomena are also in the heart of the current study. A supermassive black hole that exists in the center of each galaxy and when active (AGN), heats the gas and throws it outside the galaxy and a merger with another galaxy. Both can be triggers for compaction or to influence each other. 

\section{Satellite galaxies}
\label{sec:intro-sat}
In the lifetime of galaxies, the lion's share of galaxies becomes satellites of another bigger central galaxy and even merge with it.  This process is a crucial factor of the structure evolution in the universe  \cite{Press_Schechter1974, White_Rees1978, Bardeen_Bond1986, Lacey_Cole1993}. We define Satellite Galaxy (SG) as a galaxy in a virial dark-matter halo of a bigger galaxy. We observe SGs by the projected distance between SG and its central galaxy. For example, we observe satellite galaxies around the Milkey Way. These observations not always fitting our theoretical or numerical simulation models, as we do not observe enough SGs around the Milkey Way as we expect from theoretical models as can be found on missing SG problem \cite{Klypin1999, Moore1999, Simon_Geha2007}.
Also, as discussed earlier, it has a lead rule in halo mass quenching of galaxies \cite{Dekel_Lapiner}.  Therefore it is crucial to understand SG evolution to complete the theory of galaxy formation and evolution.

On the Quenching aspect: The purpose halo mass quenching described at \S\ref{sec:intro-Bi-modality} cannot explain the full quenching phenomena. Fig \ref{fig:Wuyts2011_gal_main_seq} shows that there are quenched galaxies well below a stellar mass of $10^{10.5}M_\odot$,  as halo mass quenching occurs on galaxies at above $10^{10.5}M_\odot$ it means another quenching process exists \cite{Wuyts2011}. This process is called environmental quenching, and it is identified as satellite quenching \cite{Peng2010, Peng2012, Wetzel2013, Kova2013, vandenBosch2008}.

Hence, satellite galaxy evolution and quenching mechanism are important to the galaxy formation and evolution research. 

\begin{figure}[H] \centering
\includegraphics[width=0.5\linewidth]{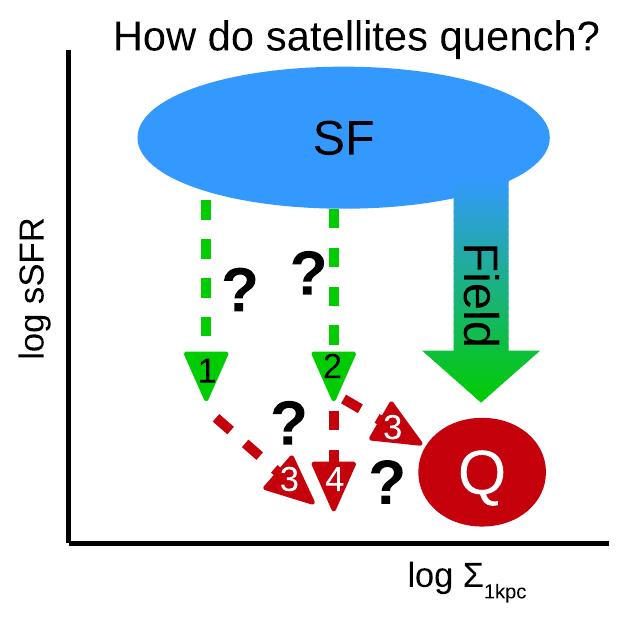}
\includegraphics[width=\linewidth]{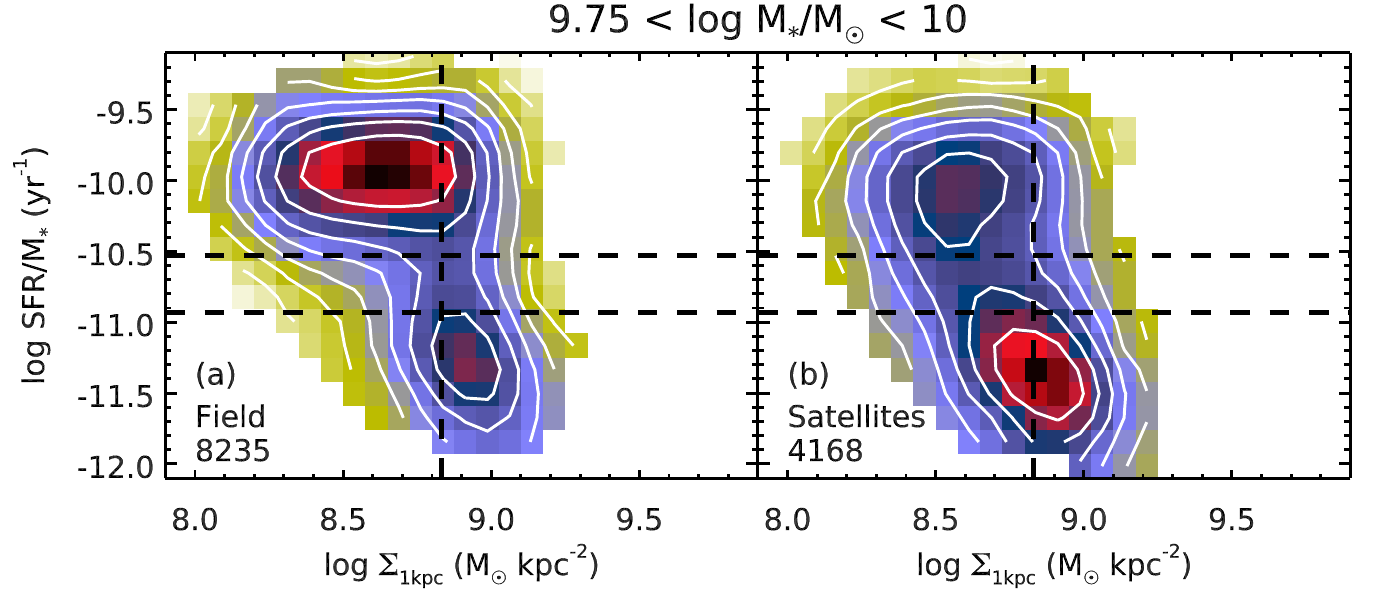}
\caption{\small \textbf{\protect\citeA{Woo2017} Satellite Galaxies proposed evolution path.} \hspace{0.5cm}
Upper fig: Schematic diagram illustrating quenching paths in sSFR-$\Sigma_{\mathrm 1kpc}$ space at constant stellar mass.  The region marked ``SF'' refers to the ``star-forming'' population, and the region marked ``Q'' refers to the ``quenched'' population. The arrows are the proposed quenching paths for SG. \hspace{0.5cm}
Lower fig: sSFR vs. $\Sigma_{\mathrm 1kpc}$ comparing the field to satellites in mass $M_\star$ bins.  
The contours represent the logarithmic number density in each panel and are separated by 0.2 dexes.  The color scale is normalized such that dark red represents the highest number density in the panel. The sample size mentioned at the bottom left.  
Satellites have a higher proportion of quenched galaxies than the field at all masses.
Quenched galaxies have higher $\Sigma_{\mathrm 1kpc}$ than star-forming galaxies (SFG) by about 0.2-0.3 dex whether in the field or as satellites. These effects are most active for the lowest masses and decrease in significance for the most massive galaxies.}
  \label{fig:Joanna2016_cartoon}
%   \label{fig:Joanna2016_sSFRvsSigma1}
\end{figure}

Observational studies \cite{Omand2014, Woo2017}, suggests possible evolution paths. All of them include gas stripping out of the SG and later compaction of the SG \cite{Wetzel2013, Woo2017}. As can be seen at fig \ref{fig:Joanna2016_cartoon} by looking at mass bins of SGs at the lower fig, it seems that the SGs are becoming compact through their evolution and quench on the way. The upper figure shows the different suggested paths.

Additionally, it has been shown that the binned quenched fraction of SGs is mass depended. The smaller the SGs mass is, the higher is the quenched fraction of the SGs \cite{Slater2014, Wheeler2014, Wetzel2015, Fillingham2015}. Also, the quenching process dependence on the distance from the central galaxy \cite{Fillingham2018, Woo2017}, The closest the SGs are, the higher is the quenched fraction. There is also halo dependence on galaxy groups and galaxy clusters as bigger the halo mass is the higher the quenched fraction \cite{Woo2012, Balogh2016, Woo2017}
\newpage

% \textcolor{red}{should I include another paragraph with more theoretical info?\\
% - IMF of satellites\\
% - Jellyfish galaxies: interaction with gas in the halo\\ (\cite{Kiyun2019}) 
% https://arxiv.org/pdf/1810.00005.pdf\\
% - SAMS?\\
% }

\newpage
\section{Objectives of This Thesis}
\label{sec:intro-obj}
Field galaxy evolution is studied today with great care. However, one could ask, \\ 
what is the next step on galaxy formation research? \\
A common practice in physics is to study system in equilibrium and later apply disturbances to the system, allowing us to get fundamental knowledge on the system (Resonance, decay...). The analog in our case is after studying the ``normal'' field galaxy evolution track is to add disturbances to the this ``normal'' evolution and earn further knowledge regarding galaxies.

As we cannot hold a galaxy in a lab and disturb it (It is just too big), we should use naturally disturbed galaxies. Those are satellite galaxies.
% It is just too big->It is enormous, trust me. 
SGs are disturbed by variating gravitational potential slope due to the bigger galaxy gravitational potential well shape. SGs are also disturbed by densities of gas from the bigger galaxy that is near the SG and apply ``wind'' (Ram pressure) on the SG. Those two forces throw galaxies out if their ``normal'' evolution track and give us new case studies of galaxies, with hopefully, some fundamental knowledge. 
Hence, SGs allow us to learn new physics and limitations on dark-matter, stars, and gas in the universe.

Our objective in this thesis is to give a crude description of the SG evolution process. We will describe these results on the famous sSFR vs. $\Sigma_{\star}$ diagram, thus, comparing them with the ``normal'' field galaxies halo mass quenching process. 
We would provide analytic expression,  high-resolution numerical results from the VELA cosmological simulation suit and observational results to support our description.

These results would enable us to learn new inner dynamics of galaxies together with integrating the SG quenching path to the complete galaxy quenching process picture.

A by-product of this work, we will present a new rigorous, fast and simple method to analyze SG out of cosmological simulations (a.k.a a new merger tree algorithm) and will provide a broad base of SG samples for future study.
This new method allows us to study other interesting phenomena such as: 
Ultra diffused galaxy (UDG), compact elliptical galaxy (eCg), globular cluster (GC) and baryon dominant galaxy (BG) formation mechanisms, star formation bursts before full quenching (``Swansong''), Globular cluster evolution in a SG, baryon dominant galaxies formation, dark-matter halo and gas halo properties and more...
% \clearpage
%
% File: chap02.tex
% Author: Tomer Nussbaum
% Description: Numerical Introduction
%
\let\textcircled=\pgftextcircled
\chapter{Numerical Introduction}
\label{chap:n_intro}
\initial{G}alactic process timescale is long, very long, above 100 Myr or more. Therefore galactic observations give us only glance at a time with no dynamical process. 
This glance is flat as we see projections of galaxies in the sky and as many objects just cannot be observed: some types of gas and of course dark-matter. 
In order to bypass these limitations and also to allow us to try different approximations and different physical parameters and their stability, one of the essential tools for a researcher today are simulations. 
In this chapter, we describe VELA, a state of the art cosmological simulation suite used to mimic galactic evolution. This process is done numerically by graduating the initial conditions of the universe by physical dynamical terms (see \S\ref{sec:n_intro-VELA}) when the results are compared and verified by observations with recalling the simulation limitations (see \S\ref{sec:n_intro-VELAlimitation}) to determine the confidence in the simulation.
Later, these simulations results \S\ref{sec:n_intro-VELAsuite}) can be used for galactic and cosmological evolution study as presented on this thesis.

\section{VELA High-Resolution Cosmological Simulation}
\label{sec:n_intro-VELA}
The VELA suite is a high-resolution zoom-in hydro-cosmological simulation with 34 simulation runs of central galaxies around $10^{11} M\odot$ halo mass at z$\sim$2. VELA was ran using Adaptive Refinement Tree
(ART) code \cite{Kravtsov1997, Kravtsov1999PhD, Kravtsov2003, Ceverino2009}, a simulation package that includes N-body particles simulation for dark matter and stars with Eulerian gas hydrodynamics using Adaptive Mesh Refinement (AMR). VELA was first run with an N-body dark-matter particles governed by gravity, with the initial distribution function of the universe at early times and then was ran again, refined around chosen halos. The second ran used higher resolution dark matter with addition of gas and stars at the minimum mass of $8.3 \cdot 10^4\ M_{\odot}$ for dark matter, $1 \cdot 10^3\ M_{\odot}$ for stars and $1.5 \cdot 10^6\ M_{\odot}$ gas with a minimum AMR cell mass of $2.6 \cdot 10^5\ M_{\odot}$ for dark matter and stars in a cell. Most of the simulation runs were ran until $z\sim1$ with a maximum AMR resolution of 17-35 pc at all times \cite{Dekel_Lapiner, Mandelker2016}

The simulation assumes an expanding universe with the standard $\Lambda CDM$ cosmology, with the WMAP5 cosmological parameters:
$\Omega_\Lambda=0.73,\ \Omega_m=0.27,\ \Omega_b=0.045,\ h=0.7\ and\ \sigma_8=0.82$ \cite{Komatsu2009}.
Moreover, it includes mechanisms for the following physical process:
(1) Gravity  (2) Gas Hydrodynamics  (3) Stochastic star-formation 
(4) Stellar mass loss  (5) Gas cooling  (6) Photo-ionization heating  
(7) Gas recycling and metal enrichment  (8) Supernovae thermal feedback 
(9) Radiation pressure feedback

\subsubsection{Gravity}
Gravity was implemented by Newtonian gravity on the AMR with force resolution of twice the cell-size (equivalent to softening length of 34 pc in the high-resolution regions). While the high-resolution region of the dark-matter particles is a Lagrangian sphere of 2 times the virial radius centered at the main galaxy at z=1. At higher red-shifts, the same Lagrangian volume has more complicated shapes.

\subsubsection{Gas Hydrodynamics}
The hydro mechanism assumes equation of state of an ideal mono-atomic gas. The artificial fragmentation on cell size is prevented by a pressure floor, ensuring that Jeans scale is resolved by at least seven cells \cite{Ceverino2010}. 

\subsubsection{Stochastic star-formation}
Star-formation added by assuming it occurs at densities above $1\ cm^{-3}$ and temperature below $10^4\ K$. More than 90\% of stars are formed well below $10^4\ K$ when the median is $300\ K$ in cells with gas density above $10\ cm^{-3}$. A stochastic model implements the star-formation in a discrete time-steps of $\Delta t_{SF} = 5\ Myr$ by a probability of:
\begin{align}
    P_{New\ star\ particle}(cell) &=
    \min\left(0.2,\ \sqrt{\frac{\rho_{gas}} {1000\ [cm^{-3}]} } \right)
    \label{eq:met-SFR_probabilty}
\end{align}

The newly formed star particle mass is:
\begin{align}
    m_\star &\equiv m_{gas} \cdot 
    \frac{\Delta t_SF}{\tau}\simeq 0.42\cdot m_{gas}
    \label{eq:met-star_particle-mass}
\end{align}
Where $m_{gas}$ is the gas mass in the cell and $\tau$ is $12\ Myr$.
We assume initial mass function as described in \cite{Chabrier2003}.
This stochastic mechanism yields star-formation efficiency per free-fall time of $\sim$2\%, at a given resolution, this efficiency ruffly mimic the empirical Kennicutt-Schmidt law \cite{Kennicutt1998}. As a result of the universal local SFR law adopted, the global SFR follows the global gas mass. By observations, a simple universal local SFR law of 1\% SFR of molecular gas per local free-fall time fits the galactic clouds, nearby galaxies and high-redshift galaxies \cite{Krumholz2012a}. 
Runaway stars included by applying velocity kick of $\sim 10\ [\nicefrac {km}{s}]$ to 30\% of new star particles as they formed.

\subsubsection{Gas Cooling \& Photoionization Heating}
Gas cooling is estimated by atomic hydrogen or helium and by metals and molecular parts in the gas and photoionization heating implemented by the UV-background with partial self-shielding.
By using the CLOUDY code \cite{CLOUDY1998} and adding terms for a gas density, temperature, metallicity, and UV-background the cooling and heating rates are calculated assuming slab of a thickness of 1kpc. UV-background is calculated assuming uniform radiation redshift dependent model by \cite{Haardt1996}. Except for cases of gas densities above 0.1 $[cm^3]$ in which we reduce the radiation to $5.9\cdot10^{26}\ [\frac{erg}{s\cdot cm^2 \cdot Hz}]$ and enabling the gas to cool down to temperatures of $300\ K$ in order to mimic the self-shielding of gas.

\subsubsection{Supernovae: Stellar mass loss,  supernovae thermal feedback, gas recycling, and metal enrichment}
Thermal stellar feedback model \cite{Ceverino2010, Ceverino2012} releases energy from stellar winds and supernova explosions at a constant heating rate of above 40 Myr follows star-formation. 40 Myr is the typical age of the least massive stars that explode as a type-II core-collapse supernova, Heating rate due to the feedback can overcome the cooling rate, depending on the gas condition in the star-forming region \cite{Dekel1986, Ceverino2009}, in this simulation there is no artificial cooling shutdown added.

The gas recycling and metal enrichment of the interstellar medium (ISM) implemented via Type-Ia supernova and stellar mass loss computation that is added as well to the computation.

\subsubsection{Radiation pressure feedback}
The radiation pressure implemented by the addition of non-thermal pressure term to the gas pressure in regions where massive stars ionizing photons are produced and may be trapped, this radiation injects momentum to cells around massive stars younger then 5 Myr with column density above $10^{21}\ cm^{-2}$ increasing pressure in the star-forming regions \cite{Agertz2013} - Appendix B.
We assume isotropic radiation fields in a given cell with radiation pressure proportional to $\Gamma \cdot m_{\star}$, where $m_{\star}$ is the mass of the stars in the cell, and $\Gamma$ is luminosity of ionizing photons per stellar mass. $\Gamma=1036\ [\frac{erg}{M_{\odot} \cdot s}]$ chosen from STARBURST99, stellar population synthesis code \cite{Leitherer1999} with a time-averaging of over the first 5 Myr of the evolution of a single stellar population. After 5 Myr, the number of high mass stars and ionized photons declines significantly. Radiation pressure is also dependent on the optical depth of the gas in the cell, using the hydrogen column density threshold, $N=1021\ cm^{-2}$, in which above it the radiation effectively trapped and the radiation is added to the total gas pressure. This threshold is chosen to correspond to the typical column density of cold neutral clouds that host optically-thick column densities of neutral hydrogen \cite{Thompson2005}.

\section{Comparison to Observations and limitations}
\label{sec:n_intro-VELAlimitation}
The VELA cosmological simulation is considered to be one of the leading cosmological simulations. Regarding high-resolution AMR hydrodynamics and key physical processes implementation at sub-grid level. Explicitly, VELA suite traces cosmological streams that feed galaxies at high-redshift, including mergers and smooth flows, resolves VDI that governs high redshift disc evolution and bulge formation \cite{Ceverino2010, Ceverino2012, Ceverino2015, Mandelker2014}.  

As shown before, star-formation calculated by SFR efficiency per free-fall time. Although this method proved to be more realistic then previous versions, it still does not fully replicate the formation of molecules and the metallicity effect of SFR \cite{Krumholz2012a}. Moreover, the resolution does not allow the Sedov-Taylor capturing of the adiabatic phase of a supernova. Radiating stellar feedback assumes no infrared trapping, as the trapping effect should be low \cite{Dekel_Krumholz2013} based on \cite{Krumholz2012b}. In contrast to other works \cite{Murray2010, Krumholz2010, Hopkins2012}, which assumes significant trapping, and therefore puts the VELA suite with a lower radiating stellar feedback in comparison to other simulations.

Some mechanisms are missing today in the VELA suite. They include AGN feedback, cosmic-ray feedback, and magnetic fields. Nonetheless, the star-formation rate, gas fraction and stellar-to-halo mass ratio are all within the estimates derived from the abundance matching to the observations, better than other simulations \cite{Ceverino2014}. Uncertainties and other possible mismatches between observations and the simulation by a factor of 2 are comparable to the observational uncertainties.

With those caveats in our mind, we emphasize the use of the simulations to understand the qualitative features of the main physical processes that govern galaxy evolution. Later we should validate those notions by analytic models that are tested in observations and additional simulations.

Note that in this specific study of satellite quenching, having lower stellar feedback in VELA can be a virtue of the VELA simulation, as it gives us a good upper bound on the significant dynamics of the SG, with less high order effects noise.

% \textcolor{red}{??? agreement with observational scaling relations (Behroozi)}

\section{VELA suite}
\label{sec:n_intro-VELAsuite}
The VELA suite includes 34 simulation runs around $10^{11} M\odot$ halo mass at $z\sim2$. It includes central galaxies, satellite galaxies, halos, filaments, walls, clumps and more... 
Some basic details about the simulation can be seen at table \ref{tab:met-vela_basic_details}

Currently, it is one of the highest cosmological simulation suits, and as such, it is a fountain of knowledge awaits to be discovered.

% \textcolor{red}{* picture of halo with stellar projection, dm proj and gas proj of the eCg\\
% $R_{eff} vs M_{\star}$\\
% VELA mock up pictures}
% \begin{figure}[H] \centering
% \includegraphics[width=0.3\linewidth]{fig02/VELA27_z2_mock.png}
% \caption{VELA  }
%   \label{fig:VELA27_mocked}
% \end{figure}

\begin{table*}
\centering \small %\footnotesize
\begin{tabular}{@{}lccccccccccc}
\multicolumn{11}{c}{{\textbf The VELA suite of 34 simulated galaxies}} \\
\hline
Sim\ id & $R_{\textrm vir}$ & $M_{\textrm vir}$ & $M_\star$ & $M_{\textrm{gas}}$ & SFR & $R_{\textrm{eff},M_\star }$ & $\textrm cell_{\mathrm min}$ &
$a_{\textrm{fin}}$ & $z_{\textrm{fin}}$ & $\#_{\textrm{components}}$ \\
& $[kpc]$ & $[M_{\odot}]$ & $[M_\odot]$ & $[M_\odot]$ & $[M_\odot {\textrm{yr}}^{-1}]$ & $[kpc]$ & $[pc]$ &  &  &\\
\hline
\hline
01 & 58.25 & 11.2 & 9.31 & 9.17 & 2.64 & 0.93 & 18 & 0.5 & 1.0 & 55 \\
02 & 54.5 & 11.11 & 9.21 & 9.07 & 1.43 & 1.81 & 36 & 0.5 & 1.0 & 22 \\
03 & 55.5 & 11.14 & 9.58 & 8.95 & 3.67 & 1.41 & 18 & 0.5 & 1.0 & 63 \\
04 & 53.5 & 11.09 & 8.91 & 8.9 & 0.45 & 1.73 & 36 & 0.5 & 1.0 & 24 \\
05 & 44.5 & 10.85 & 8.86 & 8.71 & 0.38 & 1.81 & 18 & 0.5 & 1.0 & 25 \\
06 & 88.25 & 11.74 & 10.33 & 9.51 & 20.6 & 1.05 & 36 & 0.37 & 1.7 & 131 \\
07 & 104.25 & 11.96 & 10.76 & 9.9 & 18.14 & 2.85 & 36 & 0.54 & 0.85 & 247 \\
08 & 70.5 & 11.45 & 9.54 & 9.17 & 5.7 & 0.74 & 36 & 0.57 & 0.75 & 142 \\
09 & 70.5 & 11.44 & 10.01 & 9.46 & 3.57 & 1.74 & 36 & 0.4 & 1.5 & 137 \\
10 & 55.25 & 11.12 & 9.78 & 9.11 & 3.2 & 0.46 & 36 & 0.56 & 0.78 & 113 \\
11 & 69.5 & 11.43 & 9.88 & 9.52 & 8.94 & 2.14 & 36 & 0.46 & 1.17 & 49 \\
12 & 69.5 & 11.42 & 10.29 & 9.3 & 2.7 & 1.13 & 36 & 0.44 & 1.27 & 47 \\
13 & 72.5 & 11.5 & 9.76 & 9.55 & 4.48 & 2.48 & 18 & 0.4 & 1.5 & 79 \\
14 & 76.5 & 11.56 & 10.1 & 9.64 & 23.32 & 0.32 & 36 & 0.41 & 1.44 & 46 \\
15 & 53.25 & 11.08 & 9.71 & 8.92 & 1.35 & 1.07 & 36 & 0.56 & 0.79 & 34 \\
16* & 62.75 & 11.7 & 10.61 & 9.7 & 18.47 & 0.61 & 26 & 0.24 & 3.17 & 103 \\
17* & 105.75 & 12.05 & 10.93 & 10.04 & 61.4 & 1.36 & 34 & 0.31 & 2.23 & 202 \\
19* & 91.25 & 11.94 & 10.65 & 9.76 & 40.47 & 1.22 & 32 & 0.29 & 2.44 & 94 \\
20 & 87.5 & 11.73 & 10.56 & 9.55 & 5.55 & 1.72 & 18 & 0.44 & 1.27 & 259 \\
21 & 92.25 & 11.8 & 10.61 & 9.64 & 7.89 & 1.73 & 18 & 0.5 & 1.0 & 325 \\
22 & 85.5 & 11.7 & 10.64 & 9.45 & 12.0 & 1.31 & 18 & 0.5 & 1.0 & 60 \\
23 & 57.0 & 11.17 & 9.88 & 9.12 & 3.06 & 1.16 & 18 & 0.5 & 1.0 & 64 \\
24 & 70.25 & 11.44 & 9.94 & 9.41 & 3.88 & 1.68 & 18 & 0.48 & 1.08 & 257 \\
25 & 65.0 & 11.34 & 9.84 & 8.93 & 2.29 & 0.73 & 18 & 0.5 & 1.0 & 349 \\
26 & 76.75 & 11.55 & 10.2 & 9.44 & 9.36 & 0.74 & 18 & 0.5 & 1.0 & 57 \\
27 & 75.5 & 11.54 & 9.85 & 9.48 & 6.1 & 1.98 & 18 & 0.5 & 1.0 & 53 \\
28 & 63.5 & 11.3 & 9.27 & 9.32 & 5.54 & 2.32 & 18 & 0.5 & 1.0 & 42 \\
29 & 89.25 & 11.72 & 10.36 & 9.55 & 16.82 & 1.89 & 19 & 0.5 & 1.0 & 335 \\
30 & 73.25 & 11.49 & 10.2 & 9.37 & 2.97 & 1.43 & 18 & 0.34 & 1.94 & 296 \\
31* & 38.5 & 11.37 & 9.89 & 9.1 & 15.27 & 0.43 & 21 & 0.19 & 4.26 & 37 \\
32 & 90.5 & 11.77 & 10.42 & 9.64 & 14.86 & 2.58 & 18 & 0.33 & 2.03 & 199 \\
33 & 101.25 & 11.92 & 10.68 & 9.7 & 32.68 & 1.23 & 18 & 0.39 & 1.56 & 117 \\
34 & 86.5 & 11.72 & 10.19 & 9.66 & 14.47 & 1.84 & 18 & 0.35 & 1.86 & 248 \\
35* & 44.5 & 11.35 & 9.75 & 9.39 & 22.93 & 0.33 & 24 & 0.22 & 3.54 & 92 \\
\hline
\end{tabular}

\caption{\small \textbf{VELA Simulation Major Quantities Table:} The virial radius - $R_{\mathrm vir}$, total virial mass - $M_{\mathrm vir}$, stellar mass - $M_{\star}$, gas mass - $M_{\mathrm gas}$, star-formation rate - SFR and half stellar mass-radius - $R_{\mathrm{\mathrm eff},\ M_{\mathrm \star}}$, for the 34 VELA simulations. When $M_{\star}$, $M_{\mathrm gas}$, SFR and $R_{\mathrm{eff},M_{\mathrm{\star}}}$ are quoted within $0.1R_{\mathrm vir}$ and all of the masses $M_{\mathrm vir}$, $M_{\star}$, $M_{\mathrm gas}$ are shown in their $\log_{10}$ values. Also listed are the minimum cell size $cell_{\mathrm min}$ in the snapshot, the final simulation scale factor - $a_{\mathrm{fin}}$, and redshift - $z_{\mathrm{fin}}$, as well as the number of stellar components throughout the simulation $\#_{\mathrm{components}}$. All physical properties and $\mathrm cell_{min}$ are quoted at $z=2$, except for the five cases marked $^*$, where they are quoted as the final simulation output, $z_{\mathrm{fin}}>2$. \hspace{0.5cm}
Looking at $\#_{\mathrm{components}}$, there are numerous galaxies in the simulation with a diverse population of the central halos in its mass range.}
\label{tab:met-vela_basic_details}
\end{table*}

% \clearpage
%
% File: chap03.tex
% Author: Tomer Nussbaum
% Description: Methods used on this work
%
\let\textcircled=\pgftextcircled
\chapter{Methods}
\label{chap:methods}

\initial{E}quipped with theoretical models and numerical data we go on to define what is a galaxy? This is done by the stellar kernel merger tree, an identification \& follow-up algorithm of the dense stellar population in simulations (see \S\ref{sec:met-SCMT}). 
Then we measure the galaxies with basic analysis (see \S\ref{sec:met-properties}), forces analysis (see \S\ref{sec:met-forces}) and time events analysis (see \S\ref{sec:met-event_properties}). 
Results of these calculations and others enrolled in our pandas catalog system (see \S\ref{sec:met-cat})

\section{Stellar Kernel Merger Tree}
\label{sec:met-SCMT}
In cosmology and specifically in the field of galaxy formation, our knowledge comes from what we observe, which is mainly stars and some specific gas properties (Sadly we cannot observe dark matter). 
So, in order to compare the simulations to the real universe, we choose to compare items which we can observe, hence stellar population.
This approach is different from the usually chosen research which focuses on dark matter follow up in the simulation. As the dark matter has been proven to be the main factor for structure formation in the universe. Notwithstanding, we choose to follow the observed quantity, stars, and this approach has proven right, as, in the galactic scale, there are baryonic effects in which
an observed stellar component formation or evolution are not governed by dark matter.
It can be UDGs with no dark matter \cite{vanDokkum2018} or baryonic galaxies as we will show later in this thesis. It is also true to say that galaxies are baryon dominated in their centers, and as such the stellar population is denser than the dark matter and therefore it is better to follow-up stars to follow galaxies and smaller stellar components. (if one will follow halo evolution, even its center, the stellar component sometimes departs from the halo. Therefore one can get the wrong understanding of satellite formation), case studies are shown at \cite{Chang2013}.
One word about gas tracing, as gas is not a fully observed quantity and as it is hard to trace it in the current VELA implementation, we choose not to follow it. As except mergers and early formation, gas particles flow with the stellar component. These specific cases, are known to us and are handled each case in details.
One caveat of this method is the fact that we do not follow dark matter only halos and sub-halos, this kind of research is essential, for cross-correlation and dark matter - baryon matter relations in the universe.
However, as we choose to study observed satellite galaxies (and not dark matter only satellite) this is a caveat we choose to ``live with''. Further research comparing stellar components evolution with dark matter halo evolution can be fascinating to our understanding of the missing satellite problem, the gas ratio and dynamical limits in the universe.

We have also chosen not to limit our identification by computing if a stellar component is bound as we wish to study the simulation as a ``mock universe'' that could be compared to the real universe. If those stellar components exist in our simulation, we would like to know. Likewise, this approach of identification observed quantity can let us be versatile in the theoretical models that we will apply later.
Example for that can be the question of what one measure when one has a SG inside a disk. Does he want only the bound part of the satellite or does he want all the mass in a sphere? Because that is affecting the self-gravity calculation, or because he would like to compare it to observation. Also, in this case, gas can collide (satellite gas with disk gas), then it would not be bound to any of the galaxies in this stage. So we wish to be adaptive to many cases and choose the strategies later.

Studying stellar components enabled us to explain the formation and evolution of observed phenomena such as ultra diffused galaxies (UDGs), stellar clumps, Baryon dominated galaxies (BG), and more. It can aid us to investigate any other stellar components whether the component associated with a halo or not.

% https://arxiv.org/abs/1806.11417 stellar MT - to find where to put it \cite{Canas2018}

\subsection{Stellar Component Identification}
\label{subsec:met-Stellar_id}
Identifying stellar populations is applied by the ZANAPACK package \shortcite{Tweed2009}, a stellar density identification package based on an adaptation of dark matter density ADAPTAHOP package \cite{Aubert2004}[Appendix B]. We used ZANAPACK to find dens structures. The structures found by calculating ``star neighborhood density'' by averaging over its 40 closest neighbors (the common and default number for star particles). If a star is above $\Delta\rho_{crit}$ ($\rho_{crit}$ is the critical density in the universe) threshold it is ascribed to its densest neighbor star. By this process, the stars are connected to each other and become identified as a structure. When the densest star is the structure center. We have limited our structures to be above our resolution limit. Therefore we set min particles per structure limit to be 100-star particles (min expected is 10, 50-100 is the common use).

% first sentence here is mostly from private correspondence... not sure where it is written...
The standard approach \cite{Srisawat2013} is following the Morse Theory \cite{Jost2002} which gathers structures above the critical density $\rho_{crit}$ of the universe. By \cite{Tweed2009} this threshold is set to $80\rho_{crit}$ (Corresponds to $b=0.2$ in the FOF algorithm \shortcite{More2011}). Which gives a galaxy radii for the stellar component which is similar to $0.1R_{\mathrm vir}$

Unlike the approach described above, our purpose is to follow the gravitational potential well of each component and later resolve the component around it. We choose to look at highly dens structures (kernels of galaxies and other stellar components), as the particles in dense structures are more likely to be bound and define a stable potential well (continues through time). Then, by tracing the particles in these dense structures, we resolve the whereabouts of the potential well through time. This method has shown to be well consistent in identifying and tracing of galaxies with more than one dense region, meaning a small substructure as ex-situ or in-situ clumps. This method even proved itself with identifying UDGs, eCgs, satellite galaxies in the merger/flyby process and more. We choose a $5000\rho_{crit}$ which is similar or above $R_{\star, eff}$, the half stellar mass-radius of the stellar component.
We choose this number in the following way:
For $R_{\star, \textrm eff}$ the radius in which half of the mass of the galaxy is in, and for $M_{\star, \textrm eff}$ half of the stellar mass we get:

\begin{align}
    \frac{M_{ \star, \textrm{eff}}}{\frac{4\pi}{3}(R_{ \star, \textrm{eff}})^3} &= 
    \frac{\frac{1}{2}M_{\star,0.1R_\textrm{vir}}}{\frac{4\pi}{3}(0.2\cdot0.1R_\textrm{vir})^3} = 
    \frac{1}{2\cdot0.2^3}\frac{M_{\star,0.1R_\textrm{vir}}}{\frac{4\pi}{3}(0.1R_\textrm{vir})^3} \nonumber \\
    &\approx \frac{1}{2\cdot0.2^3} \cdot 80\rho_{\textrm{crit}} \approx 5000\rho_{\textrm{crit}}
    \label{eq:met-rho_crit5000}
\end{align}

Where the first step is using the 
$R_{\star, \textrm{eff}} \approx 0.02R_{\textrm vir}$ following \cite{Kravtsov2013,Somerville2018} and the third step
we have used the \cite{Tweed2009} threshold for resolving stellar radius of galaxies $\frac{M_{\star}}{\frac{4\pi}{3}(R_{ \star})^3}=80\rho_{\textrm{crit}}$, with the  $R_{ \star} \approx 0.1R_{\textrm{vir}}$ 
% citation why R_s = 0.1Rvir

$N=5000\rho_{crit}$ was examined with different densities, and it was found that the range $2000-7000$ seems to identify well the stellar component in the VELA simulation, inside galactic disks (fly-bys), during mergers, as clumps (ex-situ or in-situ) and more.

This approach benefits us with substantial low computational overhead by resources and time with a fully comprehensive tracing of components during the simulation.

\subsection{Merger Tree Algorithm} \label{subsec:met-MT}
Merger tree (MT) algorithm is a specific algorithm used to connect halos or galaxies between different times (two snapshots in a cosmological simulation) and identifying them as one over time component. This subsection shows how using the specific definition of the stellar component structure identification with other selective ideas we have adopted achieves an efficient, fast and accurate MT.

Following the current limitation of existing common (MTs) described in \cite{Srisawat2013, Tweed2009} There are three main difficulties left to handle:
(i) Galaxy center position stability - Center identification is switching between 2 dense regions in the galaxy through time (ii) Time resolution stability - Different time resolution produces different MTs as the processes that are shorter then the snapshot timescale are mistreated (iii) Mis-identification of Satellite flybys as mergers.

As we can see, all these major difficulties are vital for studying satellite galaxies and other stellar components, (i) and (iii) as they specifically address misidentification of satellite dynamics and (ii) as we want to use high resolution in time for fast mechanisms as baryon galaxies formation. These runs called ``Thin'', and they save the simulation data each 2-5 Myr instead of 170-240 Myr between snapshots on ``Thick'' simulation runs. 

Therefore we have designed our MT in the following way: First, we use particle tracing, this method has been proven to be robust and accurate in connecting between components at different times. We implemented important changes that manifest in the following way:

\begin{enumerate}[(i)]
\item Galaxy center position stability:
As shown in \ref{subsec:met-Stellar_id} we follow potential-well position. That is done by following kernels of galaxies. In that way we redundant the problem of misidentification of centers, as they counted as two objects, except the case they are very close to each other and then they are probably bound as one kernel of the galaxy. If they count as one object, but they are not bound, it means the case is not two components in one galaxy, and therefore this is a flyby mode, which is covered by technique (iii). 

\item Time resolution stability: 
This issue solved in 2 ways. First, we look farther in time steps, and we connect between shared components no matter if there are snapshots in which they do not appear. Second, we sew different galaxy branches as the same branch if they have the same shared particles over time. Meaning, it does not matter if the shared fraction of particles between 2 snapshots is higher than some value (Unlike the conventional approach for MTs as shown at \cite{Srisawat2013}), it does matter if over the branch there is a maximization of the shared particles. This way we add a ``smoothing'' part that creates a tree in the merger graph (As the merger tree is firstly a directed graph between many objects) in the way that the maximum of the stars will go on the same line. That way we reconnect threads and find cases objects that had disappeared on the identifier and later returned (can occur at fly-bys).

\item Satellite flybys misidentification:
Flybys can be identified in 3 ways: First two ways are using the same solutions that were described in (i) and (ii). (i) Identifying kernels of galaxies and in this way identifying satellite galaxy even inside disks of bigger galaxies. (ii) If the satellite is not identified in that time-step in the simulation, it can be identified farther in time by successor search or branch sewing. The third way is using a merit function to match two snapshots \cite{Srisawat2013}. We choose the following merit function:

\begin{align}
    \mathrm{Merit}(A,B) &= 
    \frac{\left(N_{A}\cap N_{B}\right)}{N_{B}}
    \label{eq:met-mt_merit}
\end{align}

For $N_{A}$ as the number of stars in the galaxy core on the younger event in time and $N_{B}$ the latter one. We choose this merit function as it measures the fraction of the stars in the later event from the older one. Meaning the physics of the galaxy B is influenced by A. If we would have taken $\frac{\left(N_{A}\cap N_{B}\right)}{N_{A}}$ factor in, as used in some of the MTs described at \cite{Srisawat2013}, it means that in case of ``fly-by'' there would have false identification in a stage that a satellite is identified after the occurrence in the central galaxy. Because this would zero the merit function. As we are interested in this phenomena, we did not include this merit factor.
\end{enumerate}

The main difference in this MT is that we do not check
if a particle is bound or not (\cite{AHF2004, AHF2009}), nor we check the particles in the 6-phase position and velocity space \cite{Rockstar2012, Canas2018})
Meaning we do not use an analytic physical term, we learn how much particles are ``bound by practice''!
A particle is ``bound by practice'' for MT calculation (not for physical models) if it stays in the stellar component for a long time, then it is bound by the literal definition of being bound and in-depended of complications due to using analytic terms (and it is also much faster to run!)

\subsection{Results} \label{subsec:met-Results}
We can see at fig  \ref{fig:sqm-sat_evo_22_00004} a dark matter lose from the inner part of the galaxy! Without the same loss of the stellar part. This total loss of dark matter shows the great power of tracing the stellar component and specifically the gravitational potential minimum position. It directly shows that in many cases, tracing of dark-matter particles or tracing dark matter and stars will lose the tracing of a satellite galaxy!

This phenomenon was seen here for the first time in a cosmological simulation. However, it is expected in simulations \cite{Chang2013, Macci2017} and there are observations on galaxies with no dark matter \cite{vanDokkum2018} and with SG with low-velocity dispersion \cite{Wang2018a}. It is caused due to a strong tidal force that is applied for a brief time, causing lowering the tidal radius to a very low value and by that moving the gravitational potential well of the galaxy inward. Causing stellar and dark-matter material outside of this radius to dis-attach from the satellite galaxy and also drive particles with eccentric orbits outside too (high kinetic energy). As dark-matter particles are much more eccentric in their orbit, the dark-matter is flowing outside of the galaxy even inside 0.5 kpc radius, leaving the stellar component with lower dark matter core.

This dark matter detachment caused by the fact that stellar material formed from a gas that has cooled down in the center of the potential well and therefore it is less energetic. While the dark matter might be virialized at the virial radius, but does not lose all of its energy and therefore is in eccentric orbits.

This result shows one example of a galaxy losing its dark matter and keeping its stellar material.
As this phenomenon is expected to be seen more often in eccentric orbits, it might be a reason for not seeing much satellite UDGs in simulations before, as most of the merger trees having difficulty in fly-bys identifications and as most of the halo finders are based on dark matter sub-halo identification. 
Details about this mechanism and others are to be published in our further work.

Note that these phenomena separated from the numerical effects described at \cite{vandenBosch2018}, as the stellar component well kept while the inner dark-matter is pushed out. Moreover, On this case, the baryon component supposed to hold more dark matter inside the subhalo and the SG, as the stellar component is dominating the inner core of the SG and as it is lower energy component, its stellar mass profile is less expected to change due to tidal force.

\subsection{Discussion} \label{subsec:met-discussion}
We have presented a new way to define galaxies by their nucleus component instead of the group of particles that are bound or just above thresholds. In that sense, we have shown our proposed algorithm solves the current merger tree algorithm problems as described in the comparison project of \cite{Srisawat2013}.

We have presented a case in which even a stellar merger tree would not resolve fly-bys, especially in high eccentric orbits, which can explain systemic errors in SG merger trees, these phenomena exist at least 10\% of the SG.

In order to get another confirmation, in the future, we wish to add tracing of particles that are next to the center of mass of the object and see where this center of mass goes to. When the traced particles are particles inside small sphere around the center of mass of the stellar component, those traced particles should also be consistent in several of snapshots (or all of them), in that way we can find out the ``bounded particles by practice''. This method identifies more appearances in the simulation, including lower densities stellar component that was not identified because they were beneath the density threshold. This method also including cases in which the stellar component is mixed in another component and is not identified again because the identification returns one component. We have not implemented this method yet, as the results we got are good enough without it. Even so, we strongly feel that applying this as a redundant parallel method would not just increase our data points but also increase our assurance and the percentages of correct identification and can find extremely fascinating objects in cases of disagreement between the methods.

Also, note that we have purposely implemented a few redundant methods in this merger tree in order to increase our certainty in the results and to be able to quantify it.

Finally, It is our notion that the problem of building a stable MT is solvable by applying some version of ``max network flow'' algorithms such as Ford-Fulkerson, Edmonds-Karp, and others.

\section{Galaxy Properties}
\label{sec:met-properties}

\subsection{Central Galaxies}
Central galaxies analysis made in order to be consistent with previous papers \cite{Mandelker2016, Tacchella2016a, Tacchella2016b, Dekel_Lapiner}. Therefore the centering, $R_{vir}$, and other physical quantities were kept as they were. The MT has been overridden to keep the trunk of the central galaxies. Meaning we first force the MT with the central galaxies by \cite{Mandelker2016} and later we identify all the other stellar components.

The central galaxy identification was made by choosing the biggest galaxy in the latest snapshot of the simulation and by tracing this galaxy back in time by stellar particle tracing to previous snapshots. It was also altered to follow the same galaxies in previous generations of the VELA suite, GEN2, later the galaxies where re-centered. The re-centering was done by starting with a virial radius sphere around a galactic center and then by a repeating process, finding the center of mass and taking 0.1 smaller radius sphere around the new center, then, again and again, finding the center of mass and decreasing sphere radius, until convergence.

\subsection{Stellar Radius}
Following our more ``observed'' approach, we define the stellar radius, $R_{sat}$, this radius is defined to be similar to radius definition in observations: radius that encapsulates 90\% of the stellar mass of the stellar component. We have checked other percentages as 80\% and 85\% that gave similar results or even better, but we choose to define the same as observers.
The mass of the stellar component calculation dividing into two cases:
(i) Isolated stellar component (ii) entangled stellar component.

% TRIPPLE CHECK HERE:
Case (i), as the component is isolated, the Mass of the component is the point at which the mass growth is small. We chose the point for 10\% radius increase. The mass increase is less than 10\%. We found that the smallest radius in which $\frac{\Delta \ln M_{\star}(r)}{\Delta \ln r} < \frac{1}{10}$ is a good measure for it. 
% \textcolor{blue}{maybe something about Nir's method for clumps?}
This radius is defined as $R_{slope, 0.1}$

\begin{align}
    R_{\mathrm{slope},{0.1}} &= \frac{r_1+r_2}{2}
    \quad \mathrm{when} \quad
    \frac{\Delta \ln M_{\star}(r)}{\Delta \ln r} < \frac{1}{10}
    \label{eq:met-r_slope_01}
\end{align}

Case (ii), if there are two or more entangled stellar components, the radius should difference and separate the two. The separation done by taking the radius as the distance between the center of each component to the saddle point between the two components. Accordingly this radius if exists, is smaller then $R_{slope_{0.1}}$). Specifically, it is calculated $\frac{\mathrm{d}\ln\rho}{\mathrm{d}\ln r}$ and looking for the radius (r) place in which this quantity changes its sign.
% \textcolor{blue}{A word about sav-gol smoothing and splash-back radius method}
In case of satellite inside the central galaxy, the satellite radius will be its own, but the radius of the central galaxy may not change at all, as in its scale the $\frac{\mathrm{d}\ln\rho}{\mathrm{d}\ln r}$ will not change its sign. We call this radii $R_{1st\ \rho_{min}}$

Final radius calculation: The stellar component radius is set by taking the minimum of these two calculations, and then, similar to observations, we take the radius that holds $90\%$ of the stellar mass. So finally we get $R_{\mathrm{sat}}$ (stellar radius):

\begin{align}
    R_{\mathrm{sat}} &= R_{90\%}\left( \mathrm{Min}
    \left(R_{\textrm{slope},{0.1}}, R_{\textrm{1st }\ \rho_{\textrm{min}}}    \right) 
    \right)
    \label{eq:met-r_sat}
\end{align}
 
This method is consistent to \cite{Tweed2009, More2011} definition of radii. Note that this method breaks the generic and adaptable MT algorithm we used before and choose a method to define a galaxy. Other methods that can use here are a specific 6-phase decomposition \cite{Canas2018, Rockstar2012} and bound particles calculation \cite{AHF2009, AHF2004} or any other specific choice.

% \textcolor{red}{<HERE SHOULD COME PLOT OF THE RADII>}

\subsection{Derived Masses and Radii}
$R_{\mathrm vir}$ is the radius in which a sphere
around the galaxy center encompasses over-density of 
$ \Delta(z)=(18 \pi^2-82 \Omega_{\mathrm \Lambda}(z)-39 \Omega_{\mathrm \Lambda} (z)^2)/ \Omega_{\mathrm m} (z)$ , 
where $ \Omega_{\Lambda}( z ) $ and $ \Omega_{m} ( z ) $ are the cosmological parameters at $ z $  \cite{Bryan1998}. 
The $R_{\mathrm{eff}, M_{\mathrm{\star}}}$ is the radius that includes the half the stellar mass of the stars included in the stellar radius of the galaxy. When in central galaxies stellar radius is set to be $0.1R_{\mathrm vir}$ due to compatibility with previous papers.

In the same way, $R_{0.9\mathrm{cold\ mass},M_{\mathrm{gas}}}$, $R_{0.5\mathrm{cold\ mass},M_{\mathrm{gas}}}$ are the radii that includes 0.9 and 0.5 of the gas mass of the gas include in the stellar radius of the galaxy, when again, the central galaxies stellar radius is set to be $0.1R_{\mathrm vir}$.

$M_{tot, R}$ is defined to be the total mass include in a sphere with a R radius.
also, in the same method, $M_{\mathrm{dm}, R},\ M_{\star, R},\ M_{\mathrm{gas}, R}$ are defined as the dark matter, stars and gas masses include in a R radius sphere.

Other terms that shown here are $\mathrm{cold\ gas}$ and $\mathrm{young\ stars}$ when $\mathrm{cold\ gas}$ is gas with a temperature below $1.5\cdot 10^4\ K$ and $\mathrm{young\ stars}$ are stellar particles that formed in less than $100\ Myr$ since the snapshot time.

\subsection{Main Galactic Quantities}
Star-formation rate (SFR) is calculated in the following way:
\begin{align}
    SFR\left(R\right) &= \left\langle \nicefrac{M_{\star}
    \left(t_{age}<t_{max}\right)\ }{\ t_{max}}\right \rangle _{t_{max}}
    \label{eq:met-sfr_calc}
\end{align}

where $M_{\star}\left(t_{age}<t_{max}\right)$ is the mass of stars younger than $t_{max}$ within a sphere of radius R. The average $\left\langle \cdot\right\rangle _{t_{max}}$ is obtained for $t_{max}\in\left[40,80\right]_{Myr}$ in steps of $0.2\ Myr$ in order to reduce fluctuations due to a $\sim5\ Myr$ discreteness in stellar birth times in the simulation. The $t_{max}$ in this range are long enough to ensure good statistics. more info in \cite{Tacchella2016a, Tacchella2016b}

Surface stellar density $\Sigma_{\star, r}$ calculated as the stellar Mass encapsulated in a sphere with radius r divided by $\pi \cdot r^2$.
\\
Concentration calculated as $R_{sat}/R_{eff}$, $R_{orbit}$ is the distance between the centers of the satellite galaxy and the central galaxy. In the same way, $v_{orbit}$ is the difference in the velocities of the two.
Last, $v_{\mathrm circ}$ is calculated by the virial theoram: $v_{\mathrm circ}=\frac{Gm}{r}$, and specifically at the virial radius $v_{\mathrm vir}$, $v_{\mathrm vir}=\frac{GM_{\textrm vir}}{r_{\textrm vir}}$

% \subsection{Shape}

\section{Computing Forces}
\label{sec:met-forces}

To answer the causes of quenching we should ask ourselves what are the main dynamical process that acts on a satellite galaxy, will be shown at \S\ref{chap:osq}, the primary processes are: ram pressure from the surrounding gas in the halo, tidal forces inflicted by the halo gravitational potential and self-gravity of the satellite itself.

The conventional way to study these forces, mostly by semi-analytic models but also in simulations, is to find the $R_{tidal}$, $R_{ram\ pressure}$, which are the radii where the tidal force is equal to the self-gravity force and the ram-pressure is equal to the self-gravity force.
As these approximations assume a quasi-static equilibrium which we will further show that it might not be the usual case on satellites,
we have chosen to expand the technique that was used on the gas component at \cite{Simpson2018}.
We have tried to get a better approximation by measuring the forces as an average force on the gas and stellar component. As the gas is far from being spherical, we have tried different methods, and we finally found that $R_{0.9,\ cold\ gas}$ spherical radius is a good fit for the gas component and $R_{sat}$ for the stellar component. 

We have measured the forces in the following way:

\subsection{Ram Pressure Force}
Ram Pressure calculated as described at \cite{Simpson2018, Gunn1972}:

\begin{align}
    f_{\mathrm{ram}}\left(r\right) &= 
    \frac{P_{\mathrm ram}}{\Sigma_{\mathrm{cold\ gas}}\left(r\right)} \cdot \hat{v} = 
    \frac{ \rho_{\mathrm gas,\ environment} \cdot v_{\mathrm orbit}^2}
    {\Sigma_{\mathrm{cold\ gas}}\left(r\right)} \cdot \hat{v}
    \label{eq:met-fram}
\end{align}

$\rho_{\mathrm gas,\ environment}$ is the gas in front of the satellite when $\rho_{\mathrm gas,\ environment} = \frac{\Delta M_{\mathrm gas}}{\Delta V}$ of a shell in the halo of the central galaxy between $R_{\mathrm orbit}\pm R_{\mathrm sat}$

\subsection{Tidal Force}
% regular tidal force, isothermal tidal force, DD tidal force outside

Tidal force per one solar mass calculated following \cite{Dekel_Devor2003} 

\begin{align}
    f_{\mathrm{tidal}}\left(r\right) &= \left(\alpha - 1 \right)
    \frac{GM_{(\mathrm{cen},\ R_{\mathrm orbit})}}{R_{\mathrm orbit}^3} \cdot \overrightarrow{r}
    \label{eq:met-ftidal}
\end{align}

Where $M_{\mathrm{cen},\ R_{\mathrm orbit}}$ is the total mass encapsulate in a sphere with radius $R_{\mathrm orbit}$ around the central galaxy center, $R_{\mathrm orbit}$ is the distance between the satellite and the central galaxy and r is a chosen radius of the satellite galaxy. $\alpha$ is measured as the average density slope of the halo $\alpha(r)=-\frac{\mathrm{d}\ln\ \overline{\rho}}{\mathrm{d}\ln\ r}; \overline{\rho}=\frac{M_{\mathrm cen}(r)}{\frac{4\pi}{3}r^3}$  and $G$ is the gravitational constant.

\subsection{Satellite Self-Gravitational Force}
The SG gravity force per one solar mass can be described by Newton's gravitational force as:

\begin{align}
    f_{\mathrm{\mathrm self\ gravity}}\left(r\right) &= \frac{GM_{(\mathrm{sat},\ r)}}{r^2} \cdot \hat{r}
    \label{eq:met-fself_gravity}
\end{align}

where $r$ is the radial distance to the center of the satellite center, $M_{(\mathrm{sat},\ r)}$ is the total mass of the satellite in a $r$ sphere, and $G$ is the gravitational force.

\section{Satellite Galaxies characteristics}
\label{sec:met-event_properties}
Satellite galaxies has a specific events in time that are important in their evolution. We've marked the following events: 
\begin{itemize} \setlength\itemsep{0em}
\item $entering\_halo\_outside$ - Last snapshot before the satellite enter the halo
\item $entering\_halo\_inside$ - First snapshot after the satellite enter the halo
\item $entering\_halo$ - Average values of in $entering\_halo\_outside$ and $entering\_halo\_inside$
\item $peri-center\ (\#i)$ - The \#i local minimum in the orbit (discrete over snapshots)
\item $apo-center\ (\#i)$ - The \#i local maximum in the orbit (discrete over snapshots)
\item $f\cdot M_{\star}$ - Fraction f of the stellar mass from entrance to the halo remained (example: $0.9M_{\star}$ will be the first snapshot the satellite galaxy has less then 0.9 of its stellar mass since entering the halo
\item $f\cdot M_{\mathrm cold\ gas}$, $f\cdot R_{\mathrm cold\ gas}$ and $f\cdot \mathrm{SFR}$ - Fraction f of the cold gas mass, cold gas effective radius and SFR from entrance to the halo remained.
\end{itemize}

The time measured in two ways:
First, $t_{\mathrm halo}$, the time of the satellite inside the central galaxy halo in Gyr.

Second, $t_{\mathrm dyn}$, the time from the first peri-center normalized by the travel time between the first peri-center to the first apo-center.
\begin{align}
    t_{\mathrm dyn} &= \frac{t_\mathrm{H} - t_{\mathrm H, \ 1st\ peri-center}}{|t_{\mathrm H,\ 1st\ apo-center} - t_{\mathrm H,\ 1st\ peri-center}|}
    \label{eq:met-t_dyn}
\end{align}
When $t_{\mathrm H}$ is the time from the big bang, $t_{\mathrm H,\ 1st\ peri-center}$ is the time of the first peri-center and $t_{\mathrm H,\ 1st\ peri-center}$ is the time of the first apo-center.

$R_{\mathrm orbit}/R_{\mathrm vir}$ is the distance of the satellite from the central galaxy, $R_{\mathrm orbit}$ divided by the current virial radius of the central galaxy, $R_{\mathrm vir}$.

\section{Catalog Summary}
\label{sec:met-cat}
The calculated quantities described above and others, compiled with additional quantities from other papers of the HUJI cosmological group are available in one catalog system. This catalog holds simulation snapshots meta-data, central galaxies quantities, halo properties, SGs occurrences, and time-series data, Merger Graphs and Merger Trees, Mergers table and more...
Additional in preparation work includes component density plots (dark-matter, stars, gas, cold gas and more...), mock-up observational pictures, 3D-model and a 3D-animation model for each stellar component.

The central galaxies catalog includes 254 quantities for 1120 occurrences of the 34 central galaxies in the Vela suite. 
It includes masses of dark matter, stars, young stars, gas, and cold gas. Additionally, it includes SFR, different angular momentum, a 3d ellipsoid fit of the galaxy and more. All of those quantities available with different radii around the central galaxy and in the halo.

The Stellar Components analysis, include between 65 to 382 different quantities over 162189 occurrences of the 4369 stellar components identified by our merger tree algorithm.
It includes masses of dark matter, stars, young stars, gas, and cold gas. Moreover, it includes SFR, forces, numerically integrated forces... All of those quantities available with different radii around the SG.
The Stellar components time-series analysis includes 4369 stellar components evolution paths over the simulation lifetime and includes notable events in the life of the stellar component as described at \S\ref{sec:met-event_properties}.

Other catalogs are available and would be described on the relevant in preparation paper.

The catalog themselves are available at a fast, robust and easy to use research tool called pandas, a python library for data science.
\begin{figure}[H] \centering
\includegraphics[width=\linewidth]{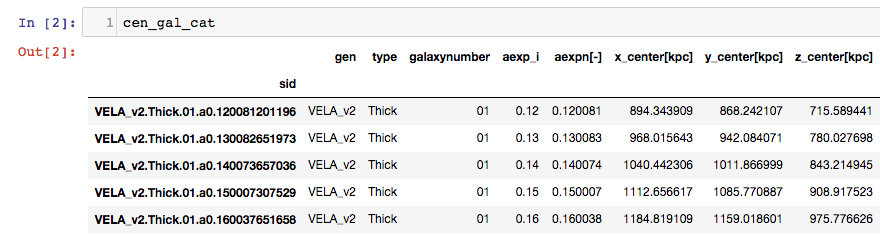}
    \caption{\small Central galaxy catalog example}
    \label{fig:pandas_example}
\end{figure}

% \clearpage
%
% File: chap04.tex
% Author: Tomer Nussbaum
% Description: Satellite Quenching results
%
\let\textcircled=\pgftextcircled
\chapter{The Satellite Quenching Process}
% where & when: Nun path, Sat story
\label{chap:sqm}

\initial{L}et's get our hands dirty and see how do satellite quench in the simulations. First, we will describe our selected satellite sample (see \S\ref{sec:sqm-sample}) following with few examples (see \S\ref{sec:sqm-evo_examples}). Then we will show how satellite evolve with time (see \S\ref{sec:sqm-time-evo}) and by their distance to the central galaxy (see \S\ref{sec:sqm-Rorbit-evo})
relative to the central galaxy and to the halo.
All the stages above bring us to the bothering core question: ``How do satellite galaxy quench?''. We will answer this question by the famous galaxy diagram, sSFR-$\Sigma_{\star , \frac{1}{2}kpc}$ (see \S\ref{sec:sqm-nun-path}).
Last, as our answer can be seen as inconsistent with today's satellite conception, we will settle those disagreement (see \S\ref{sec:sqm-observations}.

% -------------------------------------------
\section{Satellite Galaxies Sample in the VELA Simulations}
\label{sec:sqm-sample}

\begin{table*}
\centering \small %\footnotesize
\begin{tabular}{@{}lccccccc}
\multicolumn{4}{c}{\textbf{Identified secondary galaxies in the VELA suit}} \\
\hline
$\#_{\textrm appearances}$ & All & Enter the halo & Only inside halo  \\
\hline
\hline
1     & 4369 & 528 & 917 & \\
1  <  & 2009 & 528 & 189 & \\
10 <  &  834 & 333 &  15 & \\
\hline
\end{tabular}
\caption{\small SGs statistics: Identification \& follow-up of SGs by different criterion in the VELA suit}
\label{tab:sqm-vela_sat_Numbers}
\end{table*}

\begin{figure} \centering
% figure: Satellites mass distribution, halo dist, mass ratio dist, velocities, Rperi/Rvir, quenched in each Rperi etc, Mgas/Mstar, z, concentration, cold gas
\includegraphics[width=\linewidth]{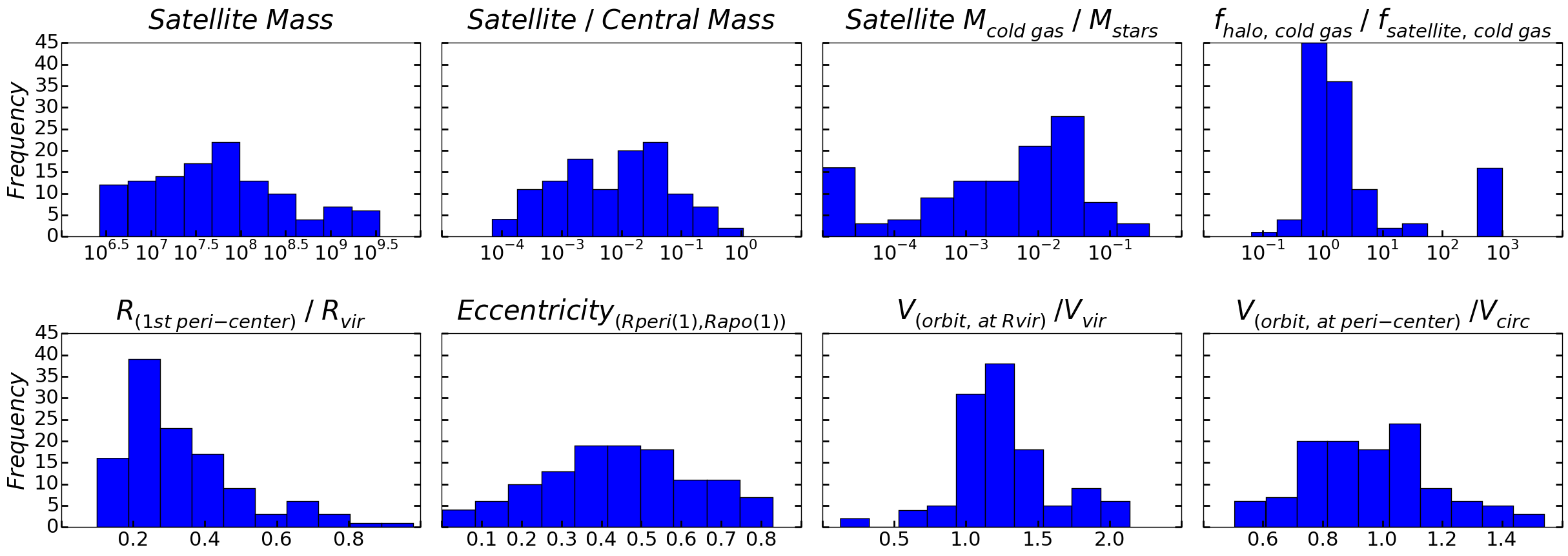}
    \caption{\small \textbf{Satellite sample statistics: Statistics of long living satellites in the halo.}
    First row (left to right): 1. SG stellar mass shows a wide range of SG masses 2. SG mass divided by the central galaxy stellar mass with a variety of ratios (but less major mergers as they introduce specific dynamics) 3. SG cold gas mass divided by the SG stellar mass 4. SG fraction of cold gas in the halo divided by the fraction of the satellite when the fraction is the cold gas mass over the stellar mass.
    Second row (left to right): 5. First minimal orbital distance from the central galaxy (peri-center) divided by the virial radius of the halo. 6. The eccentricity of the SG orbit by first peri-center and first apo-center 7. Velocity ratio at the entrance to the halo, the orbital velocity divided by the circular velocity at the halo radius. 8. Velocity ratio at first peri-center, the orbital velocity divided by the circular velocity at the first peri-center. \hspace{0.5cm} In this figure we can see the diversity of the sample, including many mass ranges and mass ratio with different orbits, from circular to eccentric orbits. }
    \label{fig:sqm-Satellite_galaxy_statistics}
\end{figure}

In order to understand the formation and evolution of satellite galaxies, we focus now on galaxies that have entered the halo and were identified by the MT in at least ten snapshots, including five inside the halo, in the way that some dynamics could be resolved. This filtering retains 214 galaxies, i.e., the majority of the galaxies that are long lived in a central galaxy halo. Among them, we select satellite galaxies with distinguished orbit, containing at least one peri-center and at least one apo-center.
This selection allows us to track the evolution of forces throughout the different stages of a SG orbit, which cleared out fast mergers (mostly major mergers) and fly-bys. 
We finally obtain a sample of 118 satellite galaxies; histograms of their main properties are shown in Fig.~ \ref{fig:sqm-Satellite_galaxy_statistics}, (The broad scope of the stellar galaxies presented at table \ref{tab:sqm-vela_sat_Numbers})

%\textcolor{red}{add Mgas rvir // Mgas apo? }

The sample covers a wide range of galaxies from $10^{6.5}$ to $10^{9.5}$ with a mass ratio of mostly less than 10\% of the main central galaxy.
Therefore includes less major mergers in the sample, as major mergers suffer great dynamical friction and hence had short dynamical time in the halo and can merge at the first peri-center. 

The distribution of cold gas to stars mass ratio  (Mcold gas /Mstars) is mostly above 1\% and is quite similar between satellite and central galaxies. We note that 18 satellites have meager cold gas ratio because they were quenched before entering the halo. 
There are three possible scenarios for this prior quenching: 
1. The SG quenched by another halo first then fell onto this halo.
2. The quenching of the SG occurred at the outskirts of the halo.
3. The SG formed on a low gas region. 
Further investigation is required.

Regarding the orbit properties, we find relatively large values of SGs with high eccentricities. That explained by the fact that ordinarily, SG enters the halo perpendicular with a filament, or just as SG entering a the halo at tangent orbit limits the possible allowed velocities (higher velocities would cause the SG to fly outside the halo (fly-by), and smaller would cause radial orbit, only few would end with a circular orbit). Those high eccentric orbits of the SGs allow them to penetrate deep in the halo at the first peri-center, reaching distances of 0.1 to 0.2 $R_{\textrm vir}$ to the center of the halo. These eccentric orbits are also manifesting by the distribution of $V_{\textrm orbit}/V_{\textrm vir}$ at $R_{\textrm vir}$, which shows that most of the satellites enter the halo with a higher velocity.
Note that for $V_{\textrm orbit}/V_{\textrm vir}$ at the peri-center this ratio is changed due to dynamical friction. 

% \textcolor{red}{citations that show this agrees with analytic work and simulation}

\section{Case Studies of Satellite Galaxy Evolution }
\label{sec:sqm-evo_examples}

\begin{figure} \centering
\includegraphics[width=\linewidth]{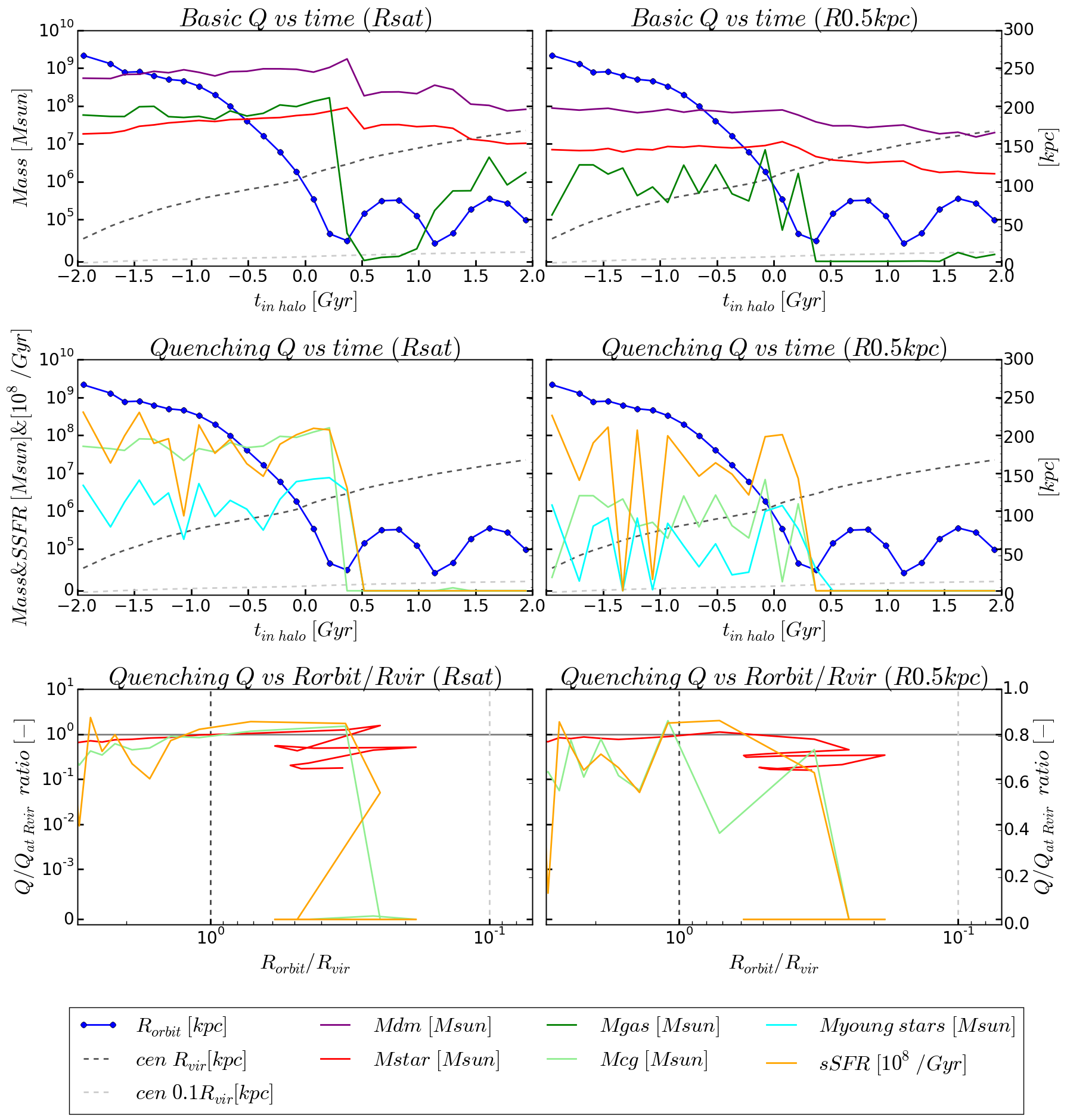}
\caption{\small \textbf{Satellite galaxy 07-041 evolution} (Total halo mass $9.5\cdot10^{11} M_\odot$, central galaxy stellar mass $5.8\cdot10^{10} M_\odot$ at halo entrance): Each plot depicts a satellite property against the time or against the distance, and splits to two columns; left column focuses on the SG enclosed in a sphere in $R_{\textrm sat}$ radius and right column focus on the  $0.5$ kpc radius sphere.
By rows, the first row describes basic galactic parameters over time in halo: $R_{\textrm orbit}$ (blue), $R_{\textrm vir}, o.1R_{\textrm vir}$ (dashed lines), dark matter mass (purple), stellar mass (red), gas (green). The second row describes additional star formation related parameters: in addition to the $R_{\textrm orbit}$ and $R_{\textrm vir}$ over time in halo, which are the sSFR (orange), cold gas mass (light green) and young stars (cyan). The last row describe important quenching and stripping parameters depending on the distance to the  central galaxy normalized by their amount while entering $R_{\textrm vir}$: sSFR, stellar mass and cold gas (with the same coloring as before).\hspace{0.5cm}
This figure present fast quenching case of a SG.
We can see the SG enter the halo at the point the blue line $R_{\textrm orbit}$ intersects with the grey dashed line $R_{\textrm vir}$. Later around the first peri-center which is the minimum of the blue line, we can see by the light green and orange lines that the SG losses its cold gas component very sharply (less than 100 Myr) and stop forming stars. Moreover, by following the red lines in both of the columns, we can see stellar heating and stripping which lowers the SG stellar mass to 10\% of its original mass}
\label{fig:sqm-sat_evo_07_00041} \end{figure}

\begin{figure} \centering
\includegraphics[width=\linewidth]{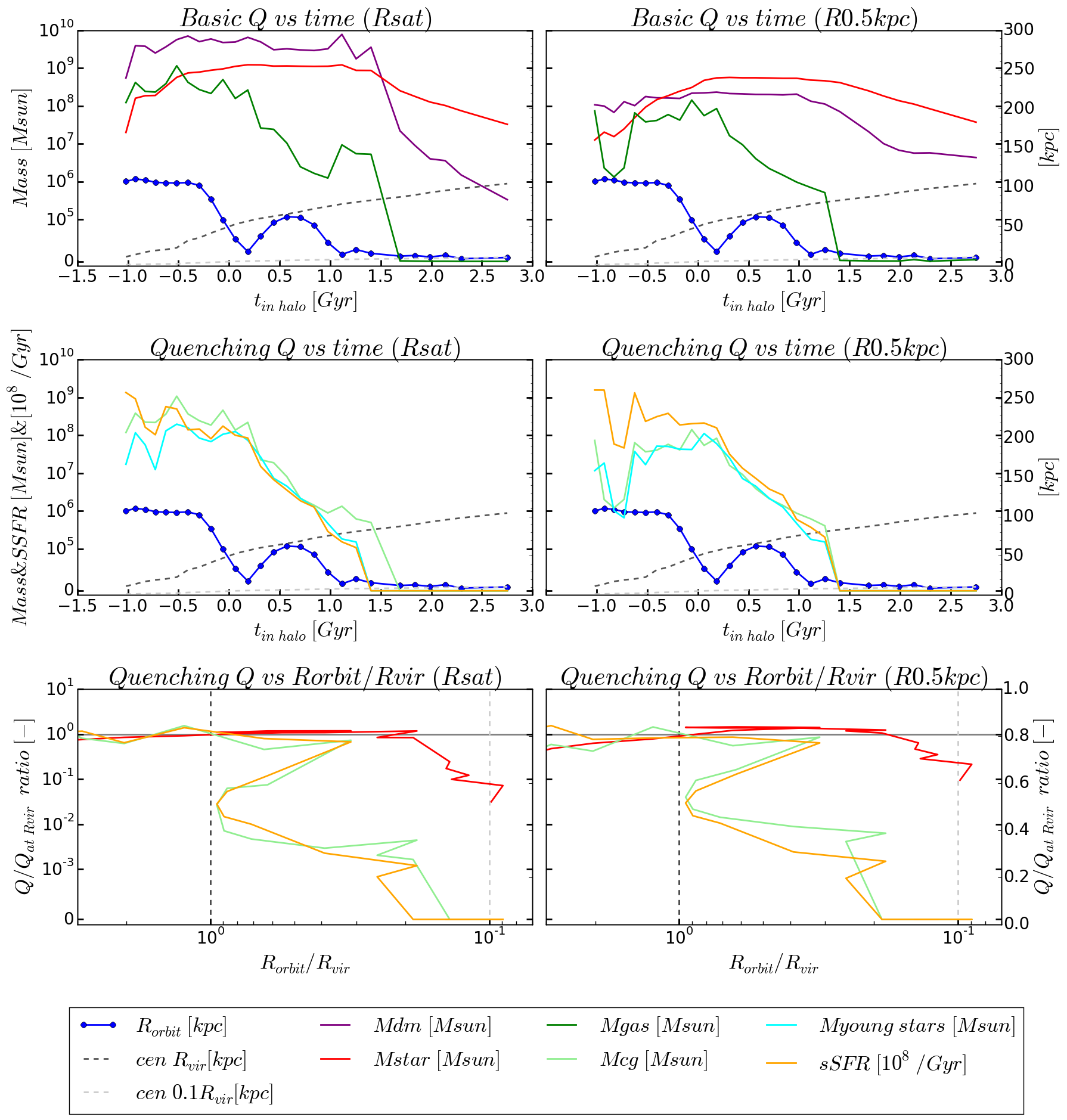}
\caption{\small \textbf{Satellite galaxy 22-004 evolution} (Total halo mass $3.1\cdot10^{11} M_\odot$, central galaxy stellar mass $1.3\cdot10^{10} M_\odot$ at halo entrance): Similar structure and colors as in caption of fig~\ref{fig:sqm-sat_evo_07_00041}. 
\hspace{0.5cm} This figure present a slow quenching, dark matter loss and eCg formation case of a SG.
Meaning the cold gas is kept in the first peri-center due to a more circular orbit and therefore a compressing mode created by the tidal force with the ram-pressure \ref{fig:osq-sat_f_evo_22_00004}. This compressing mode is inducing the star-formation which results in a very dense stellar core. Later as the SG arrives at the second peri-center, it suffers a sharp drop in gas resulting in quenching, after some decrease happened due to movement in the halo and star formation. The last stage is the total dark matter component loss and stellar loss due to tidal forces and the remaining center component of the SG which became a stellar only, elliptical compact galaxy (eCg).
} 
\label{fig:sqm-sat_evo_22_00004} \end{figure}

\begin{figure} \centering
\includegraphics[width=\linewidth]{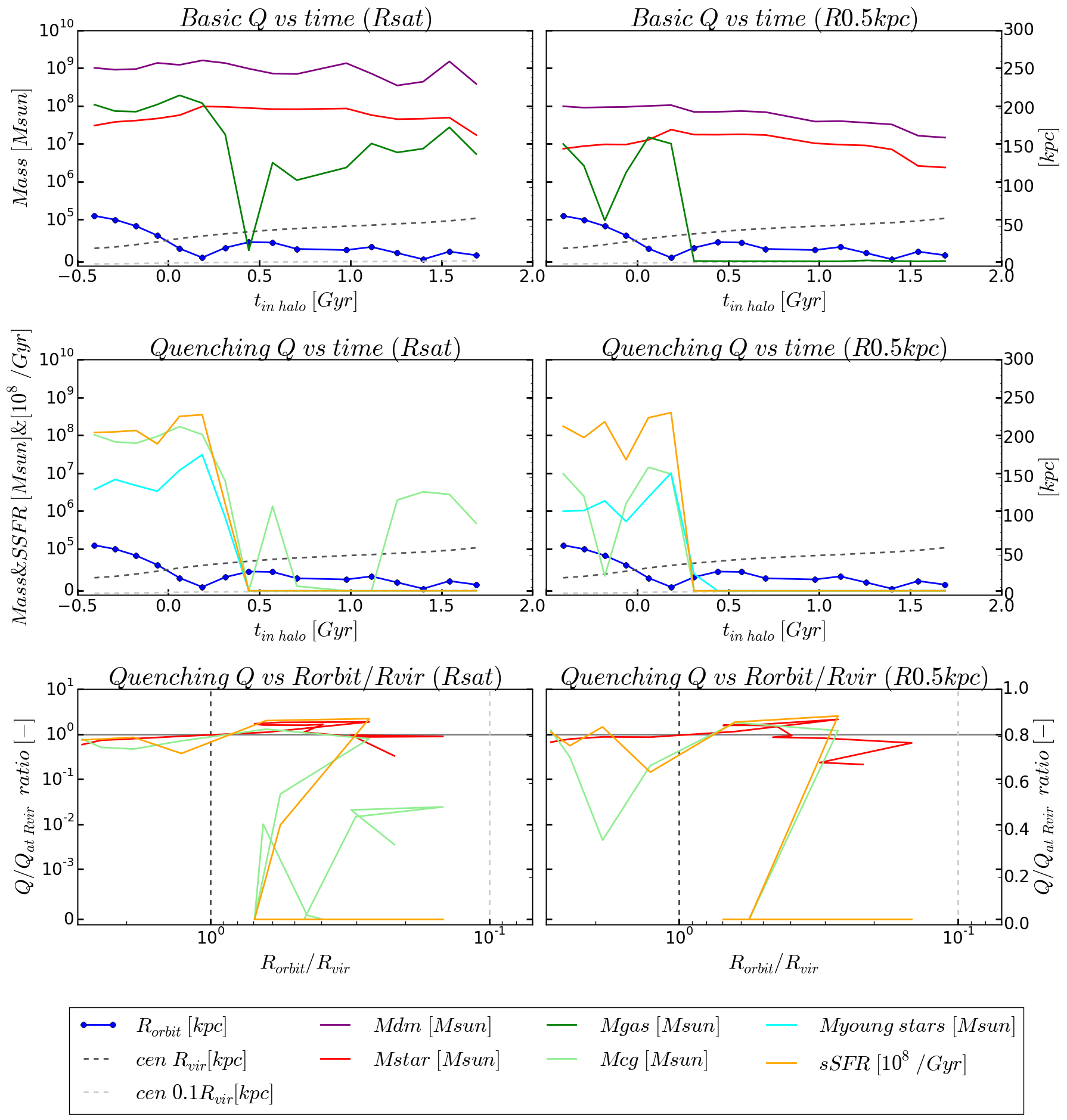}
\caption{\small \textbf{Satellite galaxy 23-007 evolution} (Total halo mass $2.4\cdot10^{11} M_\odot$, central galaxy stellar mass $3.6\cdot10^{9} M_\odot$ at halo entrance): Similar structure and colors as in caption of fig~\ref{fig:sqm-sat_evo_07_00041}.
\hspace{0.5cm} This figure present a fast quenching, and starvation, halt in gas accretion case of a SG.
We can see at the first peri-center quenching of the SG and that on a later time, 1.1 Gyr time in the halo, the SG travels through a cold gas region and do not accrete it as described at our starvation model at  \S\ref{sec:osq-starv}
}
\label{fig:sqm-sat_evo_23_00007} \end{figure}

The following figures are 3 case studies of satellite galaxies with interesting behavior that represent different aspects of the dynamics acting on the satellite galaxies:
\begin{enumerate}
\item ``Fast Quenching'' - satellite galaxy 07-041 \ref{fig:sqm-sat_evo_07_00041} is an example of the common case of SG that loses its cold gas the first peri-center and suffers a stellar stripping and heating at each peri-center.
\item ``Slow Quenching'', ``Dark-matter loss'' and ``eCg formation'' - satellite galaxy 22-004 \ref{fig:sqm-sat_evo_22_00004} is an example of a SG that loses its cold gas only at the second peri-center and looses its dark matter mass in the SG center region, at the center of the SG potential well.
\item ``Fast Quenching'' and ``Starvation'' - 
satellite galaxy 23-007 \ref{fig:sqm-sat_evo_23_00007} is an example of the common quenching at first peri-center and starvation, a halt in gas accretion.
\end{enumerate}

Fig \ref{fig:sqm-sat_evo_07_00041} at first row shows a typical case of SG evolution. A gas-rich galaxy enters the halo and loses its gas at the first peri-center and therefore stop forming stars. Initially, the stellar component increases until near the peri-center owing to star formation; it then sharply decreases (less than $100$ Myr)due to the stripping process is seen as lose of stellar mass at $R_{\textrm sat}$) and heating in the center of the SG which is seen as mass loss at $0.5$ kpc radius of the SG. Regarding the trajectory, the SG stays in a constant orbit around the central galaxy and does not merge. We can see no accretion of gas, which here is not a cold gas on later time as the gas component in $R_{sat}$ does not reach the center of the galaxy.

We can also see in fig \ref{fig:sqm-sat_evo_07_00041} second row a correlation between cold gas, sSFR, and young stars. All of them active in the same way before the first peri-center, and all of them are quiescent at after the first peri-center at the same time. It may be a trivial assumption, as they somewhat depended on each other in the simulation and calculation (SFR). However, it demonstrates why measuring one of them is enough to describe the SFR change. Note that the gas component is not correlative to star formation; therefore, we mainly focus on cold gas and not on total gas.

The last row of Fig \ref{fig:sqm-sat_evo_07_00041} shows a plot of various SG properties against the distance to the center.  We find a mass increment
at peri-center with heating at the $0.5$ kpc radius, together with a sharp cut in the star formation with a later relaxation in each apo-center, the heating and stripping also continue in each peri-center after.
No stellar change is shown between the apo-center to the peri-center.

Fig \ref{fig:sqm-sat_evo_22_00004} shows an extremely interesting case an ``Slow Quenching'', ``Dark-matter loss'' and ``eCg formation'' SG. The satellite does not quench at first peri-center, but on the second peri-center, it keeps forming stars and undergoes stellar growth in its center. Finally, we see a sharp drop in the dark matter component in the middle of the satellite whereas the stellar mass decreases less. This dark matter loss is a potentially interesting mechanism explaining how a galaxy may lose most of its dark matter component, even at its center while remaining a long-lived stellar galaxy.

Its orbit is circular, which results in less stripping, as will be seen later. Finally, 
We stress that this example is the first eCg (elliptical compact galaxy) to be found in a cosmological simulation; this is thanks to resolution and to the merger tree algorithm that we used, as explained in \S\ref{sec:met-SCMT}.

This SG is an example of an initial condition for a gas-dominated object that formed inside a filament and which, upon entering the halo, results in an eCg. This eCg formation case is similar to a specific simulated galaxy run by \cite{Du2018}.

Our tidal and ram pressure analysis presented in \S\ref{fig:osq-sat_f_evo_22_00004} shows that this situation arises in a compaction scenario, in which the tidal forces compress the gas along one direction and the ram pressure along another the opposing direction with about the same intensity. This compressive state allows the gas to stay inside the galaxy for an additional orbit, resulting in higher young stars population and thus enabling the creation of an eCg. 
%Pictures and additional visualizations can be found at \S\ref{app:app01}

Lastly, Fig \ref{fig:sqm-sat_evo_23_00007} is another example of SG fast quenching in the peri-center, with one difference, there is no 
accretion of cold gas to the satellite, This is an example for the starvation process which will is described at the Pac-Man model at \S\ref{sec:osq-starv}
Note that this galaxy is puffing-up due to tidal forces as shown by a drop in the stellar mass in the inner 0.5 kpc and later relaxation as shown by the delayed drop in stellar mass at the stellar radius sphere.

% \textcolor{red}{put in Halo total mass, stellar mass... for each sat} 

% ----------------------------------------------------
\section{Evolution of Satellites as a Function of Time} \label{sec:sqm-time-evo}
This section describes the evolution of satellite properties as a function of time. 
The evolution is stacked either against the time spent by the SG in the halo (in the observer's referential) or against $t_{dyn}$ as described in Section \ref{sec:met-event_properties}. The latter is stacking against the first peri-center and the first apo-center in order to study orbital dependencies.

\begin{figure} \centering\includegraphics[width=\linewidth]{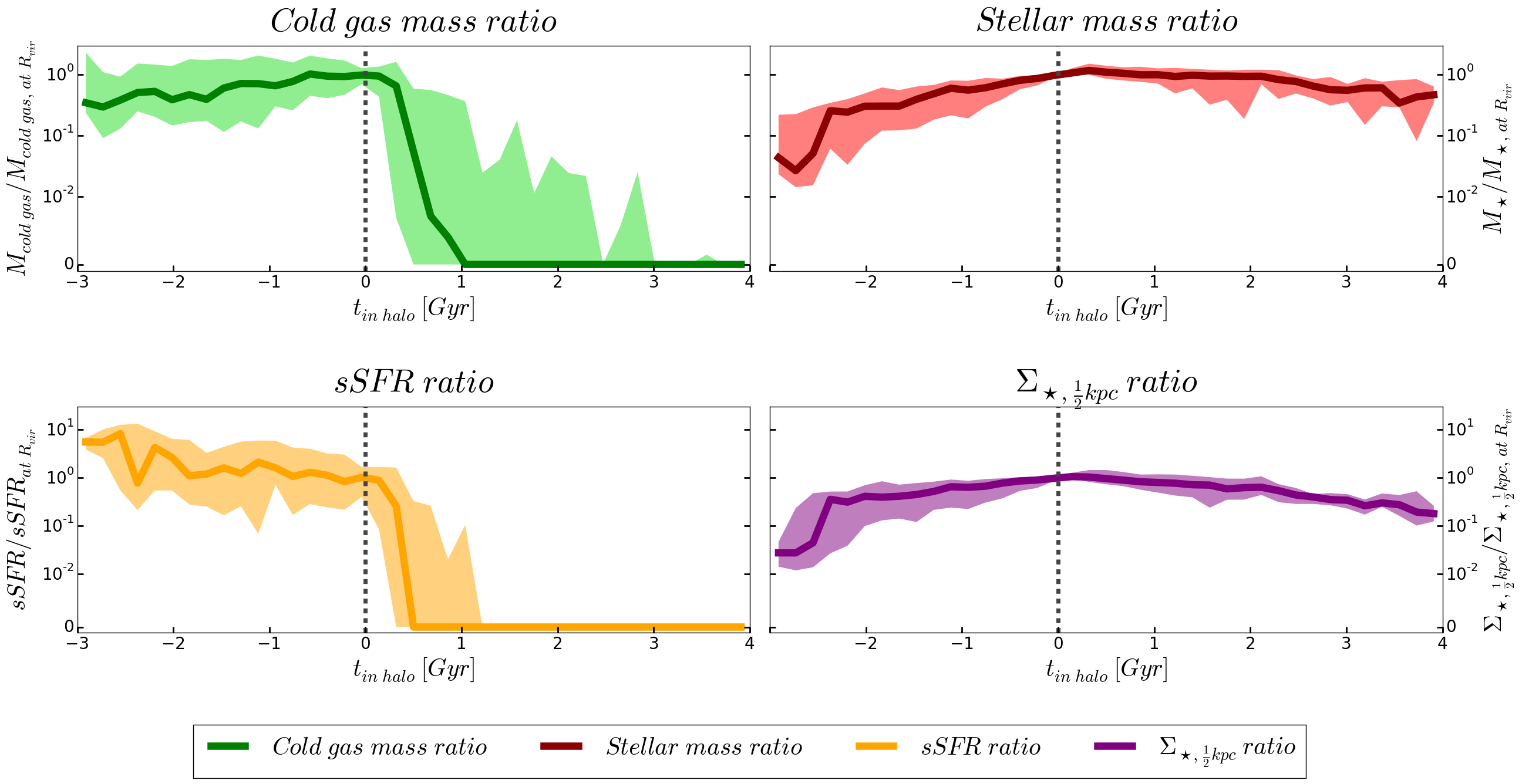}
    \caption{\small \textbf{Evolution by $t_{\textrm in\ halo}$} - A stacked plot of the complete satellite sample by time in the halo.
    Bold line represents the median while the upper and lower shade boundaries are 16\% and 84\% of the sample.    All the values are normalized to the value at the entrance to the halo. Green, orange, red and purple line represent the cold gas component, sSFR, stellar mass and the stellar surface density at 0.5 kpc radius respectively. \hspace{0.5cm}
    We can see a major event occurring for most of the SGs at approx 0.5 Gyr time inside the halo. This event involves a sharp drop in cold gas and sSFR, which are very correlated. The drop is happening in less than 100 Myr which is the simulation time resolution.
    Also, we can see show some growth in stellar mass and surface density, which shows that after this 0.5 Gyr event there is an effect of tidal stripping and tidal heating. No cold gas added to the SGs after entering the halo which suggests in general starvation is an active process. After 1 Gyr in the halo most of the SGs are quenched, thereby, there is a scale in which we can say that after it, a SG would not be star-forming.
    }
\label{fig:sqm-sat_evo_t_in_halo}\end{figure}

Fig~\ref{fig:sqm-sat_evo_t_in_halo} shows the evolution of SGs by time spent in halo, and reveals several features. First, we observe a sharp drop in sSFR which shows that overall, quenching of SGs is a rapid process. The star formation starts to drop after around $0.5$ Gyr. Even though some cold gas is remaining, By $1$ Gyr, the SGs has stopped forming stars. Note that the gas mass in any point in the halo does not increase which suggests that starvation is at least a general phenomena.
Simultaneously, we observe an increase in both the stellar mass and the inner $0.5$ kpc radius stellar mass until about $0.5$ Gyr, after which both properties start to decrease while the surface density decrease much more rapidly, which suggests a change in stripping and heating after this point and the importance of the tidal heating process.

So the raising question is what happens at 0.5 Gyr in the halo?

\begin{figure} \centering\includegraphics[width=\linewidth]{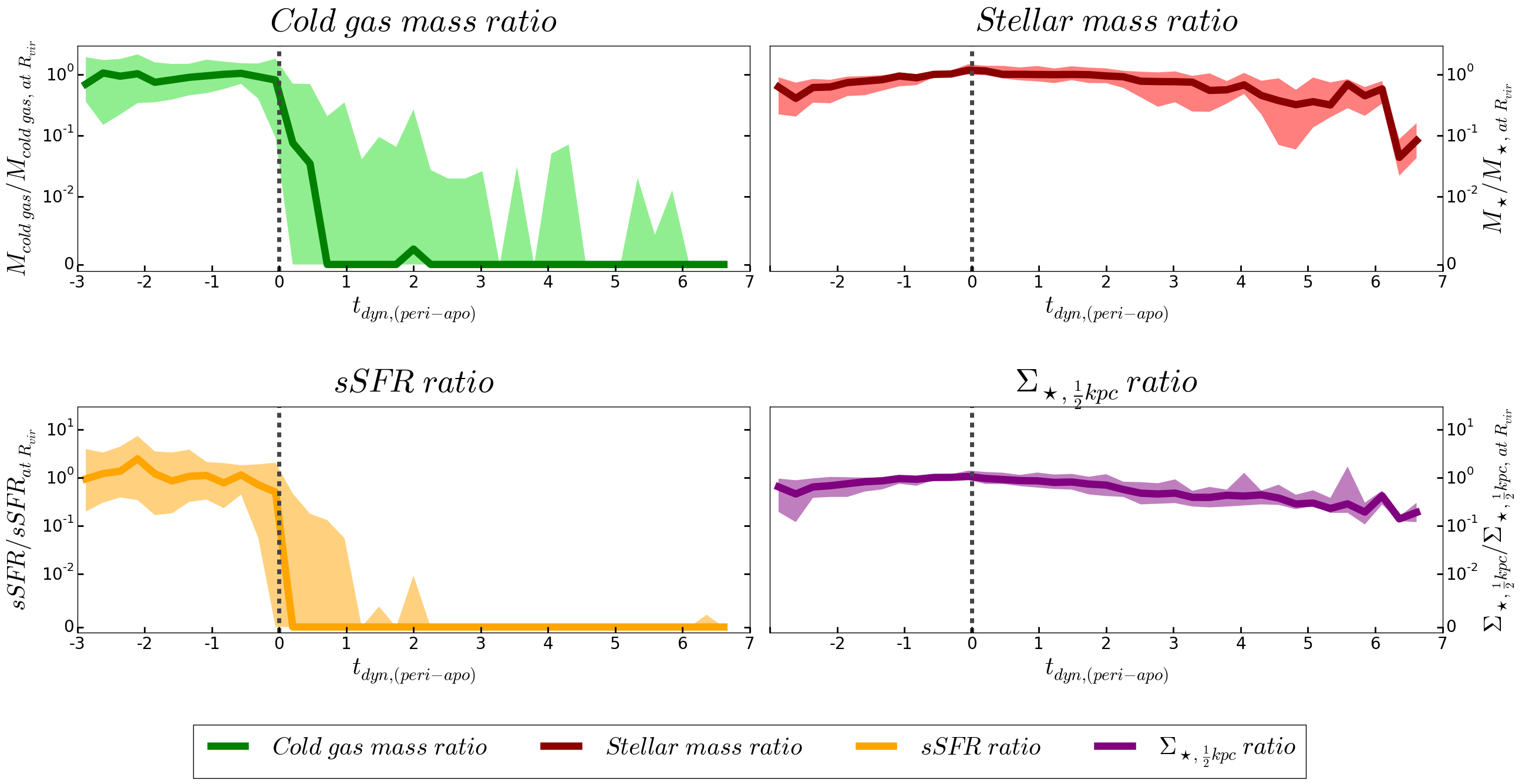}
    \caption{\small \textbf{Evolution by $t_{\textrm{dyn}\ (\textrm{peri,\ apo)}}$}  - A stacked plot of the complete satellite sample by dynamical time as defined on \S\ref{sec:met-event_properties}. Similar structure and colors as in caption of fig~\ref{fig:sqm-sat_evo_t_in_halo}. We recall that the time scale and offset are chosen such that t=0 is set as first peri-center and t=1 is set as first apo-center, by this linear approximation method, t=2 would be around the second peri-center an so on. \hspace{0.5cm} 
    We can see that most of the SG quenching occurs at the first peri-center ($t_{\textrm{dyn}\ (\textrm{peri,\ apo})} = 0$), as a sharp drop in cold gas and sSFR happens. And utmost of the SGs quench after the second prei-center ($t_{\textrm{dyn}\ (\textrm{peri,\ apo})} = 2$). Also as mentioned at Fig~\ref{fig:sqm-sat_evo_t_in_halo} we can see that as time in halo passes the stellar surface density and the stellar mass of the SG is decreasing but with increasing of its variance.
    }
\label{fig:sqm-sat_evo_t_dyn_peri-apo}\end{figure}

To answer this question we plot in Fig~\ref{fig:sqm-sat_evo_t_dyn_peri-apo} the same properties as function of $t_{\textrm dyn\ (\textrm peri,\ apo)}$. The same sharp drops are now observed at $t_{\textrm dyn\ (\textrm peri,\ apo)}$=0, suggesting that $0.5$ Gyr corresponds to the first passage time to the pericenter. For most of the SGs that still have gas when reaching first peri-center, the star formation period persists, and the cold gas survives until the second peri-center.
For the stellar components, we observe the same decays after the peri-center but not before; this suggests that after the first peri-center, the stellar process is dominated by tidal stripping and heating. 

We should note that the quenching timescale is around $100$ Myr (Fig~\ref{fig:sqm-sat_evo_t_in_halo}), which shows that the process is rapid.
To conclude, these plots imply that the main force that drives quenching is the ram pressure force
as the gas is stripped away from the inner core of the satellite. At this stage of time evolution, the stellar part is mostly intact as it decreases slightly there.

% ----------------------------------------------------
\section{Evolution of Satellites Versus the Distance to the Central Galaxy} \label{sec:sqm-Rorbit-evo}
This section describes the evolution of SG properties as a function of distance to the central galaxy.

The evolution is stacked against the SG distance to the central galaxy normalized by the virial radius, $R_{\textrm orbit}/R_{\textrm vir}$, either for all identified occurrences of the SG (in the observer's referential) or all occurrences until first peri-center.

\begin{figure} \centering\includegraphics[width=\linewidth]{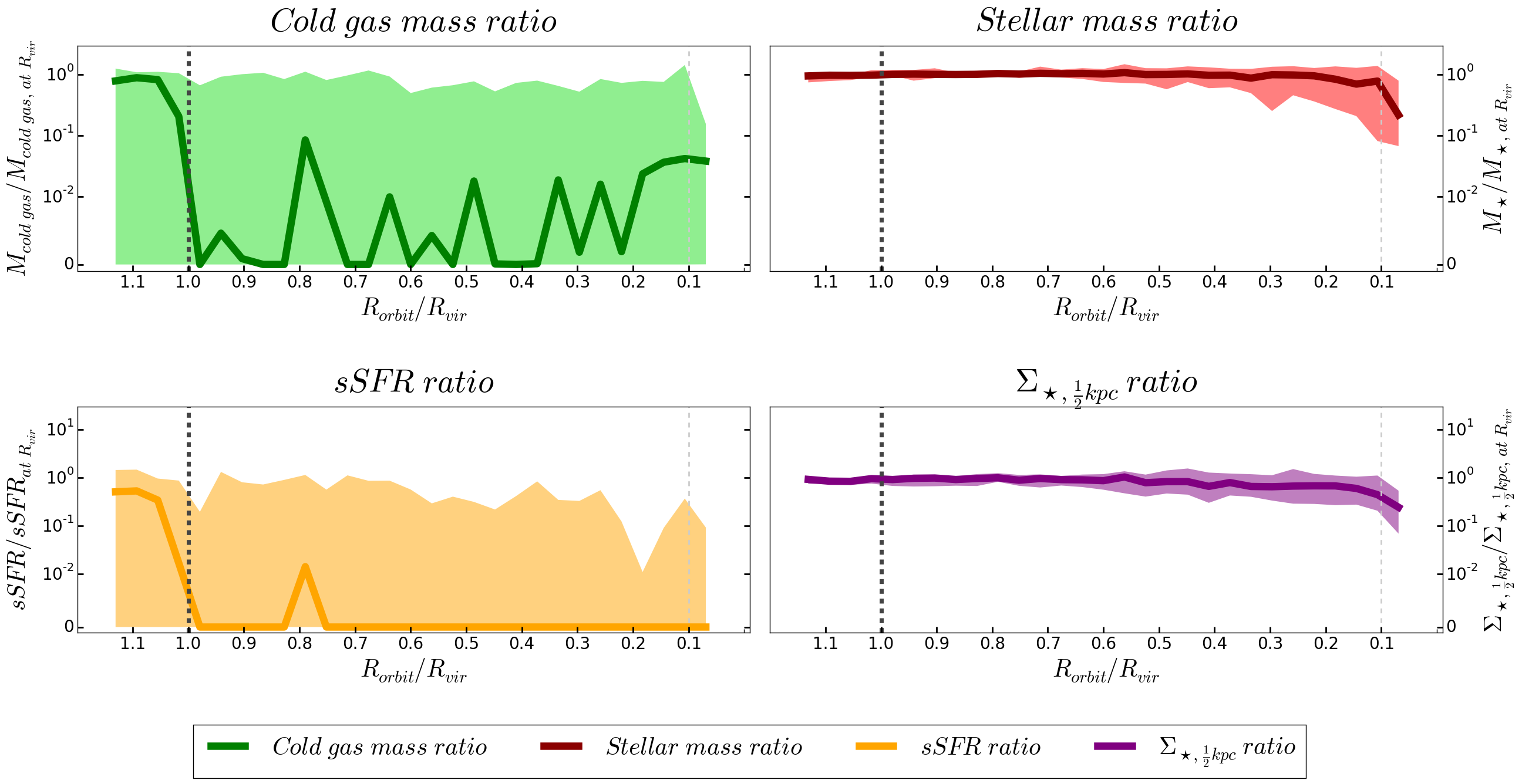}
\caption{\small \textbf{Evolution by $R_{\textrm orbit}/R_{\textrm vir}$ } 
- A stacked plot of the complete satellite sample by $R_{\textrm orbit}/R_{\textrm vir}$. Similar structure and colors as in caption of fig~\ref{fig:sqm-sat_evo_t_in_halo}. \hspace{0.5cm}
we can notice that first, most of the SG in the halo are quenched. Second, the more SG is close to the central galaxy, the more it is disrupted with high variety in disruption types, due to tidal forces, this disruption happens mostly at range which is less than $0.3 R_{\textrm vir} $. 
}
\label{fig:sqm-sat_evo_Rvir}
\end{figure}

Fig.~\ref{fig:sqm-sat_evo_Rvir} shows that most of the galaxies in the halo are quenched, as they do not form stars and do not contain cold gas.  Unlike galaxies outside of the halo, which are usually star-forming, as was shown in Section~\ref{sec:sqm-sample}. The Stellar mass ratio decreases as $R_{\textrm orbit}$ gets small, starting from about $R_{\textrm orbit} \sim 0.3 R_{\textrm vir}$. Therefore, we should find below this limit deformed stripped satellites; we also expect them to feature a stellar mass drop within their core due to heating.

Another interesting point, which contradicts the theory suggested by \citeA{Woo2017}, is that we do not observe significant stellar mass growth.

\begin{figure} \centering\includegraphics[width=\linewidth]{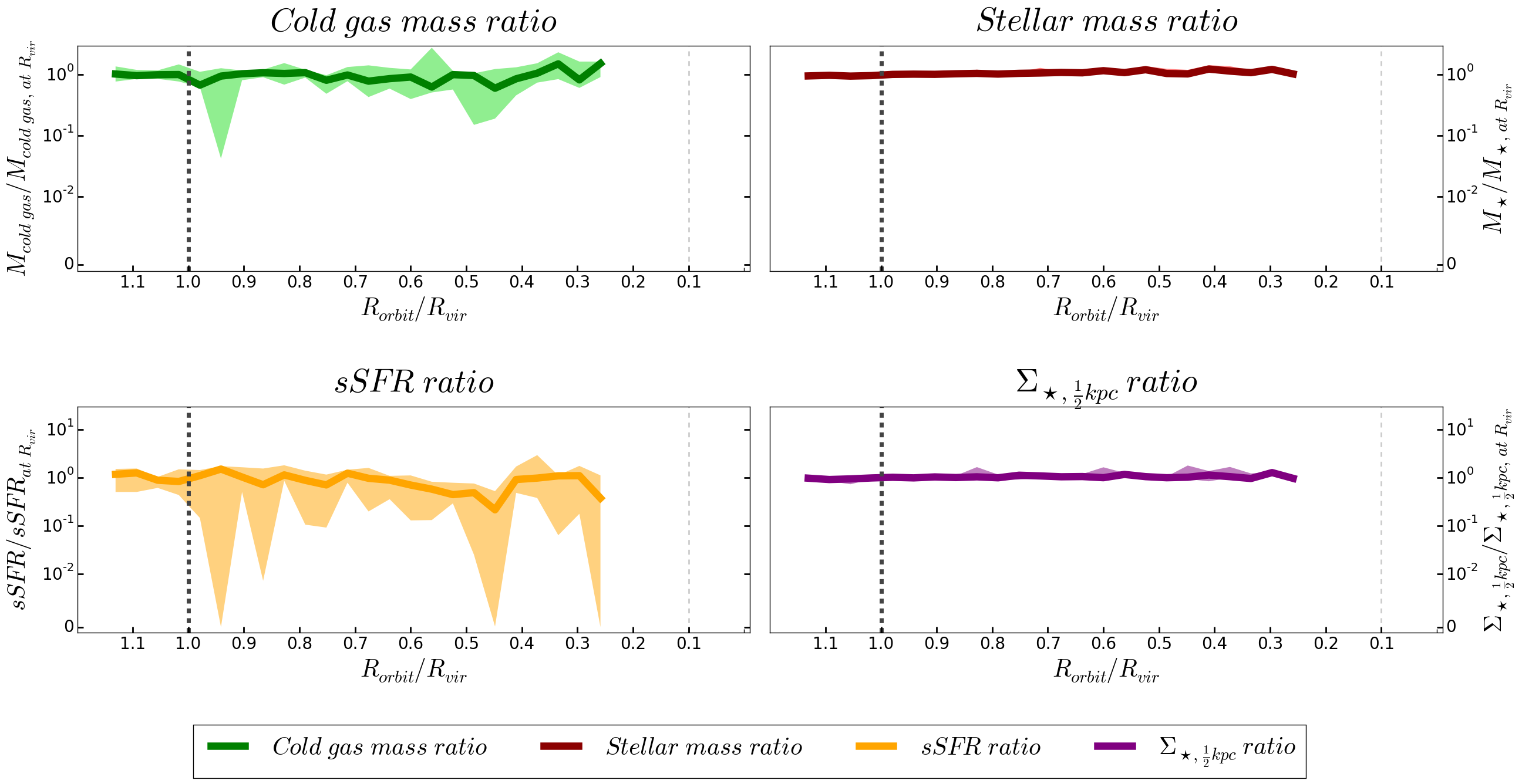}
\caption{\small \textbf{Evolution by $R_{\textrm orbit}/R_{\textrm vir}$} until first peri-center - A stacked plot of the complete satellite sample by $R_{\textrm orbit}/R_{\textrm vir}$. Similar structure and colors as in caption of fig~\ref{fig:sqm-sat_evo_t_in_halo}. \hspace{0.5cm} We can see here that before the peri-center, even relatively close to the central galaxy at $0.3 R_{\textrm vir} $ there is no drop in cold gas mass, meaning there is importance to the peri-center itself instead of just the distance. We can also see that there is no significant growth in stellar mass or cold gas mass relating the distance, meaning the tidal heating or stripping are less active before the first peri-center and there is no significant gas accretion on those regions.}
\label{fig:sqm-sat_evo_Rvir_until_peri1}\end{figure}

A more thorough explanation is provided by interpreting Fig.~\ref{fig:sqm-sat_evo_Rvir_until_peri1} in terms of evolution until the first peri-center. We observe no major evolution of the SG properties until the first peri-center. Most changes viewed in Fig.~\ref{fig:sqm-sat_evo_Rvir} occur due to the post pericenter effect, including the appearance of disrupted quenched satellites. We, therefore, conclude that any quenched disrupted galaxy around halo has already passed at least once around the peri-center of its orbit. 

Regarding the evolution of the stellar mass and cold gas, we have already seen that after the first peri-center, neither the gas nor the stellar masses grow; it drops for the gas mass. We also note that Fig.~\ref{fig:sqm-sat_evo_Rvir_until_peri1} shows that even before the first peri-center, no significant growth observed either which again suggests starvation mechanism is active.

% --------------------------------------------
\section{Nun Path \texorpdfstring{(\nunimg) }{} - Evolution Track by star formation vs compactness}
\label{sec:sqm-nun-path}
Now that we have better understood the evolution of the satellites with respect to time and location, one remaining point is to study their evolution on the general diagram of specific star formation (sSFR) vs. inner surface density ($\Sigma_{\star, \frac{1}{2} \textrm kpc}$). This diagram will allow us to address our original question, namely, how do satellites evolve and quench with respect to the central galaxies.

\begin{figure} \centering
\includegraphics[width=\linewidth]{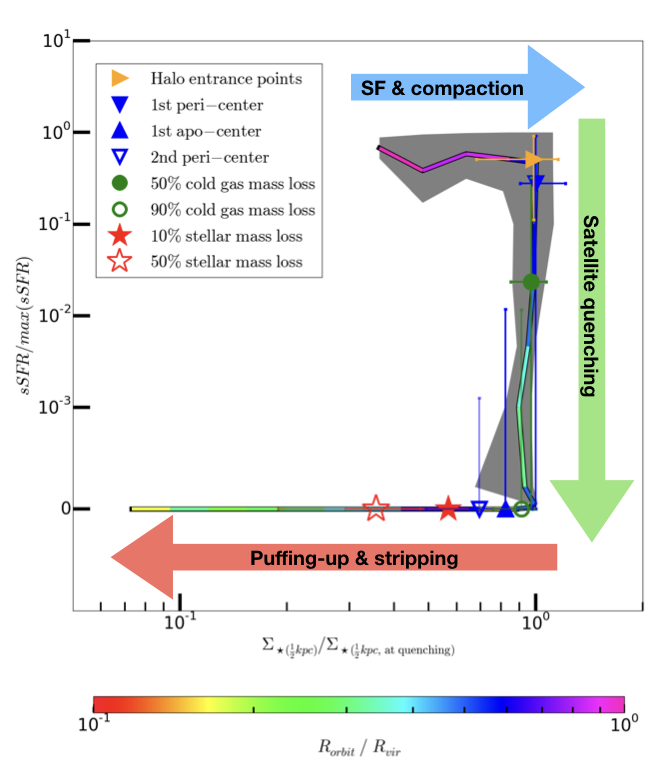}
\caption{\small \textbf{Nun (\nunimg) path - SGs Evolution tracks in the diagram of sSFR-$\Sigma_{\star , \frac{1}{2}kpc}$} - Stacked picture of 89\% of satellite galaxy evolution tracks.
Each track is normalized by its maximum star formation rate and by its average surface density at quenching. The solid line shows the average track of the satellite population and the gray area correspond to the 16\% and 84\% percentiles. The solid line colored as a function of the satellite distance from the center normalized by the virial radius of the central galaxy.
\hspace{0.5cm} The SG evolution track looks like the Hebrew letter Nun (\nunimg) which is divided to 3 major evolution stages: (1) Halt in gas accretion with SG compaction at high sSFR as the SG keeps forming stars 
(2) Gas removal and rapid drop in the sSFR at the peri-center of its orbit within the host halo (3) Stellar heating and stripping that may lead to coalesce with the halo center, an ultra-diffuse galaxy (UDG), a compact elliptical galaxy (eCg) or a globular cluster (GC). }
\label{fig:Median_cartoon}
\end{figure}

We show in Fig.~\ref{fig:Median_cartoon} satellite galaxy evolution tracks in the diagram of sSFR-$\Sigma_{\star, \frac{1}{2} \textrm kpc}$. In this representation, the track of a satellite evolution has the same shape as the Hebrew letter Nun (\nunimg), and divided into three major stages of evolution:
(1) Halt in gas accretion with SG compaction at high sSFR as the SG keeps forming stars 
(2) Gas removal and rapid drop in the sSFR at the peri-center of its orbit within the host halo
(3)  Stellar heating and stripping that may lead to coalesce with the halo center, an ultra-diffuse galaxy (UDG), a compact elliptical galaxy (eCg) or a globular cluster (GC).

% (1) continuous star formation \& compaction, without new gas accretion (2) Sharp quenching (3) puffing-up \& stripping.

The first stage lasts until reaching the first peri-center and ``delays quenching''. Then, at the peri-center region, the cold-gas sharply drops due to ram-pressure. Last by tidal forces and depending on the orbit type and the mass ratio between the satellite and central galaxy, either a merger, a UDG, a compact elliptical galaxy (eCg) or even a globular cluster (GC); a short explanation is provided in Chapter \S\ref{chap:osq}, and a longer one in a future article.

\begin{figure} \centering
\includegraphics[width=0.45\linewidth]{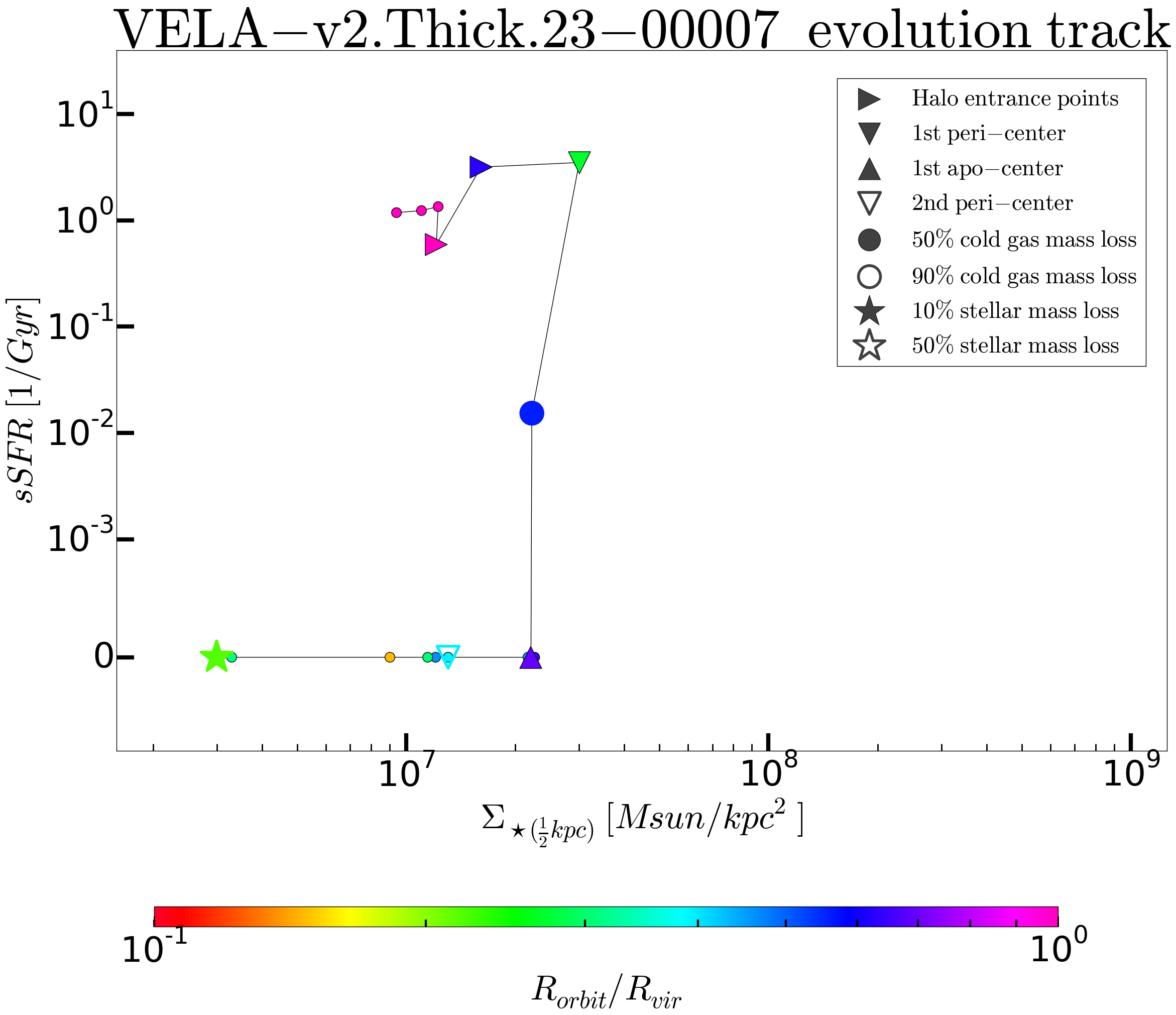}
\includegraphics[width=0.45\linewidth]{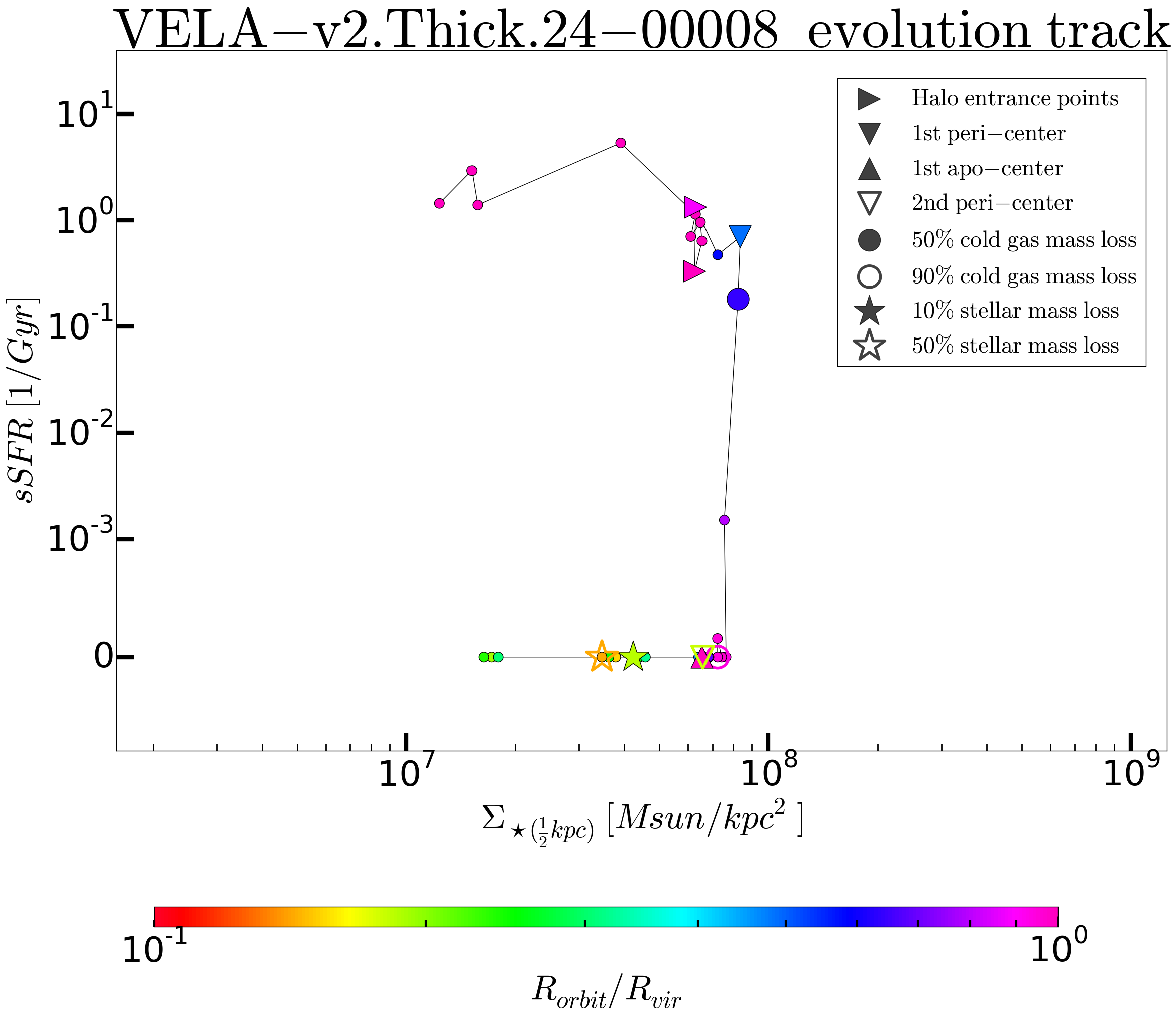}

\includegraphics[width=0.45\linewidth]{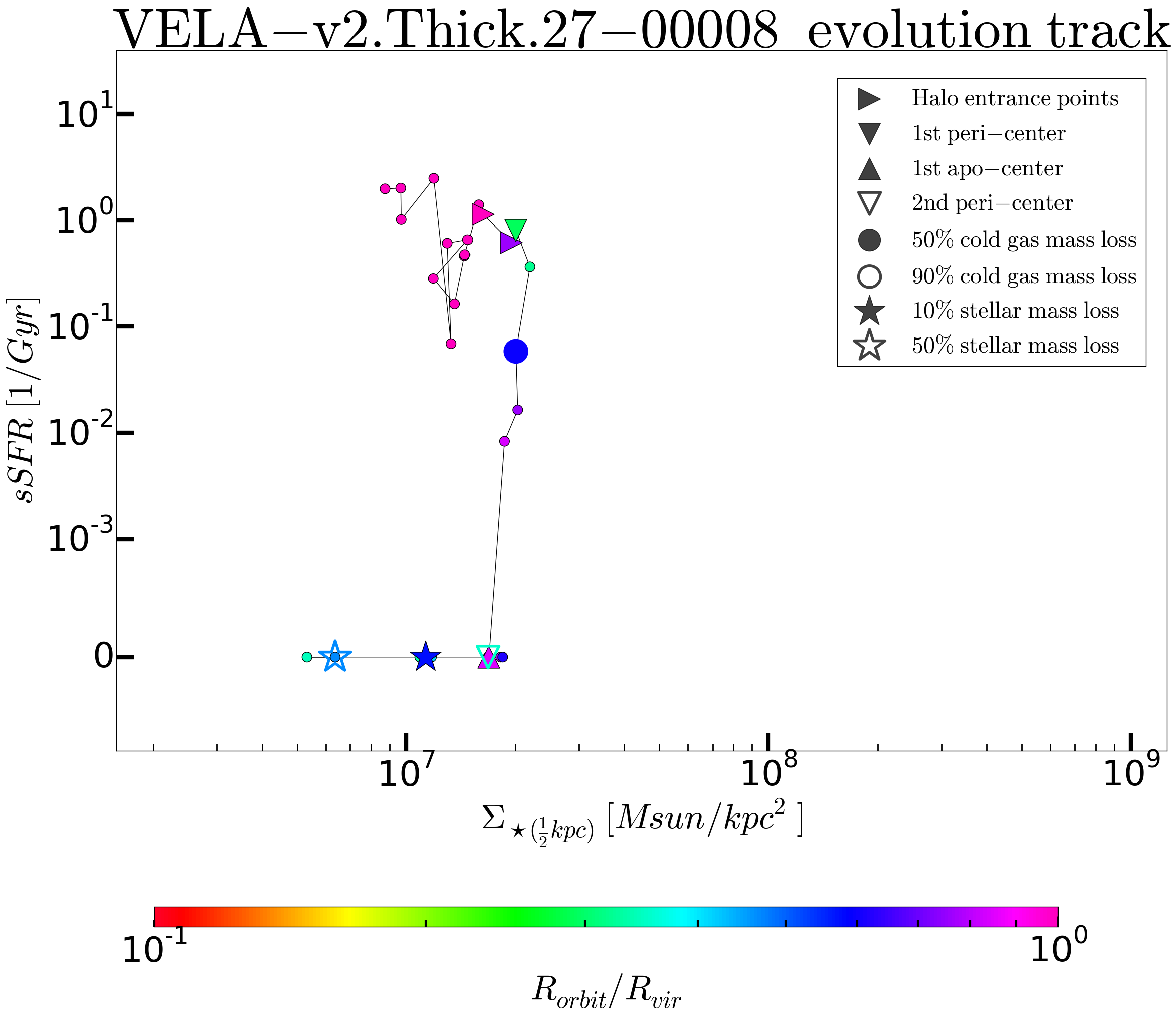}
\includegraphics[width=0.45\linewidth]{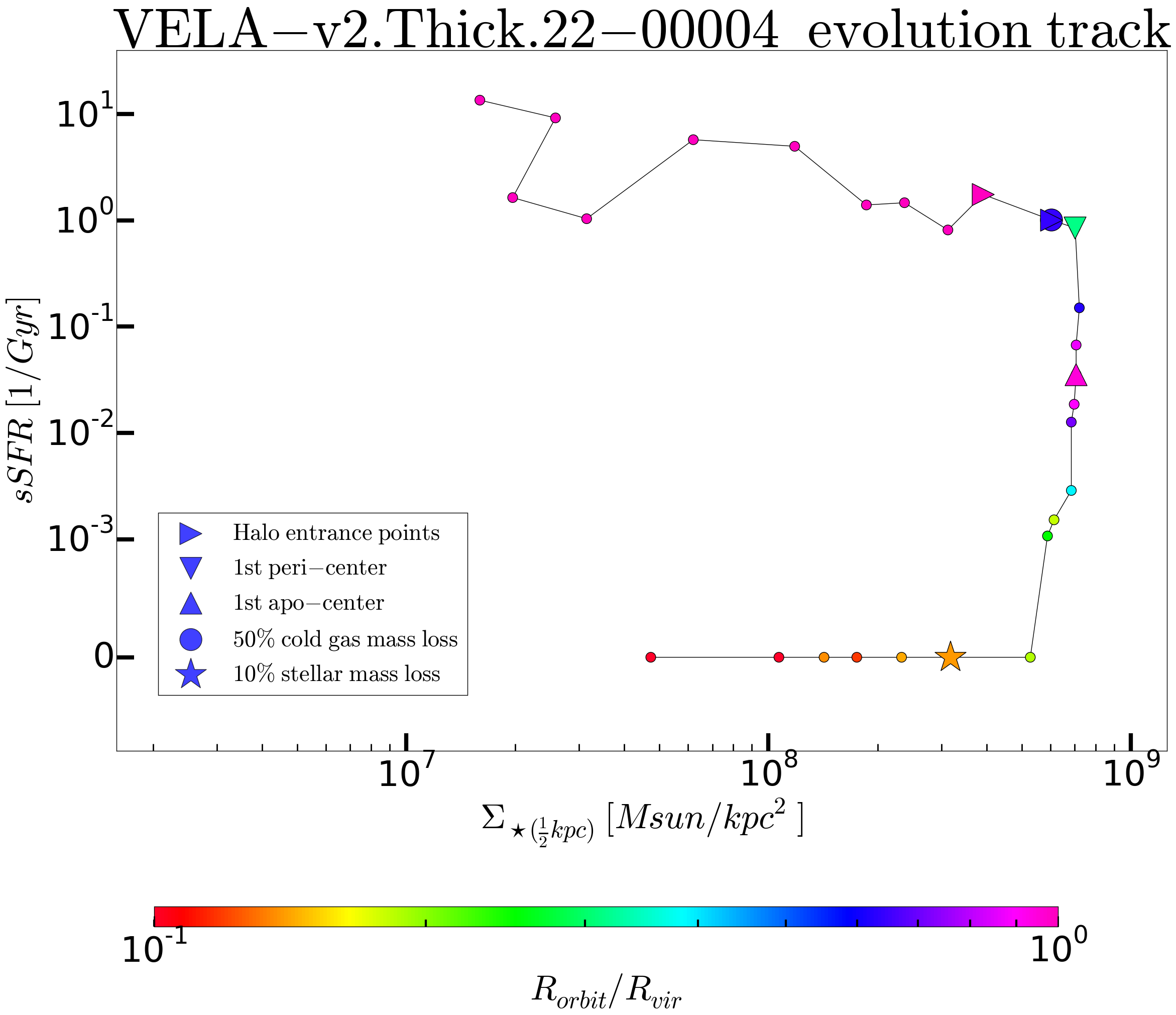}
\caption{\small \textbf{4 Satellites Galaxies in the Diagram of sSFR-$\Sigma_{\star, \frac{1}{2} \textrm{kpc}}$} - Each plot represent the total evolution of a SG from birth, first identification until last identification. Major events in the SG lifetime are marked with a specific marker. Halo entrance event (last outside the halo and first inside the halo), first and second peri-centers, first and second apo-centers and amount of masses loss, cold gas and stellar mass.\hspace{0.5cm} We can see that a distinguish Nun path can be observed in each one of the SG, Sharp drop in sSFR due to ram-pressure and then a significant stellar mass loss, when this whole mechanism is independent from the stellar surface densities.}
\label{fig:Median_cartoon_samples} \end{figure}

Beyond the population averages, we show in Fig.~\ref{fig:Median_cartoon_samples} not normalized evolution tracks of four different individual satellites. They are reasonably similar to the stacked trajectories, which confirms that deviations from the average normalized trajectory are small. Moreover, we do not find any correlation between the quenching event of the satellites and their surface density. This lack of correlation is in contrast with field galaxies, whose evolution track significantly depends on their surface density \cite{Dekel_Lapiner, Tacchella2016a}.

\begin{figure}[H] \centering
\includegraphics[width=0.50\linewidth]{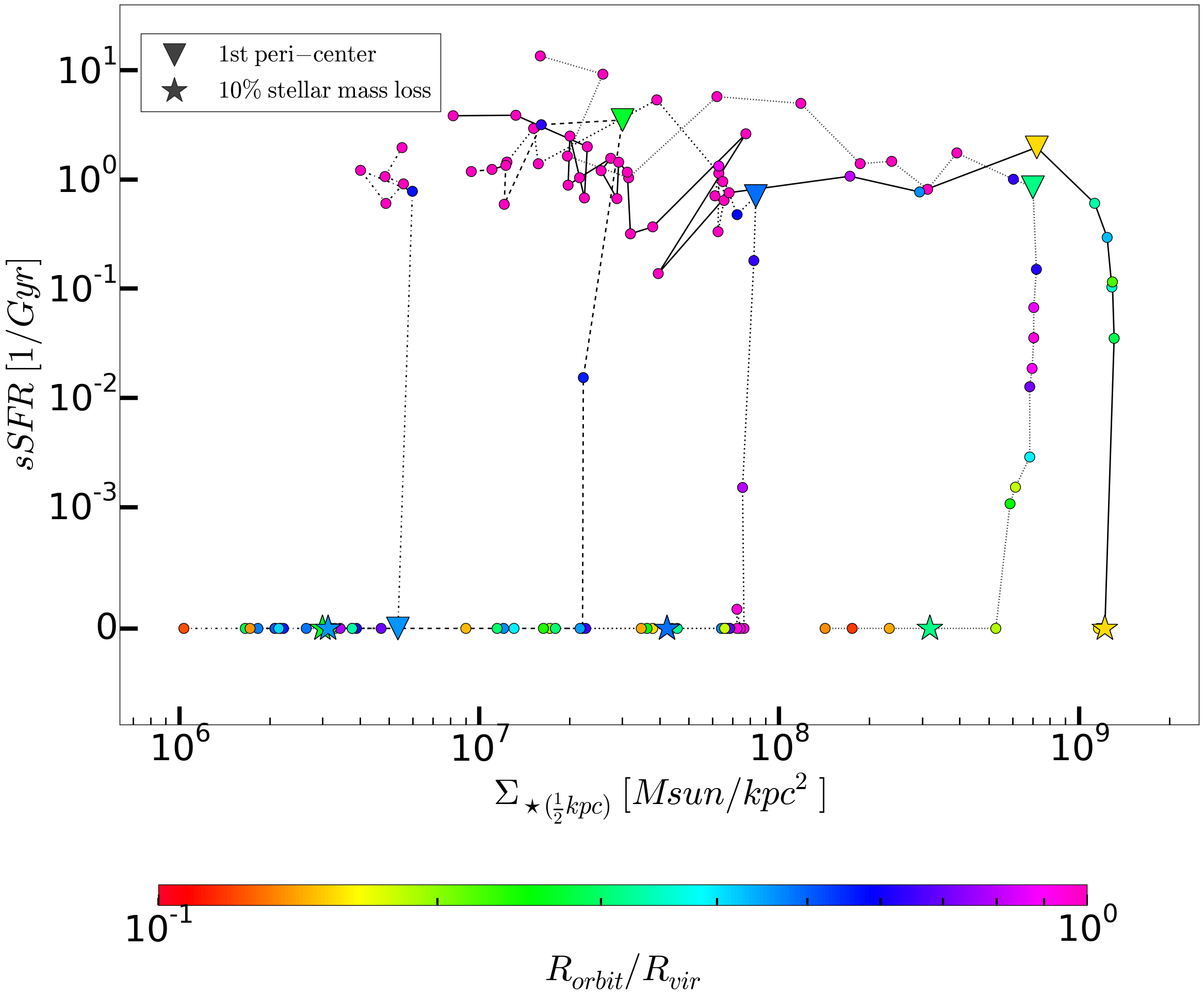}
\includegraphics[width=0.40\linewidth]{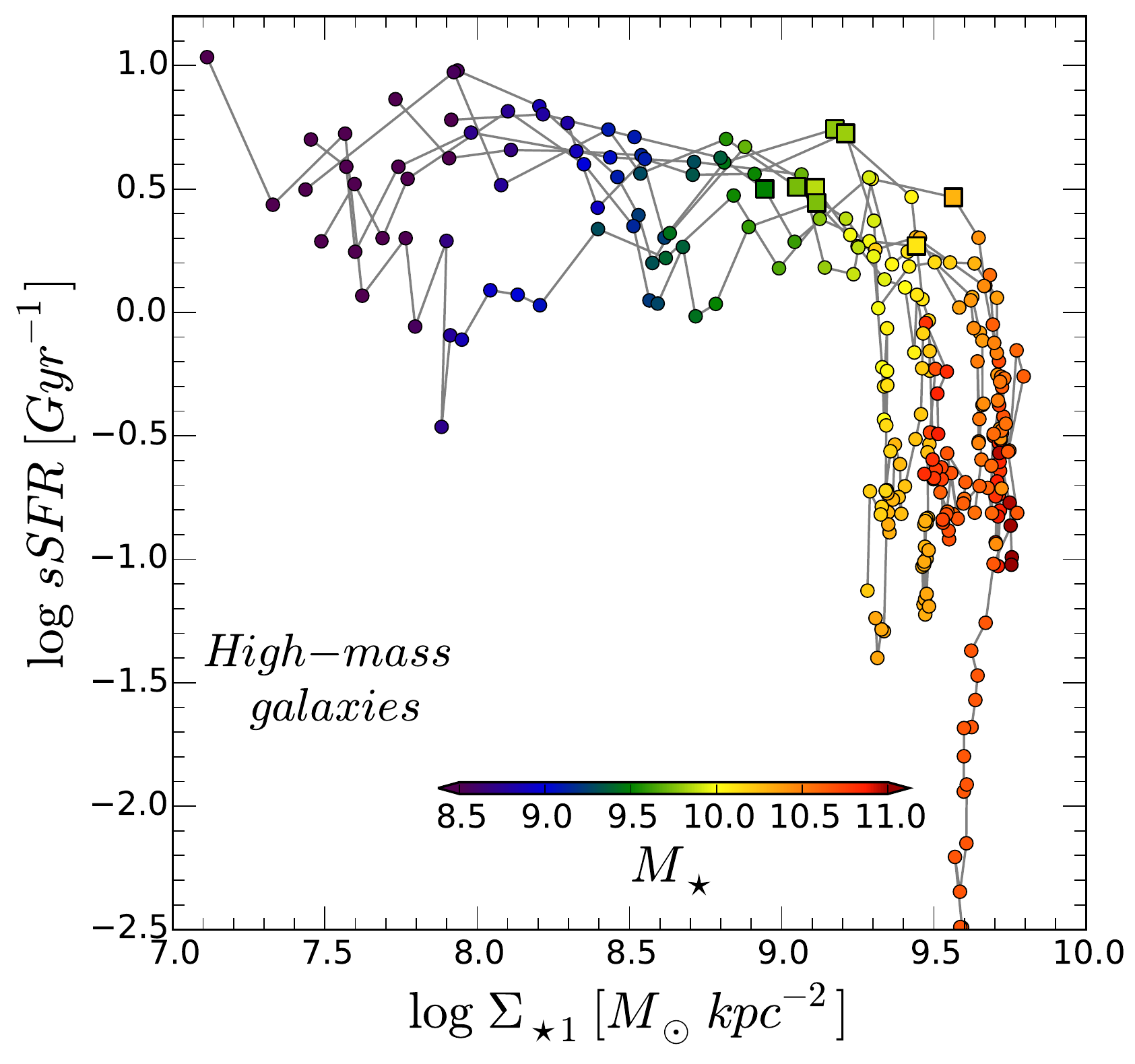}
\caption{\small \textbf{Satellites and field galaxies evolution tracks in the diagram of sSFR-$\Sigma_{\star}$} 
Left fig: SG evolution track, starting from the first event in the halo, marked with first peri-center life event and 10\% stellar mass loss life event. Colored by normalized distance to the central galaxy.
Right fig: Field galaxy evolution path presents the typical halo mass quenching process. The coloring here is by the field galaxy stellar mass. \hspace{0.5cm}
These to figures are summarizing the full quenching process, below $10^{10.5}$, quenching can be caused by SG quenching process as described in this thesis and above, whenever it is a field or SG the galaxy would quench at that mass scale when the later difference would be additional stripping on the case of SG. }
\label{fig:satellits_vs_field_quenching}
\end{figure}

This lack of correlation best illustrated by Fig.~\ref{fig:satellits_vs_field_quenching}, which shows evolution tracks for Satellite and field galaxies. Notwithstanding the field galaxies with stellar surface density above $10^{10.5}$ quench by halo mass quenching process as described at \S\ref{sec:intro-Bi-modality}, at SG in its mass range, the quenching is unrelated to stellar surface density. We, therefore, deduce from Fig.~\ref{fig:satellits_vs_field_quenching} that galaxies with stellar surface density smaller than $10^{10.5}$ quench mostly based on their orbit, and more specifically on their peri-center. Above this value, satellite quenching should occur via the same mechanism as field galaxies by halo mass quenching process \S\ref{sec:intro-Bi-modality}. Halo mass quenching is based on stellar density and AGN, which are an internal process. Therefore, it should not be affected by the environmental mechanisms of the halo. In the best case scenario, this quenching process should be even enhanced.

\section{Comparison to Observations} 
\label{sec:sqm-observations}
\begin{figure} \centering
\includegraphics[width=0.7\textwidth]{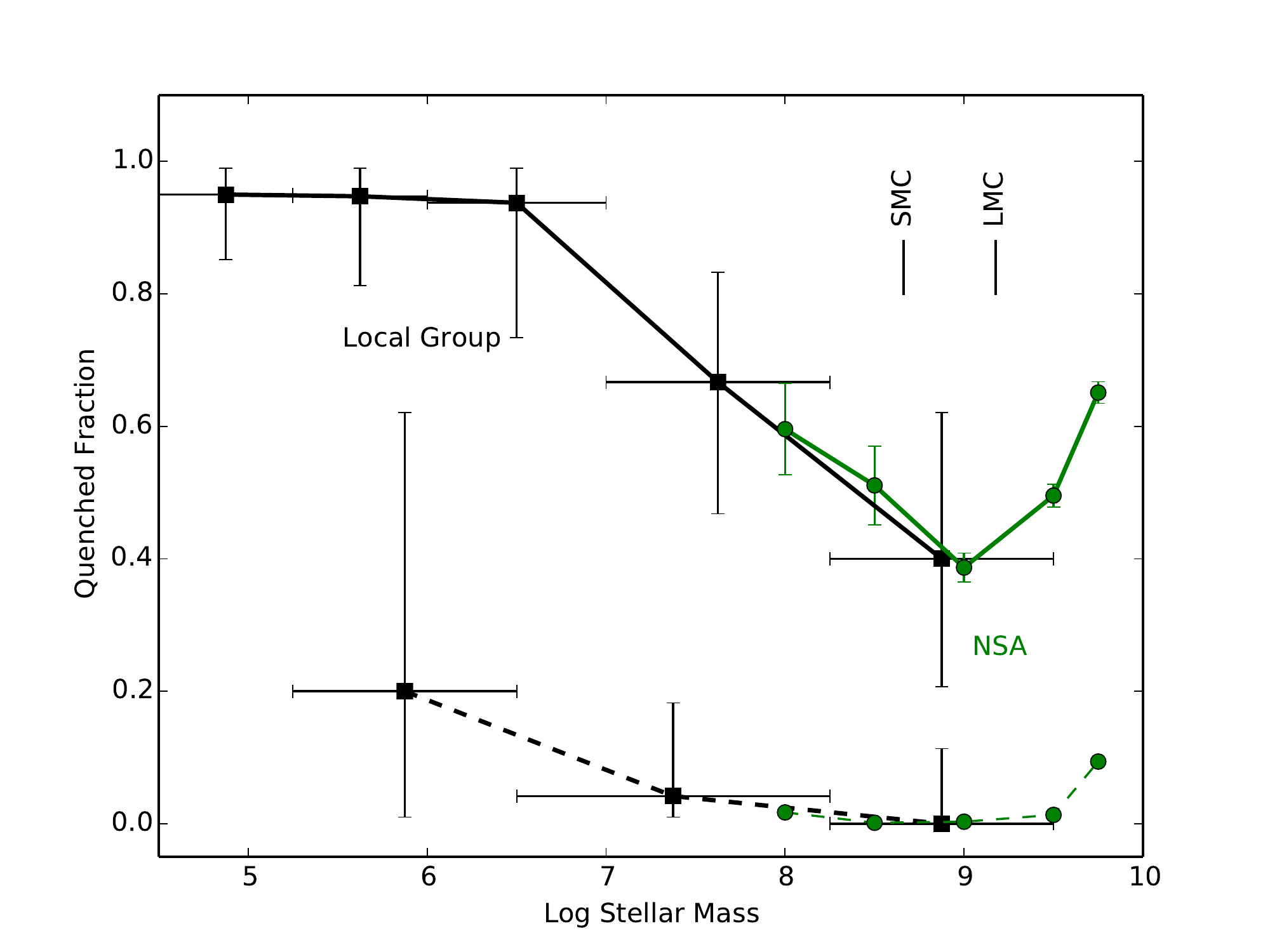}
\caption{\small Fraction of quenched satellites as a function of galaxy stellar mass (solid line), along with fraction of quenched field galaxies (dashed line). The data comprise three samples: dwarfs in the Local Group (black squares), more massive satellites from the NSA catalog after correction for contamination (green circles).\hspace{0.5cm} \protect\citeA{Slater2014}: There is a clear transition near $10^7-10^8\ M_{\odot}$ from nearly ubiquitous quenching of satellites at low mass to much lower quenched fractions at higher masses.}
\label{fig:Slater2014_f1}
\end{figure}

We've Mentioned on \S\ref{sec:intro-sat} that there is a mass dependency of the SGs to fraction of quenched SGs, example can be seen on fig  \ref{fig:Slater2014_f1}. 
We stated in \S\ref{sec:intro-sat} that the SGs to fraction of quenched SGs depends on the mass,  as illustrated by Fig.~\ref{fig:Slater2014_f1}. Additionally, we said in the same section that the surface density increases after quenching, as illustrated by Fig.~\ref{fig:Joanna2016_cartoon}. How are both these observations consistent with our proposed model, which states that quenching mostly depends on peri-center properties and only mildly the SG mass?

We now look at Fig.~\ref{fig:Slater2014_f1}, which shows the fraction of quenched galaxies as a function of the stellar mass for satellites around halos of $10^{12} M_{\odot}$ as well as for field galaxies with the same mass. We can see that most of the galaxies in this mass range that undergo quenching are SGs, demonstrating that for masses below around $10^{10.5}\ M_{\odot}$ (our mass quenching criterion), quenching is regularly a SG quenching mechanism

The conventional explanation in the literature is that there is a clear transition near $10^7-10^8\ M_{\odot}$; under this range, quenching occurs faster [than with the field galaxy mechanism]. Here, we suggest that this transition depends on the time spent by the SG in the halo. As the dynamical friction is proportional to the ratio of the mass of the halo over the mass of the SG \cite{Gal_Form_and_Evo}, less massive SGs, spend more time in the halo than massive SGs as they will suffer less dynamical friction. Therefore we should normalize this observable by dynamical time spent in the halo. Moreover, as shown by the Nun path described in \S\ref{sec:sqm-nun-path} we should also include the galaxies that underwent tidal stripping and whose mass, therefore, is underestimated. We argue that both of these effects would produce the same result.

\begin{figure}[!htb] \centering
\includegraphics[width=\linewidth]{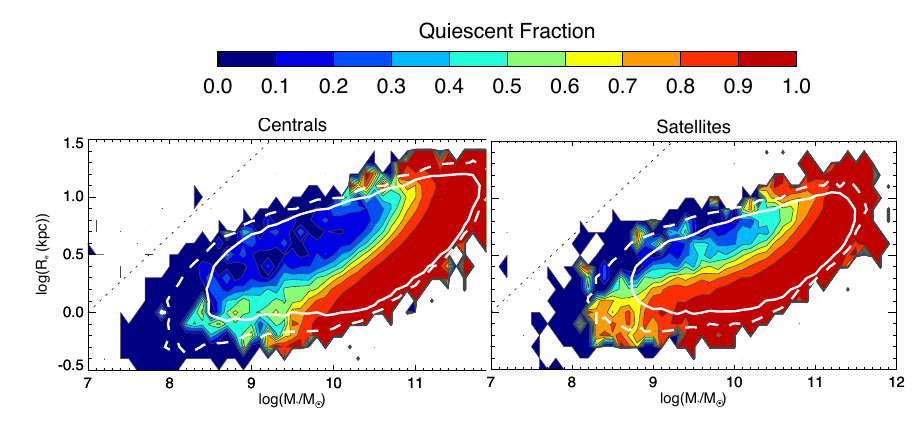}
\caption{\small The central (left) and satellite (right) galaxy quiescent fraction as a function of stellar mass $M_{\star}$ and effective radius $R_e$.  Contours are spaced every 0.1 in quiescent fraction.  The dashed line indicates constant M$_*$/R$_e^{1.5}$. Solid and dashed, white contours indicate the region where the number density of galaxies contributing to the calculation is greater than 100 per bin, and 20 per bin, respectively. This representation demonstrates that the quiescent fraction does not depend only on stellar mass, but also on galaxy structure. By \protect\citeA{Omand2014} }
\label{fig:Omand2014_f_sat_and_cen}
\end{figure}

This proposed interaction can be validated by observation, as it implies that among the quenched SGs, dense SGs should outnumber diffused ones. Indeed, dense SGs are more likely to survive several peri-centers whereas diffused are expected to disrupt tidally. Therefore, all the observed diffused SGs are at an early stage of entrance into the halo. We note that this is slightly counter-intuitive, as diffused galaxies have shallower potential wells, their surrounding gas should be less bound and therefore should be more likely to be stripped apart. Nonetheless, Fig.~\ref{fig:Omand2014_f_sat_and_cen} from \cite{Omand2014} confirms our claim: quenched fraction of SGs compared to field galaxies, contains higher fraction of dense SGs than diffused SGs. Other researchers have reported similar results; for instance, in \cite{Wang2018a, Wang2018b} the authors have shown that the bigger the bulge to disk ratio is in a SG, the more likely it is to be quenched SG.

\begin{figure} \centering
\includegraphics[width=0.5\linewidth]{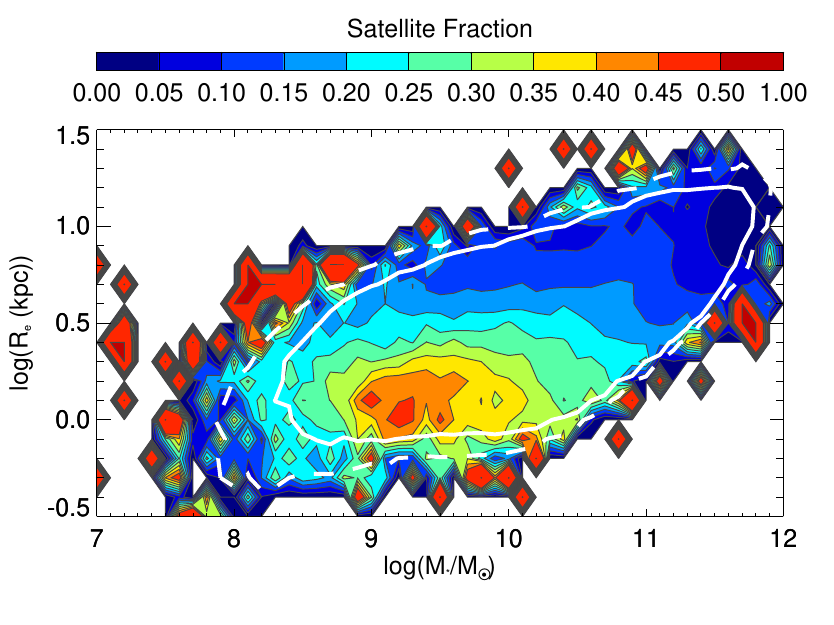}
\caption{\small The satellite fraction as a function of stellar mass $M_{\star}$ and effective radii $R_{eff}$. Differences between coloured contour levels are statistically significant within the solid white contours, which indicate the region where the number density of galaxies contributing to the calculation is greater than 100 per bin. The dashed, white contours indicate a lower number density of 20 per bin; correlations outside this region are of low statistical significance. By \protect\citeA{Omand2014} }
\label{fig:Omand2014_f_sat_on_cen}
\end{figure}

Fig.~\ref{fig:Omand2014_f_sat_on_cen} also shows that there are more dense SGs than dense field galaxies. This more extensive range of density variation is due to the tidal stripping and heating, which produce a diversity of SGs, and well predicted by our examples. It is also consistent with the work of \citeA{Hartley2015}, who showed that the power-law slopes of massive SG are systematically steeper than those of lower mass SG. They expected a higher number of low-mass SG and suggested that SG disruption could explain these results.

The extensive work of \citeA{Omand2014} also includes a model in which they take the star-forming field galaxy distribution and check different mechanisms to get the same SG quenched fraction distribution. 
If in addition to cutting from the field galaxies population part of the $R_{eff}$  in their model, they had decrease part of the $M_\star$, applying tidal heating and stripping (Shifting left the dense population). They could have had a satisfying fit between the field galaxies population after the transformation to the observed SG population. As this thesis is limited by size while the work of \citeA{Omand2014} should be explained with enough details, we did not include the full explanation here, and we encourage the reader to follow the work of \citeA{Omand2014} together with this explanation.

One last support for our claim comes from \cite{Tacchella2017}, which by age measuring techniques showed that dense SGs are older than dense field galaxies, whereas diffused SGs are younger than the diffused field galaxies. This age - density correlation is consistent with our finding that dense satellites are more resistant to tidal forces than diffused ones.

Interestingly, our proposed model solves some of the inconsistencies regarding the ram-pressure efficiency, which is sometimes too strong and sometimes too weak in SAMS. Indeed, our model predicts that the stellar time in the halo and the orbit are more important than the dark matter sub-halo time which as we have shown \S\ref{subsec:met-Results} is not always correlated.
% \textcolor{red}{SAMS citations? to ask Fangzhou}

Moreover, our model can explain the inconsistencies of Fig.~\ref{fig:Joanna2016_cartoon}. Indeed, as the graph binned by mass, some (we will claim most) of the dense quenched galaxies are remnants of bigger galaxies that were quenched and later stripped from their outer stellar sphere, keeping only their dense core to orbit the halo. Indeed we showed that the stellar component could decrease by 10\%-90\% and continue in orbit as a very compact stellar galaxy.

Finally, our model predicts that SGs with a mass ratio above $10^{10.5}$ should quench as central, which is confirmed by \cite{Gabor2014, Wang2018a}.

% \section{Discussion}
% \label{sec:sqm-dis}

% We conclude:
% 1. While looking at satellites, it is important to trace their stellar kernel.
% 2. Quenching seems to be a gas only mechanism which does not effect stars, as it strips gas from the kernels of the satellite.
% 3. The Quenching mechanism of a satellite is different from Field as it majorly caused by peri-center effects and it is not correlated with $\Sigma_{\star}$ 
% ....
% swan song
% gas bimodality

% the time evolution shows:
% 1. there are time scale, meaning observing a SG star-forming is limited by a timescale which is probably related to its orbit and the dynamical time in the halo (between 1-2 peri-centers)
% 2. the mass and stellar surface density also decrease by time, meaning the longer galaxy is in the halo the less stars it has, but the variance of the amount of stellar mass loss is getting bigger with time.
% 3. there is quenching in the peri-center, it is ram-pressure.
% 4. The process is rapid, less than 100 Myr which is the resolution limit. there is a bimodality of quenching on first peri and second, more is rare.

%\clearpage
%
% File: chap05.tex
% Author: Tomer Nussbaum
% Description: Origin of quenching
%
% \newcommand{\todo}[1]{{\color{red}\bf Tomer: #1}}
%

\let\textcircled=\pgftextcircled
\chapter{Origins of Satellite Galaxy Quenching}
\label{chap:osq}

\initial{W}hat are the causes of the quenching of satellite galaxies? We will explore the main driving forces acting on satellites (see \S\ref{sec:osq-analytical}) followed by some case studies (see  \S\ref{sec:osq-f_evo_examples}) and analysis in different regimes (as a function of time in \S\ref{sec:osq-f_time_evo} and as a function of distance to the central galaxy in \S\ref{sec:osq-f_Rorbit_evo} ). We will propose an explanation to ``starvation'' (\S\ref{sec:osq-starv}), which is the halt of gas accretion, and review additional quenching causes (\S\ref{sec:osq-sdqc}), illustrating why they are sub-dominant. Later, the results of this work are later compared to observations (\S\ref{sec:osq-obs}). 
% We close with a short discussion (\S\ref{sec:osq-dis}).

\section{Main Quenching Causes}
\label{sec:osq-analytical}
Let's begin by listing all the possible mechanisms operating on a SG: the gravitational force (both self-gravity and the tidal force), environmental gas flow forces (ram-pressure, Kelvin-Helmholtz instability, shock or gas turbulence), starvation -- i.e. the lack of gas accretion --, feedback (gas outflows due to radiation, SNs, etc), gas depletion by star formation and the influence of magnetic fields.

We will focus in this section on the tidal force, ram-pressure and self-gravity assuming that starvation is a mechanism active at all time and that the other mechanisms are sub-dominant. We will later justify those assumptions with the starvation ``Pac-Man'' model (\S\ref{sec:osq-starv})
and an analysis of possible SG quenching mechanisms (\S\ref{sec:osq-sdqc}).
%The mechanisms such as:
We wish to emphasize that since starvation is an active mechanism, no new gas enters the SG; therefore, the quenching process depends only on the gas inside the SG. This means we only need to find a mechanism to remove the gas out of the SG: once the gas is outside the SG, the SG will not accrete it again because of starvation. 
 
If we wish to choose one dominating quenching mechanism out of those three proposed mechanisms, the answer is clear. In \S\ref{chap:sqm}, we showed that the gas, and specifically the cold gas, of any SG is vanishing rapidly after entering the halo (in relation with the pericenter) while the stellar mass of the SG is kept for a much longer time. This indicates that the primary mechanism affecting the SG is acting only on the gas component. Amongst the two possible sources, one is related to gravity (the tidal force) and the other to pressure on gas (ram-pressure), so there are not many options left —- the dominant force quenching the SG needs to be ram pressure stripping. 

We now wish to quantify the different mechanisms: when, where and to what extent are they active.
%and watch when, where and how much each of them active. \\
We will describe the forces following a method described in \S\ref{sec:met-forces}. As the VELA central galaxies halos are approximately isothermal, we will approximate the tidal force as an isothermal tidal force. Using the \citeA{Dekel_Devor2003} model, this means that $\alpha$, the average density slope is $\alpha=2$ and that the maximum outward force is equal to the maximum inward force. From now on, we will use subscripts s, h, and g to refer to quantities pertaining respectively to the satellite, to the halo, and to the cold gas.

Following the derivations from \S\ref{sec:met-forces} we have 
\begin{align}
P &\equiv f_{\mathrm{ram}}\left(r_g\right) =
    \frac{ \rho_{\mathrm g,\ h} \cdot u^2}
    {\Sigma_{g,s}\left(r_g\right)}
    =\frac{ \rho_{g,h}(R) \cdot u^2}
    {\bar{\rho}_{g, s}(r_g) \cdot \frac{4}{3}r}
    \label{eq:P_eq} \\
T &\equiv f_{\mathrm{tidal}}\left(r_g\right) \approx
    \frac{GM_{h,\ R} \cdot r}{R^3} = \frac{V^2}{R^2} \cdot r_g
    \label{eq:T_eq} \\
G &\equiv f_{\mathrm{self\ gravity}}
    \left(r_g\right) = \frac{GM_{s,\ r_g}}{r_g^2} = \frac{v^2}{r_g} 
    \label{eq:G_eq}
\end{align}
where 
\begin{itemize}
\setlength\itemsep{0em}
\item $R$ is the distance from the central galaxy to the SG (previously noted $R_{\textrm orbit}$),
\item $r_\textrm{g}$ the radius of the cold gas,
\item $u$, the orbital velocity of the SG ($v_{\textrm orbit}$),
\item $\rho_\textrm{g, h}(R)$ (resp. $\Sigma_\textrm{g,h}$) the local density (resp. surface density) of the cold, halo gas at $R$,
\item $\bar{\rho}_\textrm{g, s}$ (resp. $\bar\Sigma_\textrm{g,s}$) the average cold gas density (resp. surface density) in the SG,
\item $V$ and $v$ the halo and satellite circular velocities at $R$ and $r$ respectively,
\item $M_\textrm{h,R}$, $M_\textrm{s,r}$ the total mass enclosed in spheres of radii $R$ and $r$ respectively,
\item and $G$ the gravitational constant. 
\end{itemize}
\noindent The second step between \eqref{eq:T_eq} and \eqref{eq:G_eq} uses the virial theorem, $V^2=\frac{GM}{R}$. 

Now, let us look at the ratios between the different quantities:
\begin{align}
\frac{P}{T} &= 
\frac{3}{4}\frac{\rho_{g,h}(R)}{\bar{\rho}_{g, s}(r_g)} \frac{u^2}{V^2} \frac{R^2}{r_g^ 2} 
% = \frac{3}{4} \frac{\frac{M_{g,h}}{(4\pi R_v)} \cdot \frac{1}{R^2}}{\bar{\rho}_{g, s}(r_g)} \frac{R^2}{r_g^2} \frac{u^2}{V^2} = 
% \frac{3}{4} \frac{\frac{M_{g,h}}{(4\pi R_v)}}{\bar{\rho}_{g, s}(r_g)\cdot r_g^2} \cdot \frac{u^2}{V^2}
\label{eq:PoT_eq} \\
% {\frac{M_{g,s}}{\frac{4\pi}{3} r_g^ 3} }
%  = \frac{\frac{M_{g,h}}{(4\pi R_v)}}{\frac{M_{g,s}}{r_g}} \frac{u^2}{V^2}  \label{eq:PoT_eq} \\
%  & = & \frac{1}{4\pi}\bigg(\frac{M_{g,h}}{M_{g,s}}\bigg)^{\frac{2}{3}}\frac{u^2}{V^2} = \frac{1}{4\pi}\frac{f_{g,h}}{f_{g,s}^{\frac{2}{3}}} \bigg(\frac{M_{h,R}}{M_{s,\ r_g}} \bigg)^{\frac{2}{3}}\frac{u^2}{V^2}
%  = \frac{1}{4\pi}\frac{f_{g,h}}{f_{g,s}^{\frac{2}{3}}}\frac{u^2}{V^2}\frac{R^2}{{r_g}^2}
%  \nonumber \\
\frac{P}{G} &= \frac{3}{4}\frac{\rho_{g,h}(R)}{\bar{\rho}_{g, s}(r_g)} \frac{u^2}{v^2} = \frac{3}{4}\frac{\rho_{g,h}(R)}{\bar{\rho}_{g, s}(r_g)} \frac{V^2}{v^2} \frac{u^2}{V^2}
% = \frac{3}{4}\frac{\frac{M_{g,h}}{(4\pi R_v)} \cdot \frac{1}{R^2}}{\bar{\rho}_{g, s}(r_g)} \frac{V^2}{v^2}\frac{u^2}{V^2} = 
\label{eq:PoG_eq} \\ 
% = \frac{\frac{M_{g,h}}{4\pi \cdot R_v^3} \bigg(\frac{R_v}{R}\bigg)^2}{\frac{M_{g,s}}{r_g^ 3}} \frac{V^2}{v^2}\frac{u^2}{V^2} =
% \frac{1}{4\pi} \frac{\frac{M_{g,h}}{R_v}}{\frac{M_{g,s}}{r_g} }  \frac{V^2}{v^2}\frac{u^2}{V^2}\frac{{r_g}^2}{R^2}
% =
% \frac{f_{g,h}}{4\pi} \bigg(\frac{M_{h,R}}{M_{s,\ r_g}}\bigg)^{\frac{4}{3}}
% \bigg(\frac{R_v}{R}\bigg)^2 \frac{u^2}{V^2} 
% =\frac{f_{g,h}}{4\pi}\frac{u^2}{V^2}\frac{{R_v}^2 R^2}{{r}^4}
\frac{T}{G} &= \frac{V^2}{v^2}\frac{r_g^ 2}{R^2}
= \frac{M_{h,R}}{M_{s,\ r_g}} \frac{r_g^3}{R^3} = \frac{M_{h,R}}{R^3}\cdot\bigg(\frac{M_{s,\ r_g}}{r_g^3}\bigg)^{-1}, 
\label{eq:ToG_eq}
\end{align}
\noindent where we again used the virial theorem $V^2=\frac{GM}{R}$. 

% Where we have used the isothermal approximation for gas in the halo, following \shortciteA{Zinger2018}: $\rho_{\textrm g,h}(R) = \frac{M_{g,h}}{(4\pi R_v)} \cdot \frac{1}{R^2}$, the average density definition $\bar{\rho}_{g, s}(r_g) = \frac{4\pi}{3} \frac{M_{g,s}}{r_g^ 3}$ and the virial theorem $V^2=\frac{GM}{R}$.
% \textcolor{red}{we should update here. first member to secound is just Isothermal desnsity of gas in the halo \cite{Zinger2018}}Where we've used the isothermal approximation for densities at the first steps of \eqref{eq:PoT_eq},\eqref{eq:PoG_eq},\eqref{eq:ToG_eq} where $R_v$ represents the virial radius and $M_{g,h},\ M_{g,s}$ are the gas masses in their $R,\ r_g$ spheres. Then we use $M_g = f_g*M_s$ when $f_g$ describes the cold gas fraction to the total mass. later we use the isothermal approximation $M \propto R^3$ and the virial theorem $v^2=\frac{GM}{r_g^ 2}$ at the last steps.

Some conclusions can be already drawn:
\begin{itemize}
\item $\nicefrac{P}{T}$ \eqref{eq:PoT_eq} -- from the term in $\frac{R^2}{r_g^ 2}$ we can conclude that as long as the SG is not too close to the center, the dominant force will be the ram pressure. Indeed, $R \gg r_g$ for most of the SG orbits. The most probable case in which $R \sim r_g$ is in a major merger (the SG has a comparable mass to the central galaxy), or on very eccentric orbits in which, as we will claim, $r_g$ drops rapidly as the $\frac{u^2}{V^2}$ factor is very high. 
The second important term $\frac{u^2}{V^2}$ differentiate different cases: 
(1) $\frac{u^2}{V^2} \approx 1$, which corresponds to a circular orbit -- in this case, the tidal inward force is opposing ram pressure so in the case where they are roughly equal, the SG gas can be compressed --,
(2) $\frac{u^2}{V^2}>1$, which corresponds to an eccentric orbit -- in this case, the ram-pressure and the tidal force are acting in the same direction, so the stripping is increased --, 
(3) $\frac{u^2}{V^2}<1$, which is an infrequent orbit -- in this case, the tidal force becomes more significant than the ram-pressure, but this orbit happens mostly in major mergers, which have unique dynamics that are beyond the scope of this thesis.
The third factor  $\frac{3}{4}\frac{\rho_{g,h}(R)}{\bar{\rho}_{g, s}(r_g)}$ is negligible with respect to $\frac{R^2}{r_g^ 2}$, as its value is less than 1 when $R$ is bigger than 0.1-0.3 $R_{\mathrm vir}$ and as inside this range $\rho_{g,h}(R) \approx \bar{\rho}_{g, s}(r_g)$ and still $r < 0.1\cdot R_{vir}$. Note that as $\rho_{g,h}(R)$ is redshift-dependant, the gas fraction in the halo is dropping with time. This means that the ram-pressure strength is decreasing with time, and hence, The proportion of SGs that are slowly quenching? will increase with time.
From this we conclude that, for SGs, ram pressure is the dominating force expelling the gas when the tidal force can oppose or support the stripping depending on the eccentricity of the orbit.

\item $\nicefrac{P}{G}$ \eqref{eq:PoG_eq} -- the remarkable factor here is $\frac{V^2}{v^2}$, which is proportional to $\bigg(\frac{M_{h,R}}{M_{s,r_g}} \bigg)^{\nicefrac{2}{3}}$ at the entrance to the halo since  halos are scale-free systems. Later, as gas is removed from the SG, $r_g$ becomes smaller, $M_{s,r_g}$  becomes smaller as well, and $\frac{V^2}{v^2}$ becomes even more prominent. The factor $\frac{V^2}{v^2}$ shows that SG quenching depends on the halo mass to satellite mass ratio. 
The second factor,  $\frac{u^2}{V^2}$, as mentioned on the previous item, involves the eccentricity: higher eccentricity causes high ram-pressure while ram-pressure is low on circular orbits. This can determine if the SG will quench fast or slowly. Third, the term $\frac{\rho_{g,h}(R)}{\bar{\rho}_{g, s}(r_g)}$ is negligible as mentioned earlier, with some redshift dependency that would cause higher quenching fraction at high redshift.

\item $\nicefrac{T}{G}$ \eqref{eq:ToG_eq} -- we can see here that there is a dependency between two ``average densities'': the halo density at $R$ and the SG density at $r_g$ (The same analysis can also be done for $r_s$, the stellar component radius). If the SG is in a region where the halo density is higher than the SG density, the tidal force will be stronger than gravity. However, as the SG is denser in most of its orbit region, gravity is stronger than the tidal force.
Therefore, stripping occurs mainly 
(1) during major mergers for which both of the densities are equal, 
(2) for diffuse SG, which have lower densities -- for example when a UDG enters the halo or after a SG becomes a UDG due to tidal heating at the pericenter --, 
(3) for SGs that are in very eccentric orbits and therefore enter the very dense region of the central galaxy. We have shown that ram-pressure is stronger than the tidal force in these cases, which means that this effect acts mostly on stars. It explains why dense SGs survive more the diffused SG as shown in \S\ref{sec:sqm-observations}. \end{itemize}

\section{Case Studies of Satellite Galaxy Force Evolution}
\label{sec:osq-f_evo_examples}
The following figures highlight three case studies of SGs seen in \ref{sec:sqm-evo_examples}.

\begin{figure} \centering
\includegraphics[width=\linewidth]{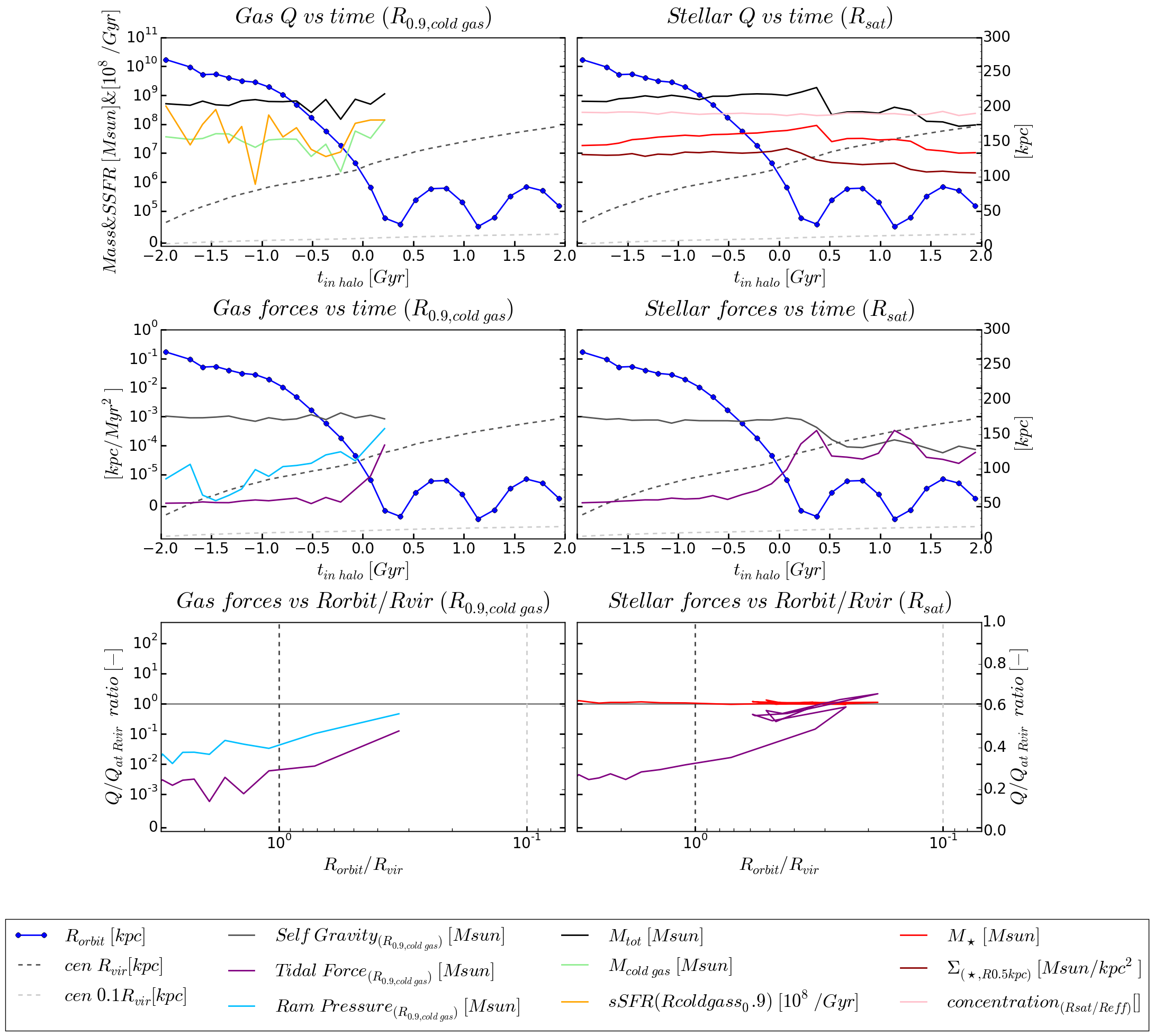}
\caption{\small \textbf{Satellite galaxy 07-041 forces evolution:}
Each plot depicts a satellite property against the time or against the distance, and splits to two columns; the left column focuses on SG gas properties enclosed in a sphere of $R_{0.9\ cold\ gas}$ radius; the right column focuses on SG stellar properties inside $R_{sat}$.
The first row describes basic galactic parameters over time in the halo: $R_{\textrm orbit}$ (blue), $R_{\textrm vir}, 0.1R_{\textrm vir}$ (dashed lines); on the right panel (gas properties) the total mass (black), the sSFR (orange), and the cold gas mass (light green): on the left panel (stellar properties) the total mass (black), the stellar mass (red), the surface density (dark red) and the concentration (pink).
The second row describes the forces related parameters: in addition to the $R_{\textrm orbit}$ and $R_{\textrm vir}$ over time in halo, there is also the ram-pressure force (light blue), the tidal force (purple), and the self-gravitational force (black).
The last row describe significant quenching and stripping parameters depending on the distance to the central galaxy normalized by their amount while entering $R_{\textrm vir}$: the ram-pressure, the tidal force, the self-gravity and the  stellar mass (with the same coloring as before).\hspace{0.5cm} This figure is completing Fig.~\ref{fig:sqm-sat_evo_07_00041} and shows the ``Fast Quenching'' case: ram-pressure and tidal force become stronger as the SG orbit is closer to the central galaxy; at the pericenter, ram-pressure becomes the dominant force and quench the SG.
On the left side we can see the tidal force effect on the system which is active on the edges of the SG: the tidal force heats the SG and strips the stellar mass. }
\label{fig:osq-sat_f_evo_07_00041}
\end{figure}

The ``Fast Quenching'' case shown in Fig. \ref{fig:osq-sat_f_evo_07_00041} demonstrates the analytic development we have shown earlier. 
Until the first pericenter, ram-pressure and the tidal force are smaller than the self-gravitational force, which keeps the cold gas inside the SG and enables it to continue to form stars.
At the pericenter, there is a steep rise in ram-pressure and tidal force that expells all the gas from the SG. As can be seen, ram-pressure is the dominating force in this quenching process. Later on, at the pericenters, we can see stellar mass stripping as the tidal force becomes active. Our calculation does not exactly apply at pericenter in this example, which means that at the pericenter itself, ram-pressure and the tidal force are more dominant relative to the self-gravity, which keeps the same intensity.

\begin{figure} \centering
\includegraphics[width=\linewidth]{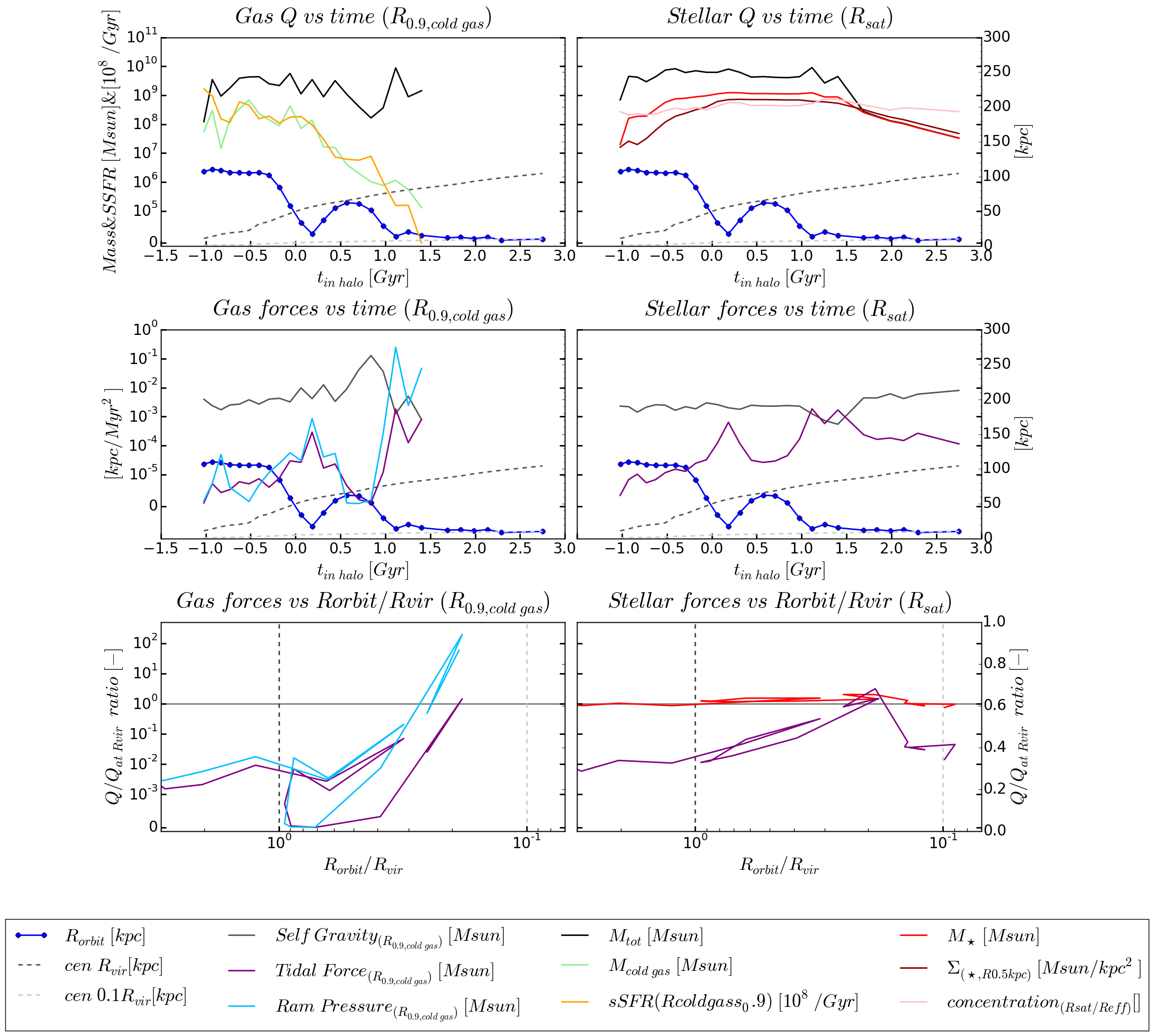}
\caption{\small \textbf{Satellite galaxy 22-004 forces evolution:} Similar structure and colors as in Fig.~\ref{fig:osq-sat_f_evo_07_00041}. \hspace{0.5cm}
This figure is completing fig~\ref{fig:sqm-sat_evo_22_00004} and presents the ``Slow Quenching'', ``Dark-matter loss'' and ``eCg formation''. We can see that since the SG orbit is circular, ram-pressure and the tidal force are of the same order at first peri-center, which keeps the cold gas inside the SG due to the compression resulting from the two forces. Later on, the SG is quenched, and again the dominating force is ram-pressure.
We can also see that the dark matter component is lost due to tidal forces as expected. }
\label{fig:osq-sat_f_evo_22_00004}
\end{figure}

The ``Slow Quenching'', ``Dark-matter loss'' and ``eCg formation'' quenching case study shown in Fig. ~\ref{fig:osq-sat_f_evo_22_00004},
This SG shows a very different behaviour than before: at the first pericenter, the ram pressure and the tidal force acting on the SG have approximately the same magnitude and are smaller than the SG self-gravity. This means that the tidal force along the velocity of the SG acts against ram-pressure. Because that in the isothermal regime, the maximum tidal force outwards equals to the force inwards as the satellite velocity is perpendicular to the distance of the SG to the central galaxy. Hence the tidal force inward component is acting in the opposite direction to the ram-pressure.
This case causes compaction of the SG gas, enhancing star-formation and keeping the gas inside the SG.

As can be seen, this SG orbit is mostly circular so the SG continues to form stars. Looking at the apocenter one can see that even though the tidal force and ram-pressure are much smaller than the self-gravity, there is a loss of gas. This can be a result of star formation or of assuming only one spherical gas component, which is not the case of SG.
% can be seen in \ref{app:app01}.

Later in the SG evolution, a clear case of ram-pressure gas removal at the second peri-center occurs when ram-pressure becomes again the dominating force. We can further see a powerful tidal force at second and third peri-centers, which is responsible for stripping the stellar and dark matter components of the SG, leaving only the SG kernel after the third peri-center.

\begin{figure} \centering
\includegraphics[width=\linewidth]
{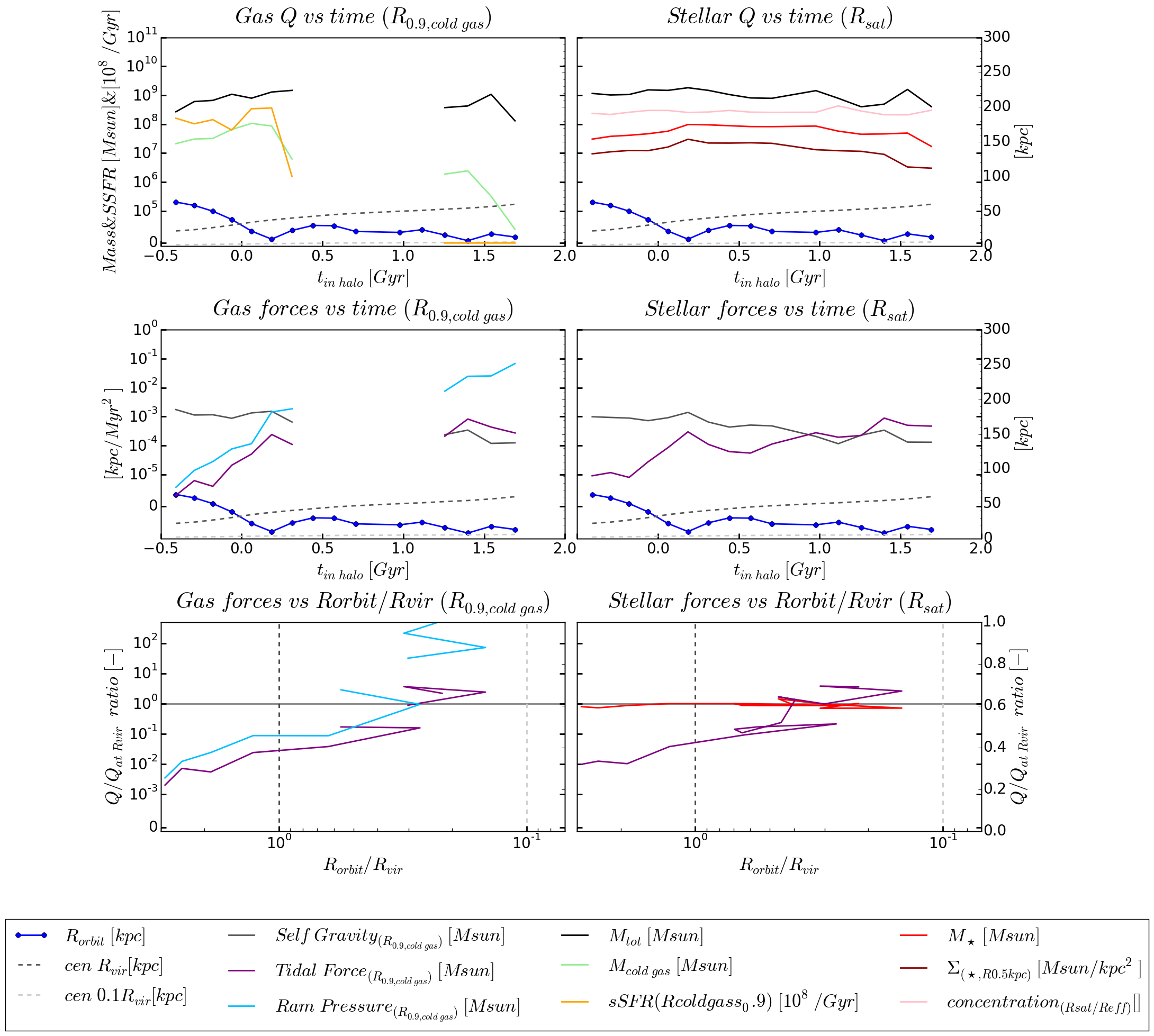}
\caption{\small \textbf{Satellite galaxy 23-007 forces evolution:} Similar structure and colors as in Fig.~\ref{fig:osq-sat_f_evo_07_00041}. \hspace{0.5cm}
This figure is completing fig~\ref{fig:sqm-sat_evo_23_00007} and
presents a ``Fast Quenching'' and ``Starvation'' case study.
As in Fig.~\ref{fig:osq-sat_f_evo_07_00041}, we can see a clean gas stripping at the first peri-center with ram-pressure being the dominating force. Moreover, we can observe at $t_{in\ halo}\approx1.3$ that the SG encounters cold gas in its orbit, but there is no gas accretion due to the ram-pressure as suggested in our ``Pac-Man'' model in \S\ref{sec:osq-sdqc}.}
\label{fig:osq-sat_f_evo_23_00007}
\end{figure}

The ``Fast Quenching'' and ``Starvation'' case study is shown in Fig. \ref{fig:osq-sat_f_evo_23_00007}. As in the first case study, we can see a clean gas stripping at first peri-center. The interesting part here is later on, at $t_{in\ halo}\approx1.3$, where the SG encounters cold gas. However, as the ram-pressure is higher than self-gravity, the SG does not accrete gas so there is no new gas in the SG and hence no star-formation. This illustrates the ``Pac-Man'' model discussed in \S\ref{sec:osq-sdqc}. 
In addition, we can see that the satellite puffing up stage is indeed due to the tidal force acting on the SG at the last peri-center.

\section{Force Evolution as a Function of Time} 
\label{sec:osq-f_time_evo}
This section continues the study of the combined SGs evolution by time at \S\ref{sec:osq-f_time_evo}. We will now focus on the forces acting on SGs as a function of time.

\begin{figure}
\centering
\includegraphics[width=\linewidth]{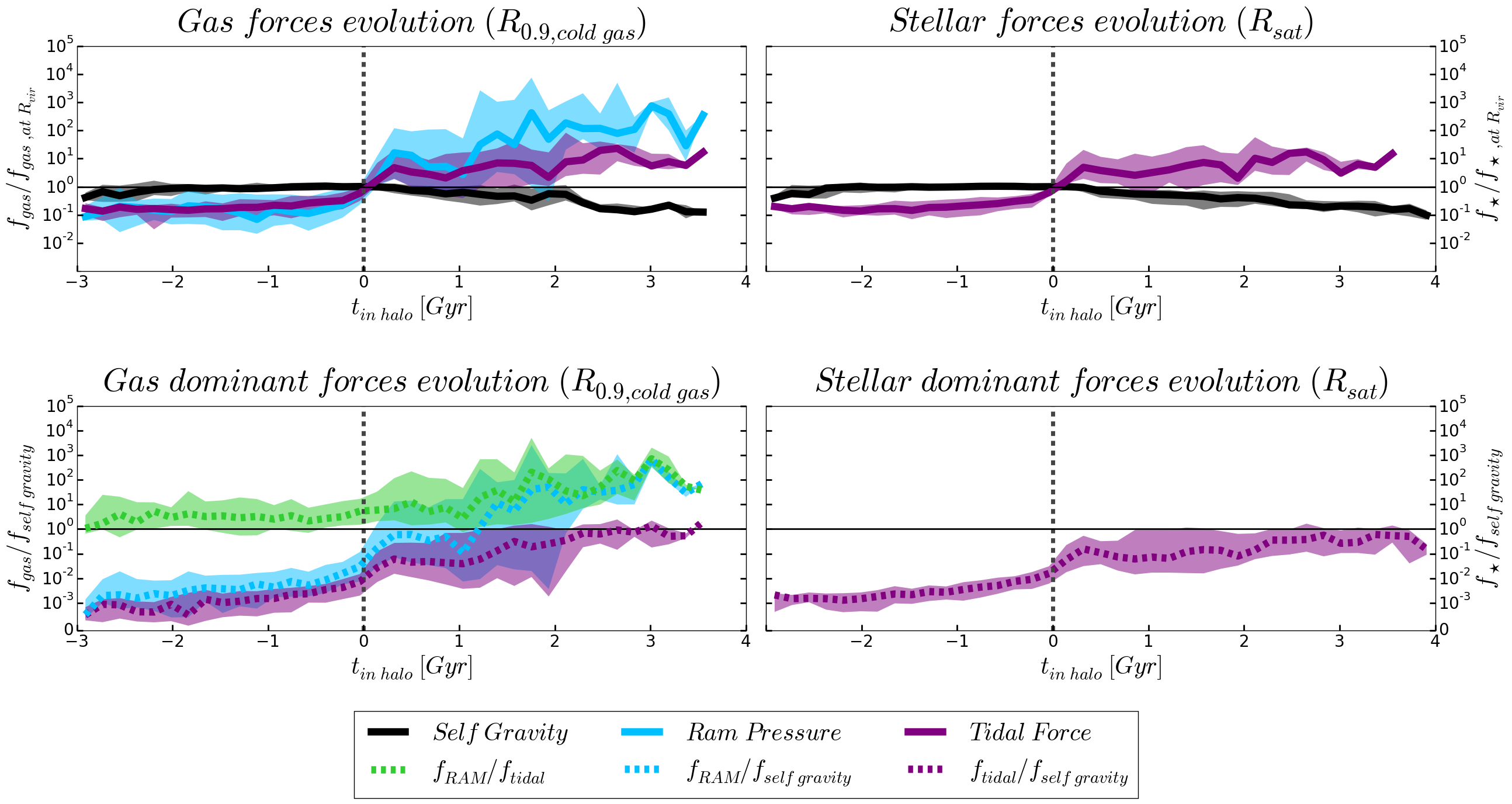}
	\caption{\small \textbf{Force evolution by $t_{\textrm in\ halo}$} - A stacked plot of the complete satellite sample by time in the halo.
	Bold lines represent the median while the upper and lower shade boundaries correspond to 16\% and 84\% of the sample.
	All the values are normalized to the value at entrance to the halo. 
	The left column describes the evolution of forces acting on the gas component and the right column describes the evolution of forces on stellar component. The upper row shows the evolution of the forces: ram-pressure (blue), tidal force (purple) and self-gravity (black). The lower row show in dashed lines the ratios between ram-pressure and the tidal force (green), between ram pressure and self-gravity(blue) and between the tidal force and self-gravity (purple)
	\hspace{0.5cm} This figure shows, following the green dashed line, that along time, ram-pressure is bigger than tidal force. Additionally, looking at the blue dashed line we can see that after a short delay in the halo, of about 0.5 Gyr, ram-pressure becomes the dominating mechanism in the SG evolution. We remind that the 0.5 Gyr is identified as the averaged peri-center point and therefore, the ram-pressure become dominating at the peri-center region. }
\label{fig:osq-sat_f_evo_t_in_halo}\end{figure}

The main result from the following figures is the apparent dominance of ram-pressure as the dominating force over tidal force while the SG is in the halo -- at peri-centers, apo-centers and between them. On the meaning of the activation of the ram-pressure, it is bigger than the self-gravity as expected, at the peri-center region but not much before.
In Fig.~\ref{fig:osq-sat_f_evo_t_in_halo} we identify the peri-center at 0.5 Gyr, as the tidal force and ram-pressure peak and as we know that the tidal force is most active at the minimum distance to the central galaxy. 

\begin{figure} \centering\includegraphics[width=\linewidth]{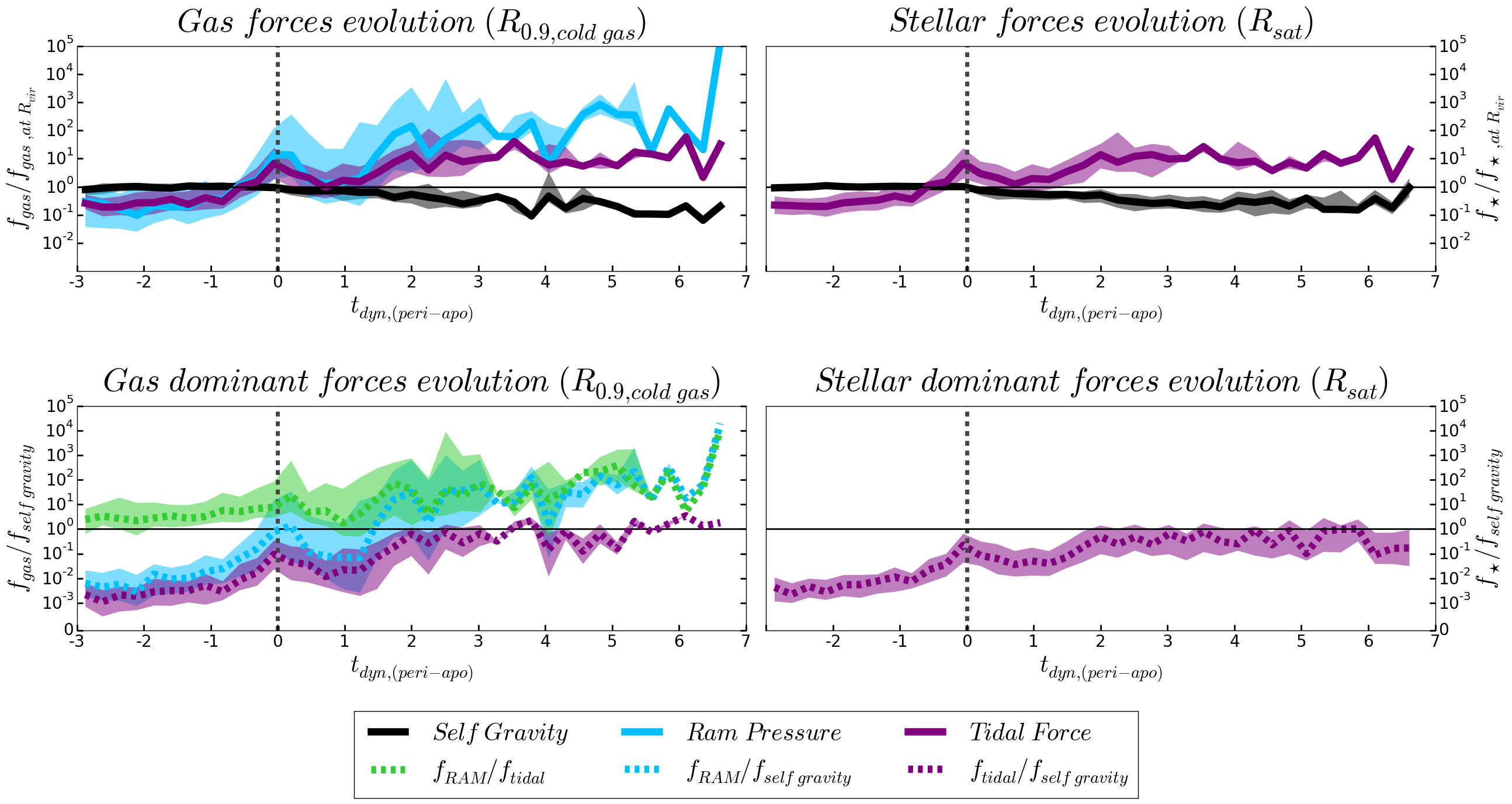} 
	\caption{\small \textbf{Force evolution by $t_{dyn\ (peri,\ apo)}$} - A stacked plot of the complete satellite sample by dynamical time as defined on \S\ref{sec:met-event_properties}. Colors and methods are similar to Fig.~\ref{fig:osq-sat_f_evo_t_in_halo}.
	Reminder: t=0 corresponds to the first peri-center and t=1 to the first apo-center so t=2 would be around the second peri-center an so on.
	\hspace{0.5cm} This figure shows the dominating region for quenching is the peri-center. A sharp peak in the forces is visible at the first and second peri-centers (t=0,2). Again, the green dashed line shows dominance of the ram-pressure at all times.}
\label{fig:osq-sat_f_evo_t_dyn_peri-apo}\end{figure}

Stacking as a function of $t_{dyn}$ as shown in Fig.~\ref{fig:osq-sat_f_evo_t_dyn_peri-apo} can organize our understanding better. We can notice that the peri-center to ram pressure correlation, which can be seen by the peaks of the ram-pressure that occurs exactly at the peri-centers ($t_dyn=0,2$). The ram-pressure correlation establishes our understanding that the ram-pressure controls the gas stripping mechanism. It is the dominant force relative to self-gravity and the tidal force in most of the events while at apo-centers the tidal force and ram pressure are roughly the same. Later we see that the tidal force rises relatively to self-gravity. However, the tidal force is always smaller or equal to the SG self-gravity as it is computed at the SG stellar component radius. The tidal force becomes significant around the second peri-center. This is probably due to tidal heating at the first peri-center, which lowers the potential well, as can be seen by the fact that self-gravity decreases over time, making gas and stars less bound over time. 

% Not sure if needed, I toke this out.
% * About starvation, as the tidal force is strong, gas cannot be accreted (this is without ram calculation) we can get from here an upper bound for gas accretion in the halo (less than 2 Gyr... from behind the satellite less then 0.5 Gyr from front and sides?)

% ------------------------------------------
\section{Force Evolution as a Function of distance to the central galaxy} 
\label{sec:osq-f_Rorbit_evo}
This section continues the study of the combined SGs evolution by distance to the central galaxy at \S\ref{sec:sqm-Rorbit-evo}. We will now focus on the forces acting on SGs as a function of distance.

\begin{figure} \centering\includegraphics[width=\linewidth]{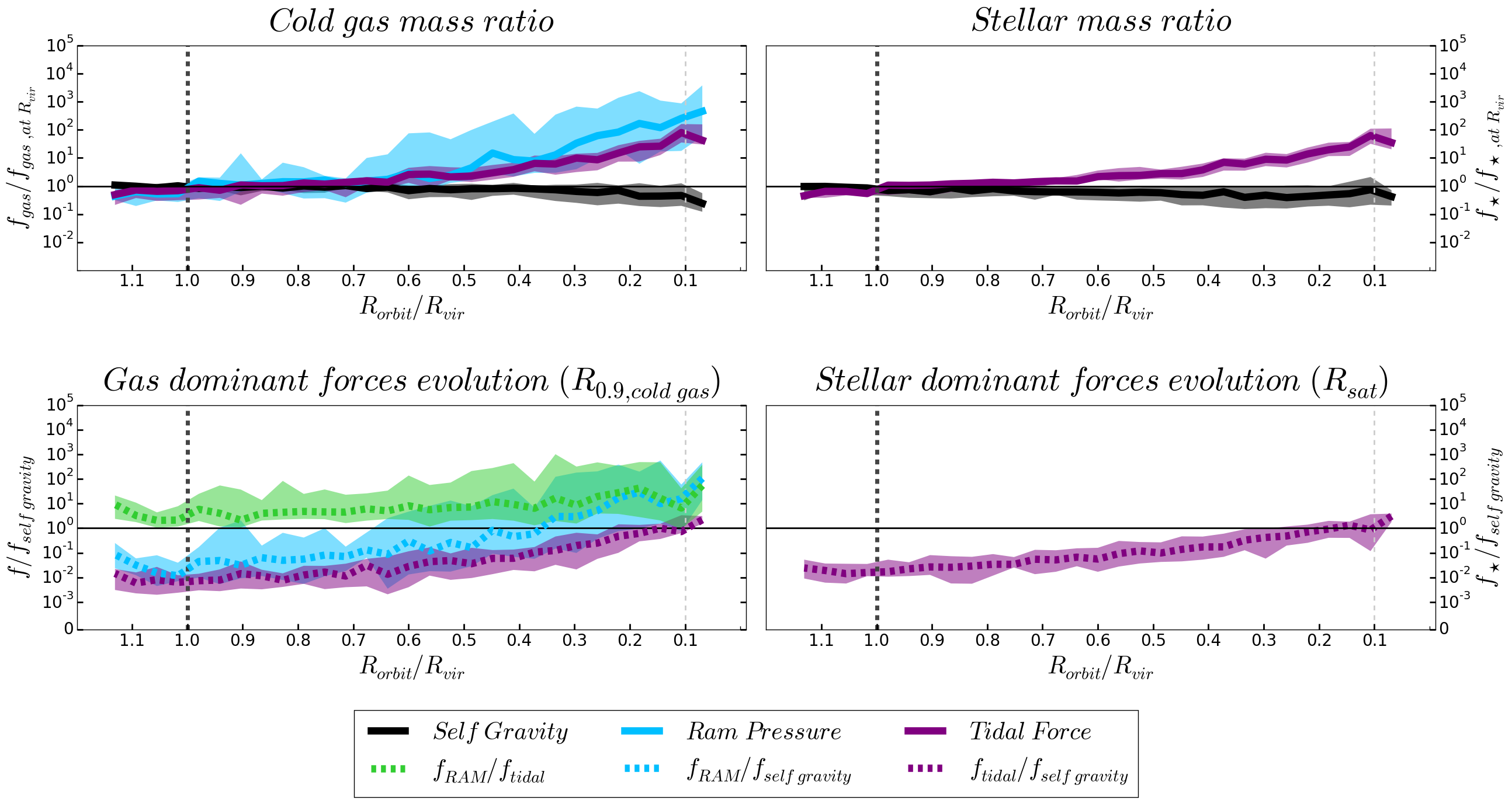}
\caption{\small \textbf{Force evolution by $R_{\textrm orbit}/R_{\textrm vir}$ } 
- A stacked plot of the complete satellite sample by $R_{\textrm orbit}/R_{\textrm vir}$. This is a stacked plot of the complete satellite sample. Colors and methods are similar to Fig.~\ref{fig:osq-sat_f_evo_t_in_halo}
\hspace{0.5cm} This figure shows, the closer the SG is to the central galaxy, the higher the tidal and ram-pressure forces are. ram-pressure is dominant also here.
Note the scatter, showing hidden parameter in the quenching process which is here the velocity of the SG.}
\label{fig:osq-sat_f_evo_Rvir}\end{figure}

Fig.~\ref{fig:osq-sat_f_evo_Rvir} presents the main feature: the closer the SG is to the central galaxy, the higher the tidal and ram-pressure forces are. We can see that ram-pressure is again always larger than the tidal force and becomes significant relatively to the self-gravity at about $0.3-0.4R_{\textrm vir}$. We also see that the scatter in ram pressure is higher, which happens mostly because ram-pressure depends on  velocity, indicating that a secondary parameter, the orbit eccentricity, is also essential for quenching. These conclusions are in agreement with the analytic study presented in \S\ref{sec:osq-analytical}.

\begin{figure} \centering\includegraphics[width=\linewidth]{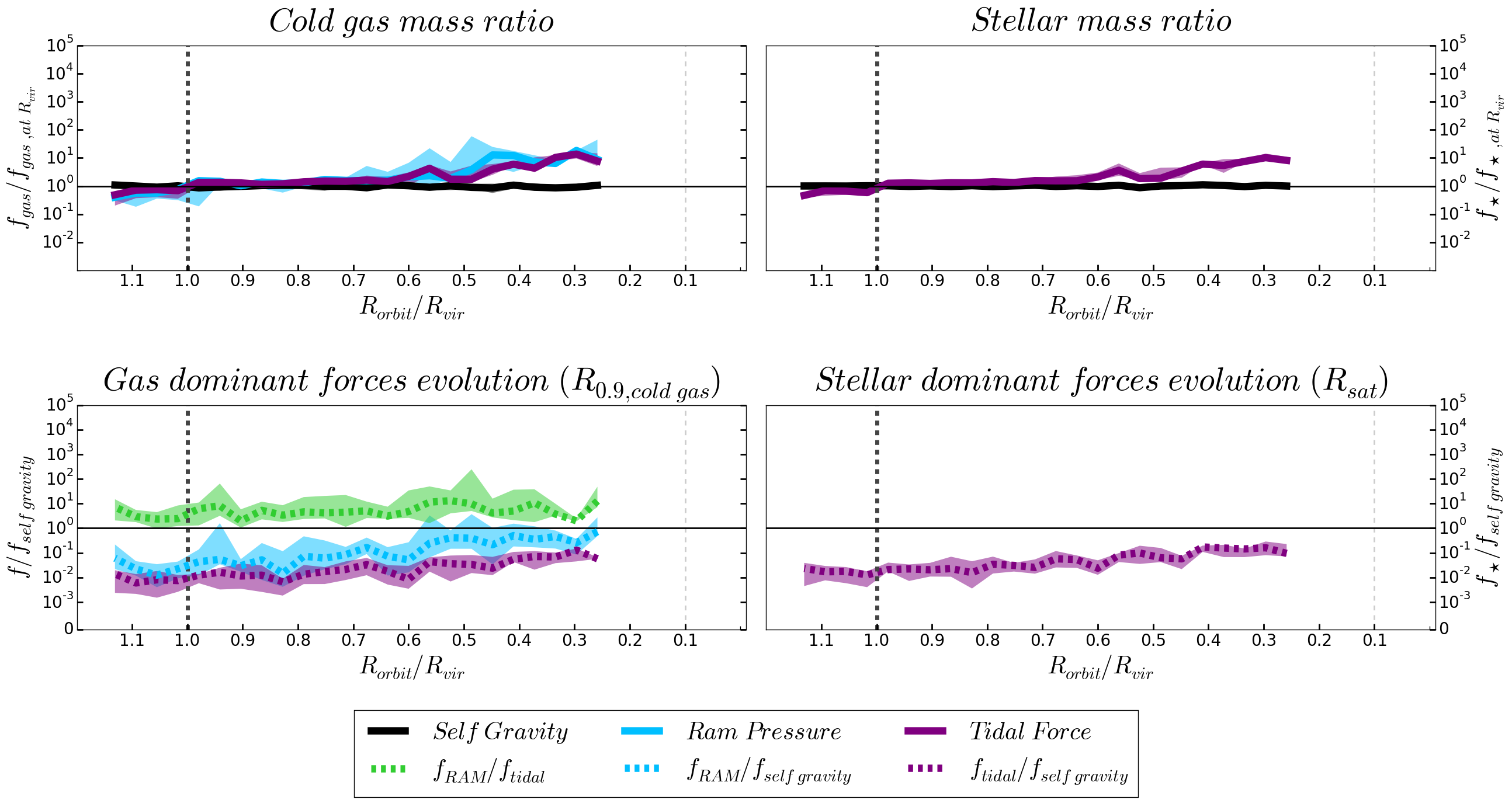}
\caption{\small \textbf{Force evolution by $R_{\textrm orbit}/R_{\textrm vir}$} until first peri-center - A stacked plot of the complete satellite sample by $R_{\textrm orbit}/R_{\textrm vir}$.
Colors and methods are similar to Fig. ~\ref{fig:osq-sat_f_evo_t_in_halo}
\hspace{0.5cm} The forces here are at most equals to the self-gravity, showing that distance from the center alone is not enough to produce high ram-pressure to quench a SG.
Another component is needed which is the velocity.
}
\label{fig:osq-sat_f_evo_Rvir_until_1peri}\end{figure}

Moreover, we can see in Fig.~\ref{fig:osq-sat_f_evo_Rvir_until_1peri} that until the first peri-center, the forces grow bigger but are still smaller than self-gravity, this time with a smaller scatter in ram-pressure, indicating that the velocities and the peri-center itself are essential to the quenching process, as galaxy can be at the same distance from the center without quenching as seen here. Another important point when comparing the plots are the regions where the tidal force is significant: before the first peri-center, the tidal force is smaller than self-gravity at all distances, while in Fig.~\ref{fig:osq-sat_f_evo_Rvir} we can notice that the tidal force is stronger below $0.2R_{\textrm vir}$, showing that SGs suffer more from tidal effects of heating and stellar stripping at later time. 

\section{Starvation}
\label{sec:osq-starv}
Gas starvation, that is a stop of gas accretion onto the SG, can be an important quenching mechanism. In case gas accretion resumes at any moment, the new gas can reignite star formation. \citeA{vandeVoort2016} show in their cosmological SPH simulation that there is no gas accretion to the SGs. We have seen in \S\ref{sec:sqm-time-evo} and \S\ref{sec:sqm-Rorbit-evo} that there was no measured gas addition in the VELA SGs. We will now show this result is robust, and that SGs starvation is an active process throughout the lifetime of SGs in the halo.

Let us take a look on the case where a halo SG meets cold gas -- akin to a Pac-Man eating points.

``Pac-Man'' accretion:
In order to accrete gas from the halo the gas velocity relative to the SG should be lower then the escape velocity of the SG. Assuming the gas in the halo is at rest, while the SG moves in its orbital velocity. We will get:

\begin{align}
\sqrt{\frac{GM_{\textrm halo}}{R_{\textrm halo}}} = v_{\textrm orbit} &<  v_{\textrm esc} = \sqrt{\frac{GM_{\textrm{satellite}}}{r_{\textrm acc}}} \Longrightarrow 1 <  \frac{M_{\textrm satellite}}{M_{\textrm halo}}\cdot \frac{R_{\textrm halo}}{r_{\textrm acc}} \propto \frac{r_{\textrm acc}^2}{R_{\textrm halo}^2}
\label{eq:osq-packman}
\end{align}
\noindent Where the first step uses the virial theorem, $V^2=\frac{GM}{R}$.
Also, the last step uses the $M_h \propto R_h^3$ relation.

Since $r_{\textrm acc} \ll R_{\textrm halo}$, it is hard or even impossible to accrete gas in this way: this kind of accretion is irrelevant.

The fact that  a SG can not accrete gas by encounter in the halo means that in order for a SG to accrete gas, the gas has to move with the SG.
The case occurs only when a SG enters the halo inside a cold gas stream from a filament to the central galaxy, which leads to the following mechanism.

``Wrapped Pac-Man'' accretion:
We assume that a SG accretes all the gas in its gravitational potential well. The well of a satellite inside a halo can be described by the tidal radius, which is the radius at which the self-gravity of the satellite is equal to the gravity of the halo: 
\begin{align}
r_{\textrm{tidal}} = \Big( \frac{1}{2} \frac{M_\textrm{s}}{M_\textrm{h}} \Big)^{\nicefrac{1}{3}} R_{\textrm orbit} = \frac{M_\textrm{s}}{\rho_{\textrm{all, halo}}}^{\nicefrac{1}{3}}.
\end{align}

Therefore, a spherical accretion gives us:
\begin{align}
M_{\textrm acc} &= \frac{4\pi}{3} r_{\textrm{tidal}}^3 \cdot \rho_{\textrm{gas, filament}} =  \frac{2\pi}{3} M_s \cdot \frac{\rho_{\textrm{gas, filament}}}{\rho_{\textrm{all, halo}}} \nonumber \\
\frac{M_{\textrm acc}}{M_{\textrm s}} &= \frac{2\pi}{3} \cdot \frac{\rho_{\textrm{gas, filament}}}{\rho_{\textrm{all, halo}}}
\label{eq:osq-Stream} \end{align}

In The VELA suite, $\rho_{\mathrm all, halo}$ is ranging between $10^{5.3} to 10^{7.5}$ with a mean value of $10^{6\pm 0.5} \nicefrac{M_{\odot}}{{kpc}^3}$
[by private correspondence with Dr. Nicolas Cornuault] so
 $\rho_{\mathrm gas, filament}\sim 10^3 \ \frac{M_{\odot}}{{kpc}^3}$
In those cases, $\frac{M_{\mathrm acc}}{M_s} \approx 10^{-3}$!
This is a negligible amount since gas is around $2\%$ of the mass of a normal galaxy.
Please notice that in this approximation, we did not include the fast decrease of the tidal radius, which in fact will leave most of the surrounding gas outside of the satellite (as we saw in our examples, the tidal radius is becoming smaller than the stellar radius after the first peri-center).
We thus conclude that gas accretion inside the halo for a SG is an extremely rare case.

\section{Sub-Dominant Quenching Causes}
\label{sec:osq-sdqc}
SGs are affected by many mechanisms. We should ask what is the influence on quenching of those mechanisms and what are their evolution compared to tidal forces and ram pressure.

First, environmental gas flow forces other than ram-pressure.
It has been shown by \citeA{Fillingham2016, Zinger2018} that  Kelvin-Helmholtz instability, shock, and gas turbulence can represent up to 50\% of the original ram-pressure in SGs. 
These additional changes to the ram-pressure are not necessary relevent in the regime studied here since we approximate a ``JellyFish''-shaped SG as spherical when computing the force. These changes introduce a small correction to the ram-pressure, which we address as a constant times the ram-pressure. This of course does not change the force comparison much. 

Feedback (gas outflows due to radiation, SNs, and others) has a crucial role in galaxy formation. However, we find its role less significant for SG evolution.  As mentioned before, the VELA suit has lower feedback than other cosmological simulations. In the case of higher feedback, the main change in the satellite evolution will probably be a shallower density profile profile, the gas content of the satellite being both puffed up and lowered. In this case where $r_{\textrm g}$ is larger and $\rho_{\textrm{cold\ gas, satellite}}$ smaller, our ram-pressure calculation induces a stronger ram-pressure. Bus as the primary mechanism of quenching is the ram pressure peak at the peri-center, the results shouldn't be that different. If there was a change, it would be through some earlier quenching just before the peri-center. In some sense, this mechanism is similar to the first environmental gas flow dynamics and we can take it into account as a factor times the ram-pressure. This result is consistent with \cite{Emerick2016}.

For gas depletion via star-formation, we can see that it does not happen  within our samples. The initial gas mass does not significantly turn into a young stellar mass. Moreover, the decrease of gas mass is correlated to the ram-pressure force.

The last additional mechanism is the effect of the magnetic fields, which we address for completion. It has been shown that galaxies have magnetic fields around them. These magnetic fields mostly influence small plasma regions, causing turbulence in them. Hence, for a SG to be affected by magnetic force, it should be very close to the central galaxy, and it will result in small turbulences which will change the gas in the SG in a very negligible way.

\section{Comparison to Observations}
\label{sec:osq-obs}
\begin{figure} \centering\includegraphics[width=\linewidth]{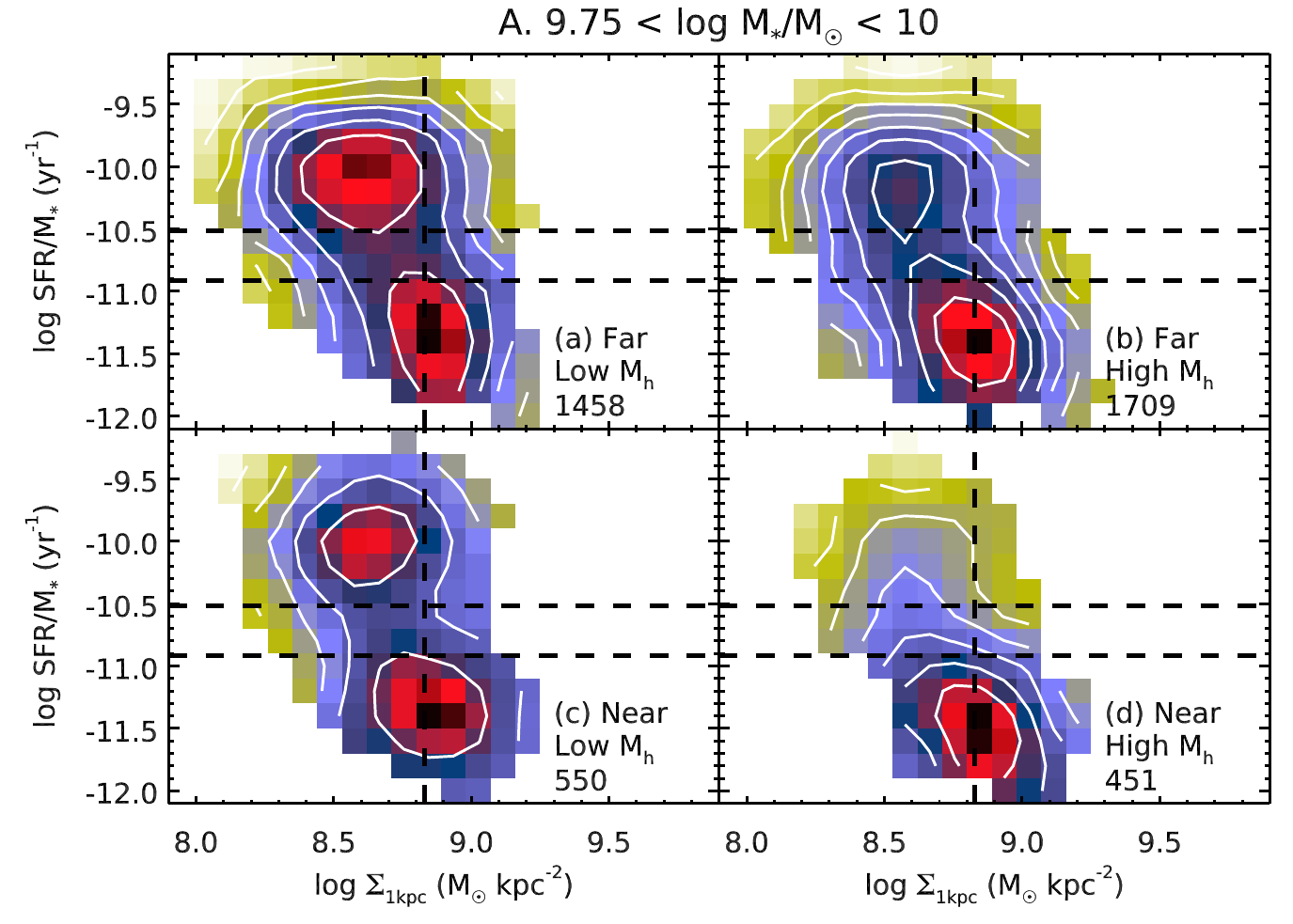} 
\caption{\small \textbf{The distribution of satellite galaxies in the sSFR-$\Sigma_{\star,\ \mathrm{1kpc}}$ plane in four panels} The panels representing different environments, for two mass bins (A and B).  ``Near'' and ``Far'' refer to log $R_{\mathrm{proj}}$ lesser or greater than -0.6 (or $R_{\mathrm{proj}} \sim 0.25 R_{\mathrm{vir}}$).  ``Low $M_{\mathrm{h}}$'' and ``High $M_{\mathrm{h}}$'' refer to haloes less or more massive than $10^{13.5}\ \mathrm{M}_{odot}$. 
The contours represent the logarithmic number density of the population in each panel and are separated by 0.2 dex. The color scale is normalized such that dark red represents the highest number density in that panel.
The sample size of each panel is indicated in the bottom corner.  The shift of the GV bridge, the lowering of sSFR for the SF satellites and the extension of quenched satellites toward lower $\Sigma_{\star,\ \mathrm{1kpc}}$ compared with the field is driven by high $M_{\mathrm{h}}$ and low $R_{\mathrm{proj}}$. \hspace{0.5cm} By \protect\citeA{Wuyts2011}, we can see the quenched SG fraction change depending on the halo mass scale and the distance to the central galaxy scale.}
\label{fig:osq-quenching_by_halo_and_distance}
\end{figure}

We have shown the significance of the SG distance to the central galaxy: the closer a SG is to the central galaxy, the stronger ram-pressure is. This means that the probability of a SG to be quenched should be correlated to its distance to the central galaxy. Note in this regard that the bias of dense old SG presented in \S\ref{sec:sqm-observations} might add an equal number of quenched SG independent of the SG position, since there are SG with apo-centers above $0.6R_{\textrm vir}$ in our sample. We nevertheless expect to observe a higher fraction of quenched SG closer to the central galaxy. As shown in Fig.~\ref{fig:osq-quenching_by_halo_and_distance} and by \cite{Woo2017, Wang2018a, Wang2018b}, the percentage of quenched SG decreases with distance.

Another important observed phenomena is the dependence of quenched SG fraction to the halo mass and SG mass.
We have shown in \S\ref{sec:osq-analytical} that there is a strong connection between ram pressure and the halo to satellite mass ratio, meaning that at fixed SG size, the bigger the halo is, the stronger the ram-pressure will be.  Therefore, a higher quenched fraction of SGs is expected to be observed on high mass haloes. Following \shortciteA{Woo2012, Balogh2016, Woo2017, Wang2018a, Wang2018b} and as shown in Fig.~\ref{fig:osq-quenching_by_halo_and_distance}), it is indeed the case. 
The SG mass dependence is harder to show as a specific analysis should be included to deal with the dense SG survivor population discussed in \S\ref{sec:sqm-observations}. 
Note also that we did not have statistics on the ratio itself and therefore could not give a direct proof of the ratio dependence.

The last factor to discuss is the decrease of the gas fraction $f_{gas}$ with time. This is a delicate subject as the hot gas in the halo can affect the quenching process. We want to note that as the gas fraction in the halo drops with time \cite{Genel2014}, and for molecular gas as notably observed by \cite{Tacconi2013, Tacconi2018, Spilker2018} , which is also reproduced in the VELA simulations, the ram-pressure strength should also decrease by time. Therefore, at later times, quenching can be slower or require more than one peri-center. This could explain the evolution of the quenching timescales shown by \cite{Guo2017} and the quenching through starvation at higher masses discussed by \cite{Fillingham2015} (since it takes more time to quench SG at lower redshift).

\let\textcircled=\pgftextcircled

\chapter{Summary and Conclusions}
\initial{W}e have presented a full description of satellite galaxy evolution track, summarized at fig~\ref{fig:Median_cartoon}.
We find that SG quenching evolves very differently then filed galaxies, through a typical path in the diagram of specific star formation (sSFR) vs. inner stellar surface density ($\Sigma_{\star, 0.5\textrm kpc}$), 
with the inner surface density defined as the density within 0.5kpc from the center of mass of the SG. 
\textbf{Three discrete phases characterize this path: 
1. Halt in gas accretion with SG compaction at high sSFR as the SG keeps forming stars 
2. Gas removal and rapid drop in the sSFR at the peri-center of its orbit within the host halo
3.  Stellar heating and stripping that may lead to coalesce with the halo center, an ultra-diffuse galaxy (UDG), a compact elliptical galaxy (eCg) or a globular cluster (GC).}
These results are in contrast to previous articles which suggested a very different mechanism \cite{Woo2017}. However, we show, that by the same observation we can see a better agreement to the proposed evolution path.

The satellite evolution track was measured on satellite galaxies at the VELA cosmological simulation at z<1. However, by analytic analysis and by observations, we suggest it describes the \textbf{broad SG quenching mechanism}, until z=0 and to the extent of at least galaxy groups.
Hence, providing a completing term to the halo mass quenching mechanism and providing a \textbf{full comprehensive description of the galaxy quenching mechanism} which can be seen at fig~\ref{fig:satellits_vs_field_quenching}.

we show that the SG evolution main drivers are ram-pressure stripping at the initial stages and tidal forces at the final stages, aided by starvation, suppression of gas accretion. Other processes: gas depletion by star formation or stellar feedback are secondary. These results are consistent with an analytic model we have developed that successfully reproduces this SG evolution.

We find there is great importance to the SG orbit in order for a galaxy to quench, and it depends on the eccentricity, peri-center. The quenching surprisingly does not depends on the stellar mass of the SG as was suggested before but instead depends on the ratio of stellar SG mass to the halo mass. Another change that was presented by \cite{Balogh2016} is a change in the mechanism around z~1. We suggest that for a quenching process depending on the fraction of gas in the halo, a quantity which decreases by time, we resolve this observed change. We also show that this halo gas fraction dependence driven from our calculation.
These parameters change the ratio of ram-pressure to self-gravity magnitudes, as ram-pressure is the dominating quenching cause, These parameters control when and where the quenching would happen.\\
Another important mechanism is starvation, a halt in gas accretion to SGs. We show it is improbable for a SG, or dark-matter sub-halo to accrete gas at the halo. Starvation can even reach farther than the virial radius, which can also imply to conformity issues. This result is beneficial to understand the SG evolution, but even more generally, it means dark-matter sub-halos which did not accrete gas earlier to entering the halo, cannot accrete gas and later form stars in the halo, therefore lowering the missing satellite problem, as some sub-halos would never accrete gas.

Tidal force is also an important mechanism. We find the orbital eccentricity relates to the 
``puffing up'' of the SG. High eccentric orbits correlated with tidal heating, a process in which the inner stellar surface density decrease. We find this change happens at peri-centers, with some delay of relaxation of the SG, no change in stellar mass is visible until first peri-center, and between apo-centers to the next pericenter. Except for tidal heating, we also see tidal stripping, a process in which a material stripped out of the SG due to tidal forces. We find this force extremely efficient in stripping dark-matter and less to the stellar matter, even though they are roughly at the same mass resolution in the simulation. We explain this phenomenon, by the fact that stars formed from a gas that has been cooled down in the center of the halo. Therefore stars which formed in the halo, are on a lower energetic level compared to the dark matter particles which have a higher energy level. This claim supported by \cite{Chang2013, Macci2017} that showed a simulated model to show these dynamics and is first time seen in a cosmological simulation on this MSc thesis. This scenario, suggests, that higher eccentric SG are more likely to lose their dark matter component. We found at least 10\% of our SG population to loses its dark matter components. Moreover, this losing of dark matter can explain the observed lacking dark matter ultra diffused galaxy (UDG), as UDGs in the halo are formed through an eccentric orbit, causing them due to tidal heating to become diffused, therefore, we would expect UDG to with less dark matter than other galaxies.

The lost of the dark matter component also explains the difficulties in the SGs simulation study. Today, many of the studies use dark-matter follow up in order to resolve SG, as the dark matter vanishes quickly, it is preferred to use stellar tracking. However, as we showed, a SG can lose more than 90\% of its mass to the central galaxy between two snapshots, therefore even with a classical stellar tracking method, one would not resolve the right orbit and consider effect as a merger. The tracking fly-bys is a systemic problem in nowadays merger tree algorithm as described at \cite{Srisawat2013}. We propose a different method which traces the potential well position of the galaxies by tracing the kernel of the galaxies, the densest star region in the galaxy and later rebuild the galaxy from it, this method with other upgrades which solves other SG systematic issues, help us resolve the total SG evolution path and deliver the results above.

Another product of this new merger tree is the finding of the formation of an elliptical compact galaxy (eCg),  from the formation at a filament, through the formation of stars in the core of the galaxy, until an eCg flying around the central galaxy. We suggest that the same mechanism as here is responsible also for globular clusters (GC) as the kernel, central part of the SG detach from the galaxy and continues to fly around the central galaxy.
This case as the two former cases, losing dark matter and UDG are systematically vulnerable by the currently used merger trees, and a solution like the proposed merger tree algorithm is needed.

We wish to emphasize the importance to compare good observations with our model. We made an effort to compare the results, but some correlations should explicitly be tested: For example, the difference of quenching by satellite to halo mass ratio or a broader analysis of SG densities, age, orbit velocity and peri-center for different masses. We also provided SG quantities by time and distance with to be compared at later studies.
An urgent call is about the measurements of dark matter at SG, as we have shown, from the tidal dynamics, SG has no NFW profile. There are observed SG with small velocity dispersion, but a further study is needed to confirm or contradict the question of dark matter in satellite dwarf galaxy, an important question which related to the dark matter fundamental characteristics.

This study raises new questions:
First, we see an existent of a bi-modal quenching process: fast quenching at the first pericenter and slow quenching on the second pericenter, as we have shown, it is the tidal force and the circular orbit that holds the cold gas in the SG for a longer time. We tried to distinguish between them by a simple role, and we are on an ongoing study in this field to find the different causes to the bi-modality in SG quenching. It is our notion that an upgrade to the quenching model is needed. Most of the studies on SG assumes quasi-static quenching process, while the quenching can be very rapid at the pericenter. We are developing a pulse model to get a more accurate measurement.\\
Second, we have seen in our SG sample, and by other articles, that a unique star-burst is occurring for some of the SG at its pericenters, we describe this event as the ``Swansong'', as the SG in a very rapid event forming a last burst of stars and later quenches. Our model shows that this is an effect of tidal force vs. ram-pressure compression at the pericenter, a further study also awaits here.\\
Third, is the question of UDG, eCg and other formation in SG evolution, we would like to extend our study also here especially as VELA is a very high-resolution simulation and the only one currently which resolves eCg and other compact galaxies and has high resolution also for the UDG.\\
Forth, at the introduction, we have discussed the importance of SG to quenching of the central galaxy of the SG, we feel a further study is needed to explain why and which SG quench the central galaxies. Dry or wet mergers? Circular or eccentric? Each one of these questions raises its limitations, which we will address at a proper article.\\
Fifth, as we presented a SG with no dark matter, a question is raising, how much dark matter less, baryon galaxies (BG) are identified in the VELA simulation, how are they form? How do they evolve? We find six different mechanisms to form staying alive BGs. These results too should be addressed.

All of the data above and more are ordered and used in our group in a fast, robust catalog system using the pandas python library for data science. 
This extensive catalog is a new infrastructure for further studies on the VELA and other simulations.
\\
\\
\\
\appendix
% \import{chapters/appendices/}{app0A.tex}
\clearemptydoublepage
%
% Apparently the guidelines don't say anything about citations or
% bibliography styles so I guess we can use anything.
\backmatter

\bibliographystyle{apacite}
\refstepcounter{chapter}
\bibliography{thesis}
\clearemptydoublepage
%
% Add index
%\printindex
%   
\end{document}